\documentclass[11pt]{article}
\usepackage{graphicx}
\usepackage{epsfig}
\usepackage{hyperref}
\usepackage[usenames,dvipsnames]{color}
\usepackage{wrapfig}
\def\lsim{\raise0.3ex\hbox{$<$\kern-0.75em\raise-1.1ex\hbox{$\sim$}}}
\def\gsim{\raise0.3ex\hbox{$>$\kern-0.75em\raise-1.1ex\hbox{$\sim$}}}

\def\mean#1{\left<#1\right>}
\def\Journal#1#2#3#4{{#1}{\bf #2} (#4) #3}

\def\EPJC{{Eur. Phys. J. C}}
\def\JPG{{J. Phys. G}}
\def\JPCS{{J. Phys: Conf. Series\ }}

\def\NIMA{{Nucl. Instrum. Methods A}}
\def\NPA{{Nucl. Phys. A}}

\def\PLB{{Phys. Lett. B}}
\def\PLC{Phys. Repts.\ }

\def\PRL{Phys. Rev. Lett.\ }
\def\PRD{{Phys. Rev. D}}
\def\PRC{{Phys. Rev. C}}

\def\ARNPS{{Ann. Rev. Nucl. Part. Sci.\ }}

\def\RPP{Rep. Prog. Phys.\ }
\def\QGP{{\color{Red} Q}{\color{Blue} G}{\color{Green} P}} 
\def\QCD{{\color{Red} Q}{\color{Green} C}{\color{Blue} D}} 

\normalsize
\begin{document}
\title{Highlights from BNL-RHIC}
\author{M.~J.~Tannenbaum 
\thanks{Research supported by U.S. Department of Energy, DE-AC02-98CH10886.}
\\ Physics Department, 510c,\\
Brookhaven National Laboratory,\\
Upton, NY 11973-5000, USA\\
mjt@bnl.gov}\maketitle
\maketitle
\thispagestyle{empty}

\section{Introduction}\label{sec:introduction}
High energy nucleus-nucleus collisions provide the means of creating nuclear matter in conditions of extreme temperature and density~\cite{BearMountain,seeMJTROP,MJTROP}.  
 The kinetic energy of the incident projectiles would be dissipated in the large 
volume of nuclear matter involved in the reaction.  At large energy or baryon densities, a phase transition is expected from a state of nucleons containing confined quarks and gluons to a state of ``deconfined'' (from their individual nucleons) quarks and gluons, in chemical and thermal equilibrium, covering a volume that is many units of the confining length scale. This state of nuclear matter was originally given the name Quark Gluon Plasma (\QGP)~\cite{Shuryak80}, a plasma being an ionized gas. 

\begin{figure}[!htb]
\begin{center}
\includegraphics[width=0.5\textwidth]{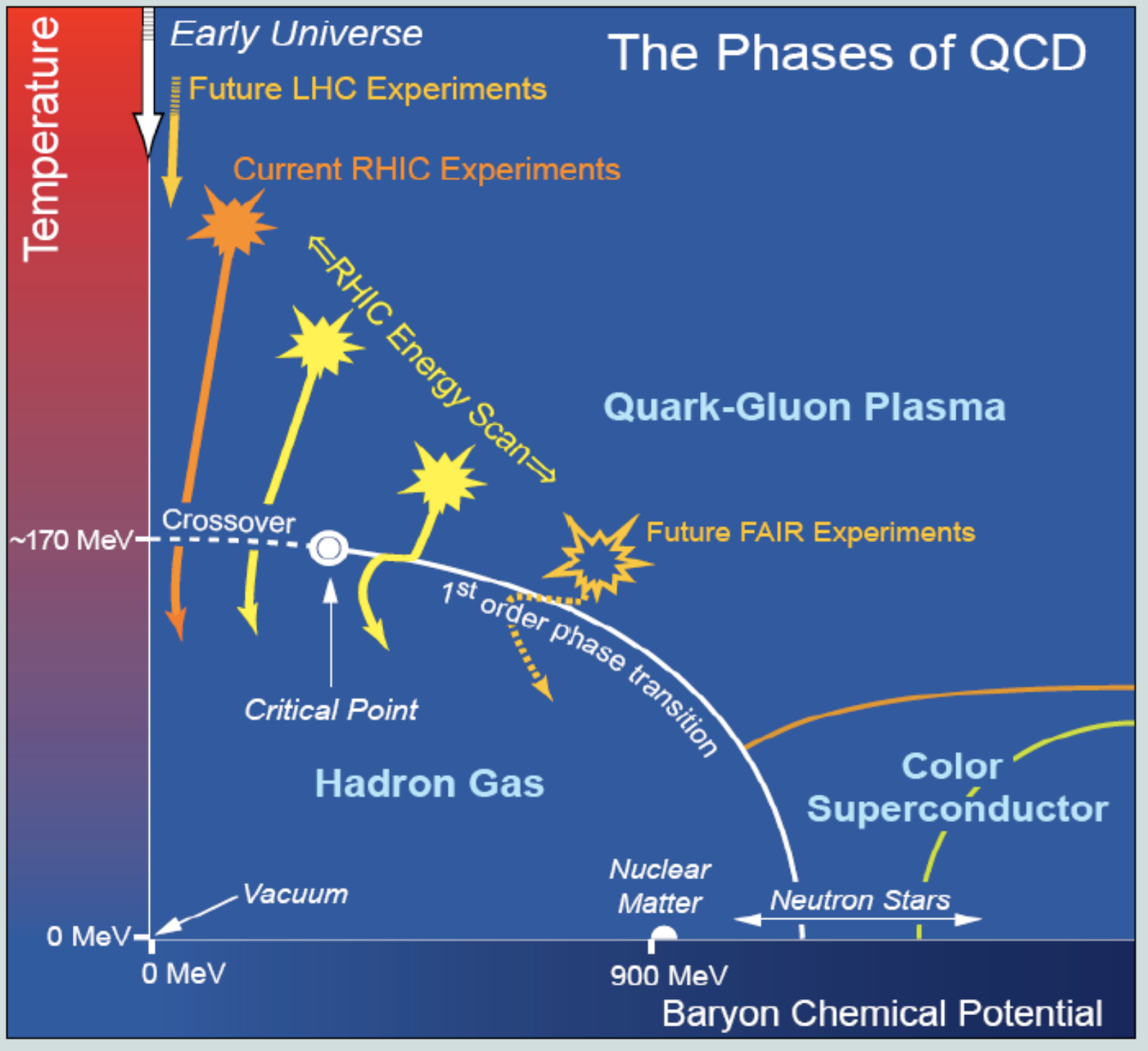}
\end{center}\vspace*{-1.5pc}
\caption[]{(left) A proposed phase diagram for nuclear matter~\cite{LRP07}: Temperature,  $T$,  vs Baryon Chemical Potential, $\mu_B$. \label{fig:phase_boundary}}
\end{figure}

 A typical proposed phase diagram of nuclear matter~\cite{LRP07} is shown in Fig.~\ref{fig:phase_boundary} together with 
the presumed trajectories of the evolution of the medium for collisions at RHIC and LHC c.m. energies, where the axes are the temperature $T$ vs. the baryon chemical potential $\mu_B$. The temperature for the transition from the Quark Gluon Plasma (\QGP) to a hadron gas is taken as 170 MeV for $\mu_B=0$ and the phase boundary is predicted to be a smooth crossover down to a critical point below which the phase boundary becomes a first order phase transition. 
Also shown are idealized trajectories for the RHIC c.m. energy scan and future experiments at FAIR which are being performed in order to find the \QCD\ critical point.

\section{Discovery of the \QGP}
The \QGP\ was discovered at RHIC, and announced on April 19, 2005. 
However the results at RHIC~\cite{seeMJTROP} indicated that instead of behaving like a gas of free quarks and gluons, the matter created in heavy ion collisions at nucleon-nucleon c.m. energy $\sqrt{s_{NN}}=200$ GeV appears to be more like a {\em liquid}. This matter interacts much more strongly than originally expected, as elaborated in peer reviewed articles by the 4 RHIC experiments~\cite{BRWP,PHWP,STWP,PXWP}, which inspired the theorists~\cite{THWPS} to give it the new name ``s\QGP" (strongly interacting \QGP). These properties were quite different from the ``new state of matter'' claimed in a press-conference~\cite{CERNBaloney} by the CERN fixed target heavy ion program on February 10, 2000, which was neither peer-reviewed nor published. 

In spite of not being published, the CERN press-release had a major effect on the press in the United States resulting in an article on the front page of the New York Times~\cite{NYT02102000}. Ironically, on this very same front page was an article announcing that the true version of the famous Italian sausage, Mortadella, would, for the first time, be allowed to be imported into the United States. A photograph of the iconic Bologna sausages appeared right next to the article about the CERN ``qgp''. Unfortunately, the first European Baloney to arrive in the U.~S. was the CERN announcement.~\footnote{It is important for the reader of these proceedings to be aware that a high official of CERN was in the audience during this talk and made no objection to this comment. Furthermore, the author has a long, positive and productive relationship with this great laboratory and has praised its many successes. Thus he feels justified in commenting on one of their rare misjudgments. See reference~\cite{MJTROP} for a detailed scientific discussion. } 

\section{What's new at BNL and the RHIC machine this year.}
\begin{figure}[!h]
\begin{center}
\includegraphics[width=0.70\textwidth]{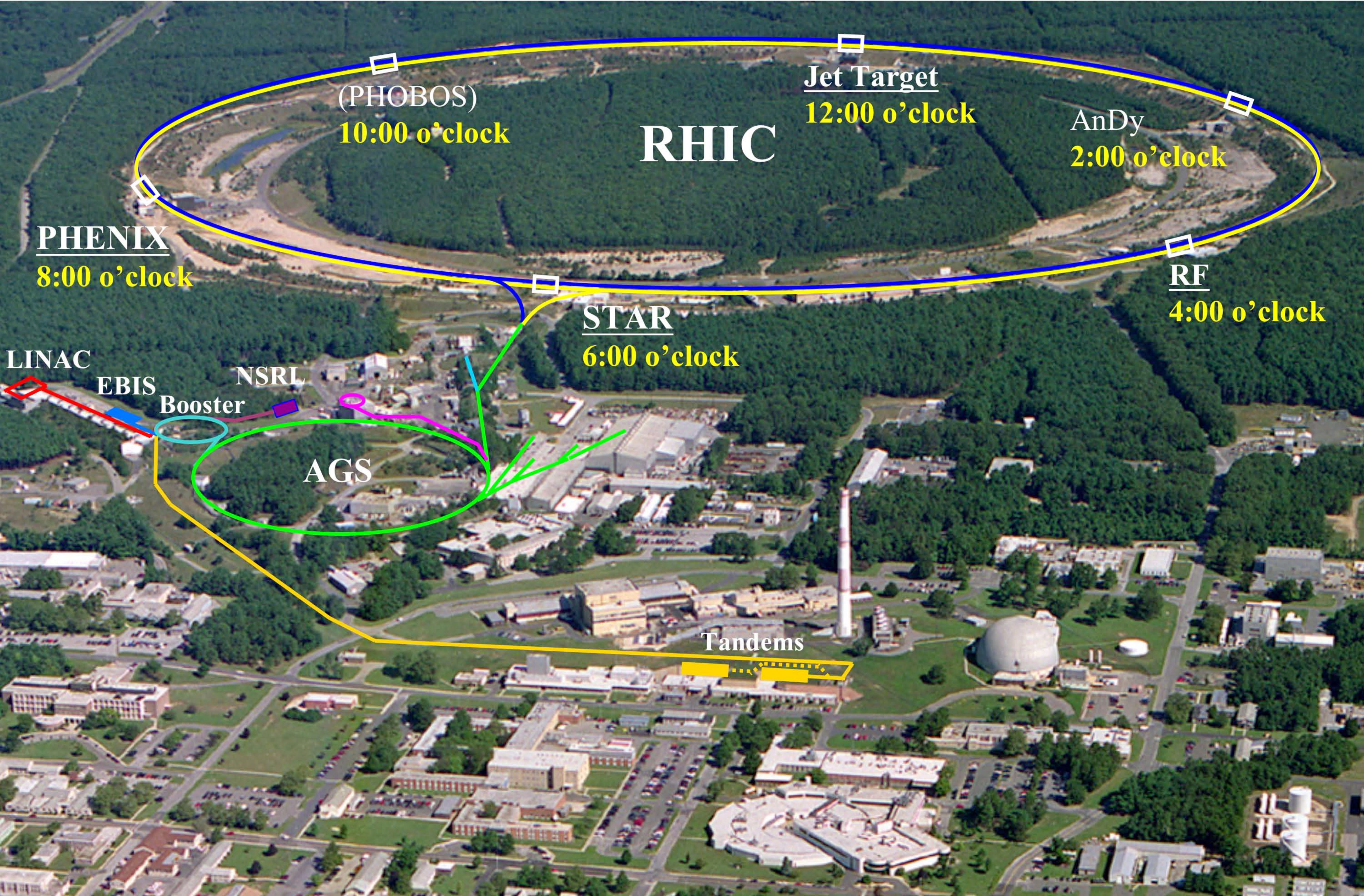}
\end{center}
\caption[]{Aerial view of RHIC/AGS facility. The two principal experiments still running are PHENIX and STAR. The LINAC is the injector for polarized protons. The TANDEM injector for Ions will be replaced by the Electon Beam Ion Source (EBIS) in fall 2011. The dome is the decommissioned High Flux Beam Reactor while the enclosed ring on the lower right center is the National Synchrotron Light Source (NSLS).}
\label{fig:RHICoverview}
\end{figure}

\begin{figure}[!h]
\begin{center}
\includegraphics[width=0.47\textwidth]{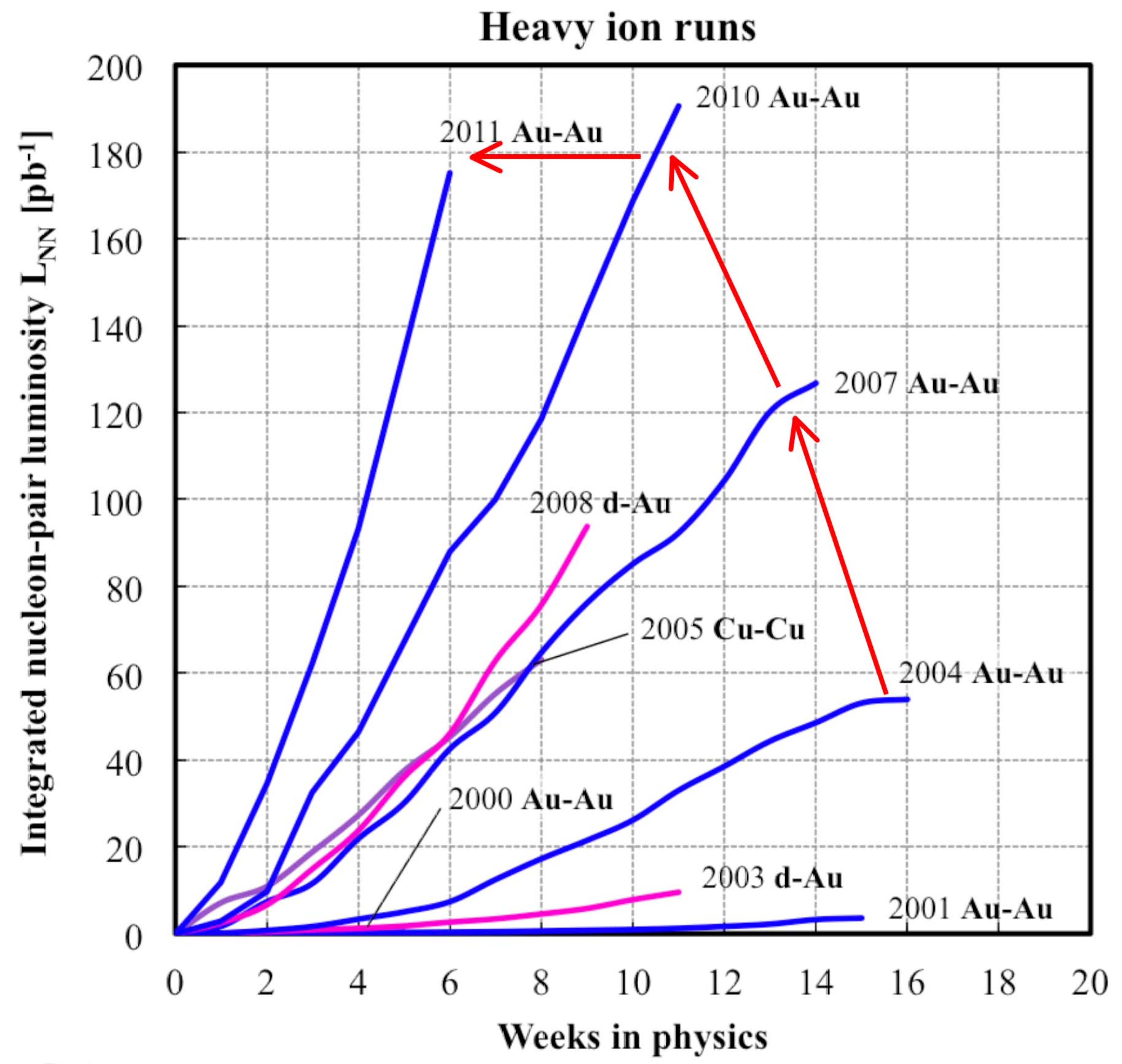}\hspace*{1pc}
\includegraphics[width=0.47\textwidth]{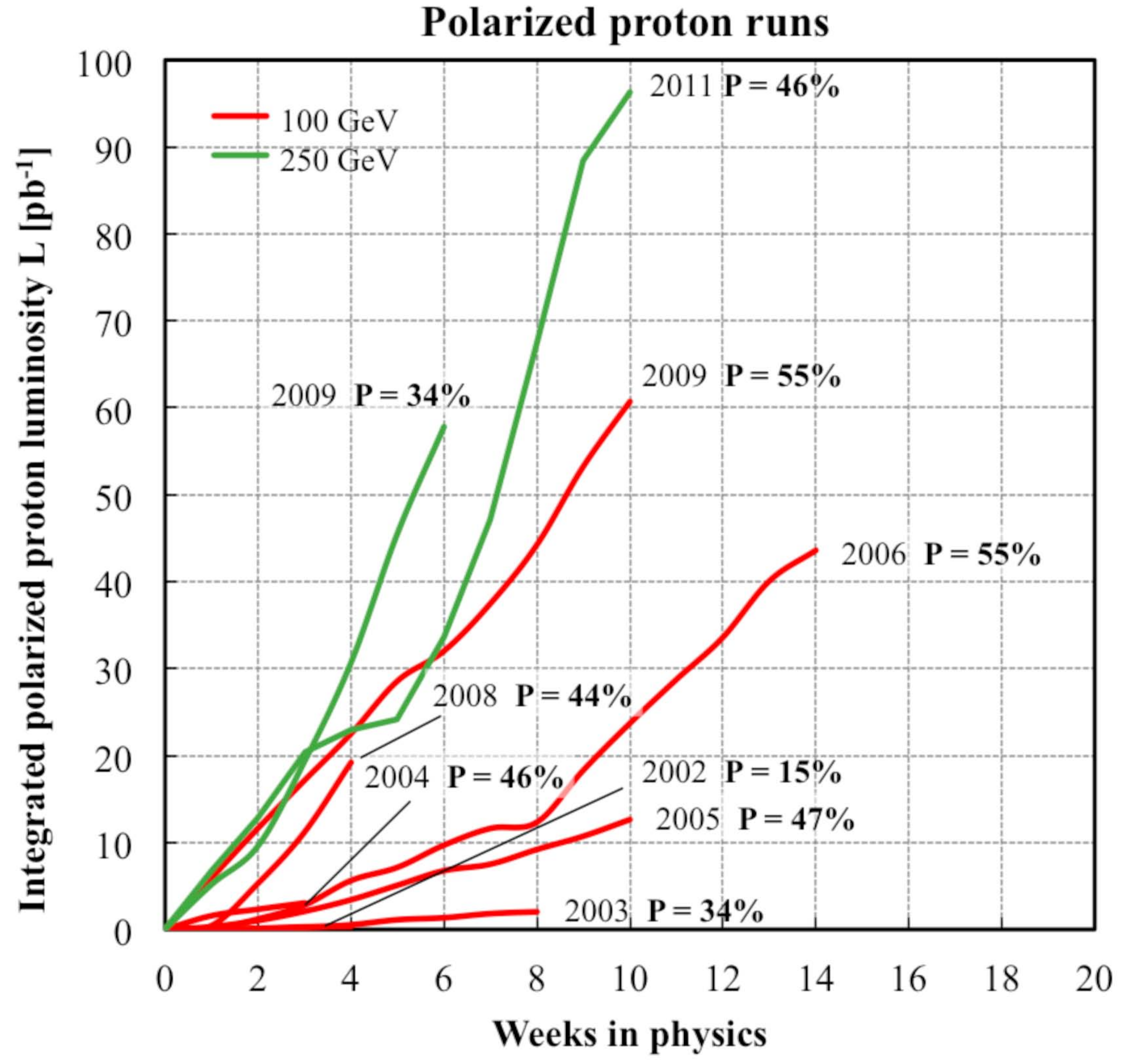}
\end{center}
\caption[]{a)(left) Au+Au performance. b) (right) polarized p-p performance. Courtesy Wolfram Fischer.}
\label{fig:RHICperf}
\end{figure}

RHIC (Fig.~\ref{fig:RHICoverview}) is one of only two Heavy Ion Colliders in the world, the other being the CERN LHC.
RHIC is composed of two independent rings of superconducting magnets, 3.8 km in diameter (see Fig.~\ref{fig:RHICring}a, below). RHIC can collide any species with any other species and so far has provided Au+Au, d+Au, Cu+Cu collisions at 12 different values of nucleon-nucleon c.m. energy $\sqrt{s_{NN}}$. For the past two years, an Au+Au energy scan has been performed with Energy/beam 3.85, 5.75, 9.8, 13.5, 19.5, 31.2, 100 GeV/nucleon. Also RHIC is the world's first and only polarized proton collider. The performance history of RHIC with A+A and polarized p-p collisions is shown in Fig.~\ref{fig:RHICperf}. 
At present RHIC operates at 15 times design luminosity for Au+Au and has shown a factor 2 progress in integrated luminosity $\cal{L}_{\rm int}$ per week from Run-4 (2004) to Run-7 to Run-10 to Run-11. This year, 3 dimensional stochastic cooling was introduced for the Au+Au collisions which will improve the storage lifetime. Future improvements in the longitudinal profile, i.e. smaller diamond size without bucket migration, will be made with increased longitudinal focusing from new 56 MHz storage r.f. now under construction, with commissioning planned for Run-14. 

A significant improvement for the upcoming Run-12 will be the replacement of the 40 year old Tandem Van de Graaff injector  with an Electron Beam Ion Source (EBIS). A 10 A electron beam creates the desired charge state(s) in a trap within a 5 T superconducting solenoid. This is then accelerated through the RFQ and linac and injected into the AGS Booster (Fig.~\ref{fig:RHICoverview}). All ion species including noble gases, uranium and polarized $^3$He are available. A uranium cathode has been received in preparation for a pilot U+U run in 2012. Commissioning of the EBIS started during early 2011 and it has already operated and supplied He$^+$, He$^{2+}$, Ne$^{5+}$, Ne$^{8+}$, Ar$^{11+}$, Ti$^{18+}$ and Fe$^{20+}$ for the NASA Space Radiation Research Laboratory (NSRL) at BNL~\cite{NSRL}.

\subsection{BNL's Superconducting Magnet Division}
All of the upgrades to the RHIC machine mentioned above involved superconducting magnets built or developed in BNL's Superconducting Magnet Division. The Magnet Division also developed the RHIC machine superconducting magnets (Palmer magnet~\cite{CBAmagnetsNIMA235}) which are the basis for the other post-Tevatron machines such as HERA and the LHC. 
\begin{figure}[!h]
\begin{center}
\includegraphics[width=0.47\textwidth]{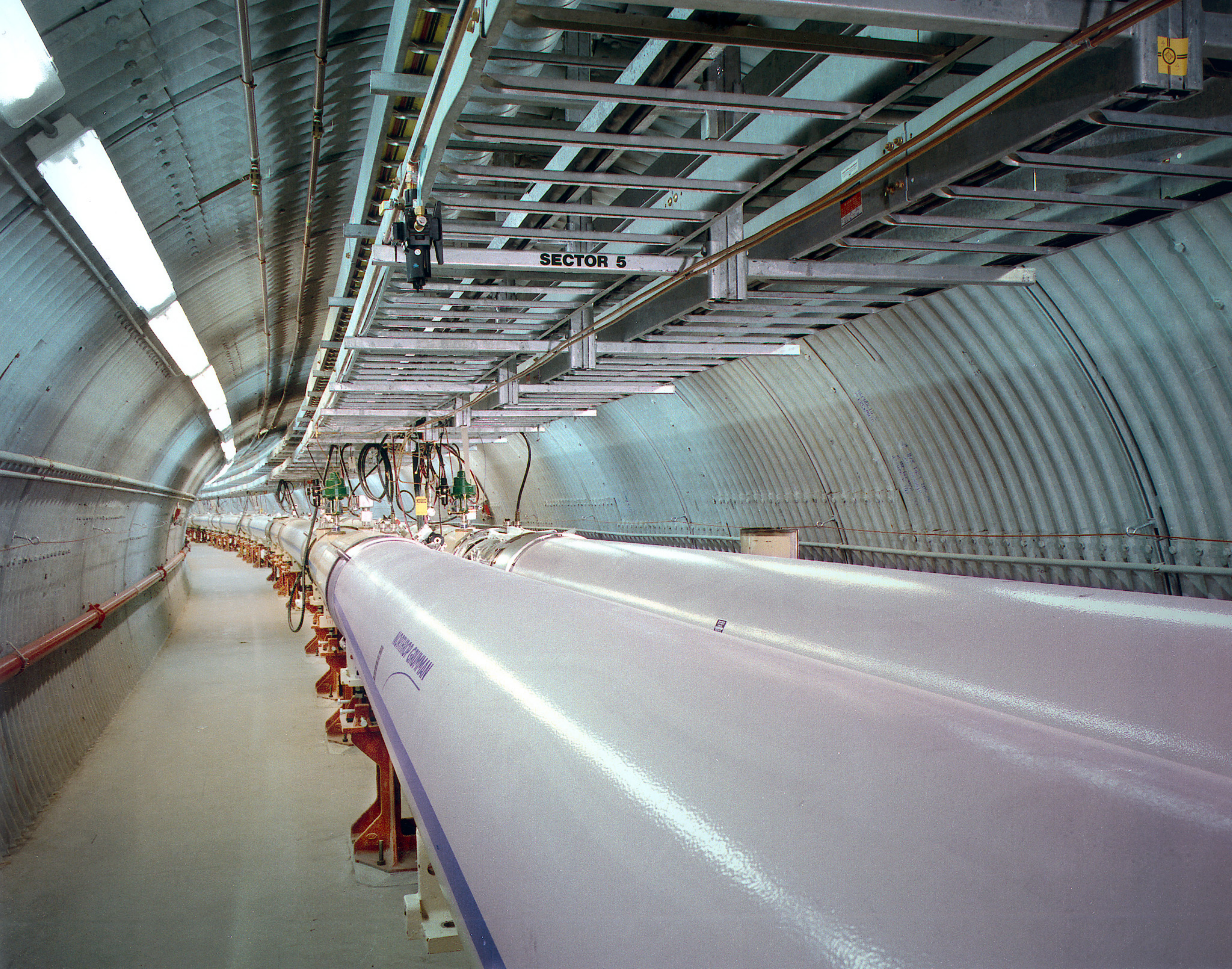}\hspace*{0.1pc}
\includegraphics[width=0.51\textwidth]{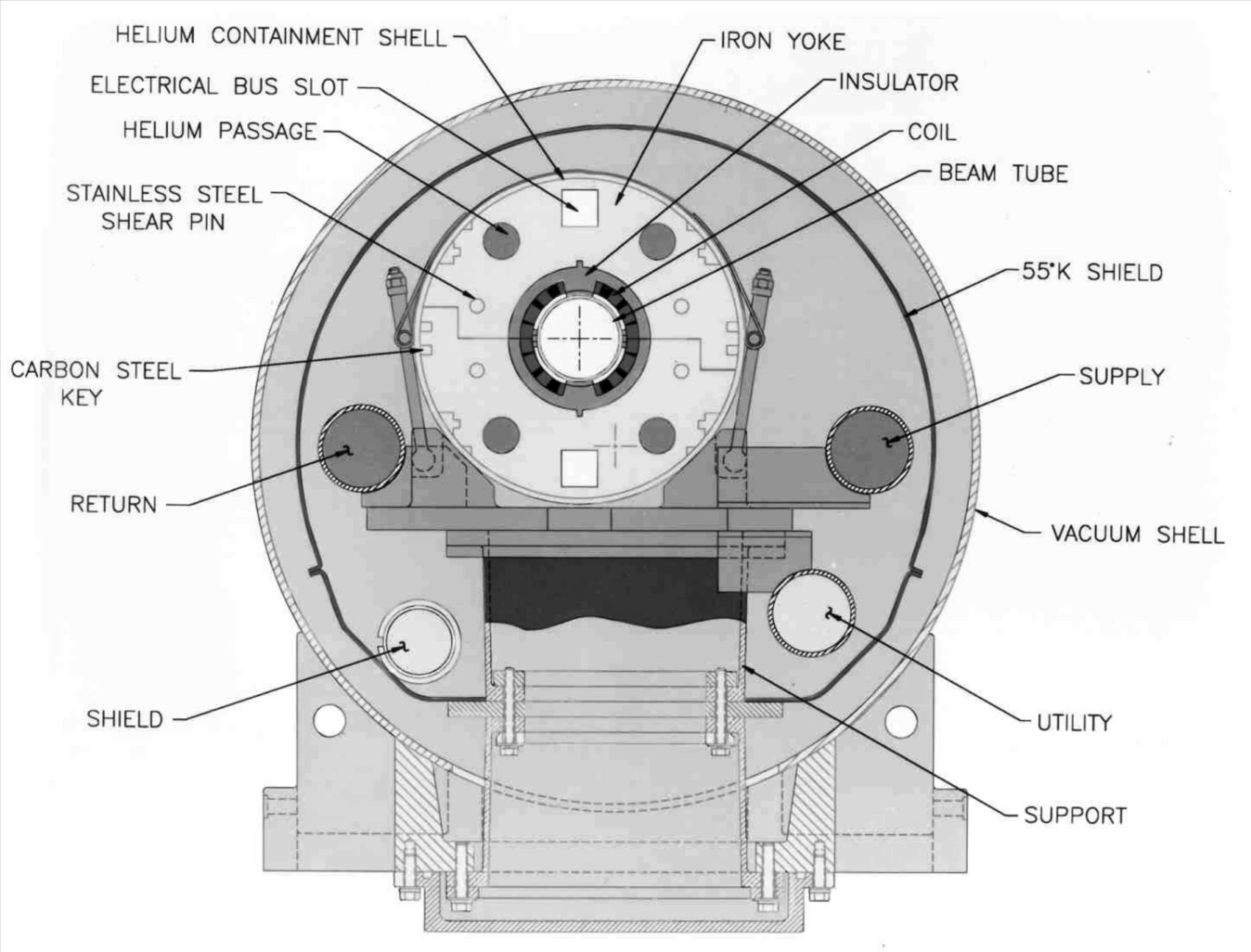}
\end{center}
\caption[]{a)(left) Photo of the RHIC machine composed of two independent rings, with a total of 1740 superconducting dipole, quadrupole and corrector magnets.  b) (right) Cross section of RHIC dipole.}
\label{fig:RHICring}\vspace*{-1.0pc}
\end{figure}
The RHIC dipole design (Fig.~\ref{fig:RHICring}b) is based on a relatively large bore (80 mm inner diameter), single-layer ``cosine theta" coil, wound from a (partially) keystoned, kapton-insulated, 30-strand Rutherford-type cable, arranged in coil blocks with intervening copper wedges, in order to meet the stringent field quality specifications, and mechanically supported by a laminated, cold steel yoke encased in a stainless steel shell. The shell contains the helium and is also a load bearing part of the assembly. 

\begin{figure}[!hbt]
\begin{center}
\includegraphics[width=0.45\textwidth,angle=1.0]{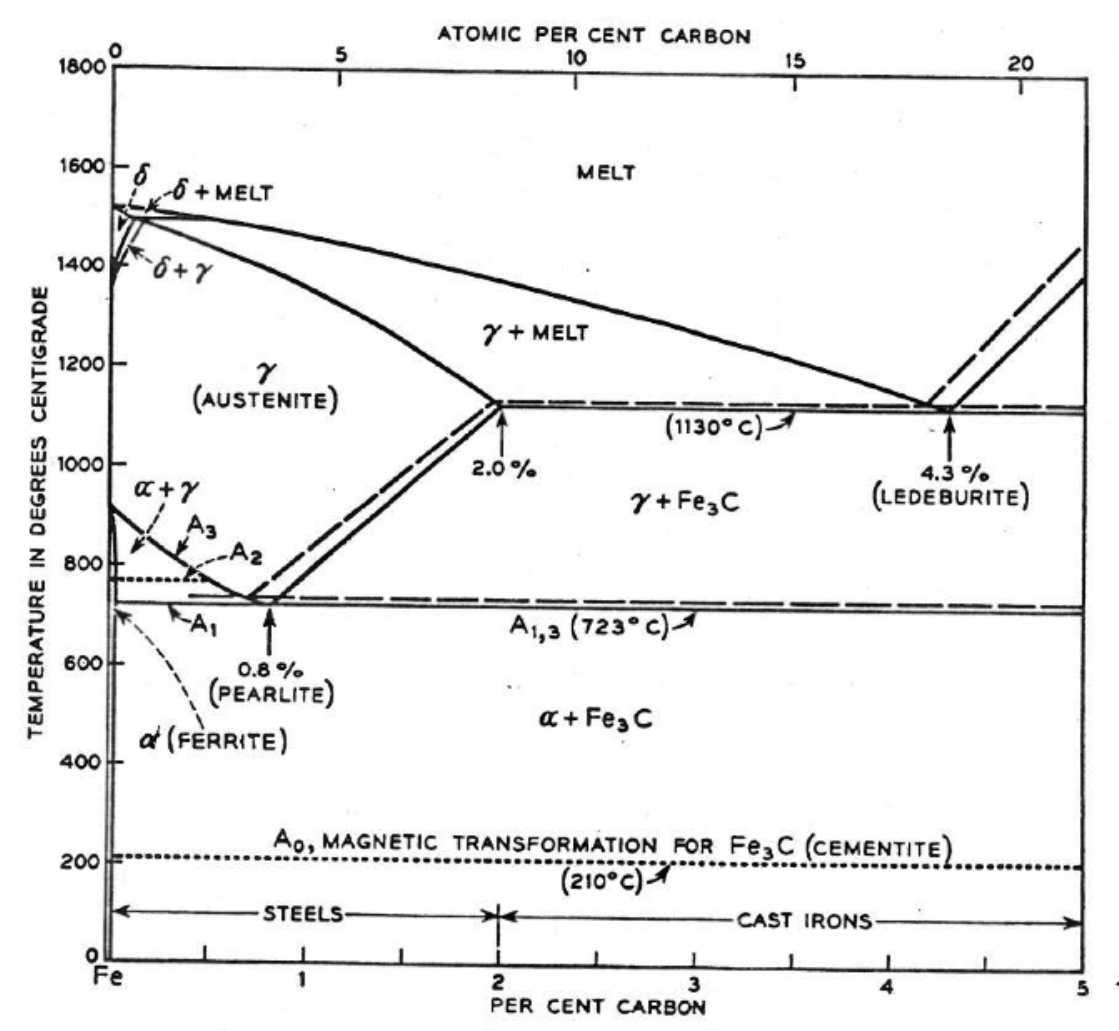}
\end{center}\vspace*{-2.0pc}
\caption[]{Phase diagram of iron-carbon alloys~\cite{Bozorth}. \label{fig:FeCphasediagram}}
\end{figure}

It is amusing to note that in order to build the RHIC magnets for the purpose of making the \QGP\ and studying its phase diagram, it is important to understand another phase-diagram, that of Fe+C, i.e. magnet steel (Fig.~\ref{fig:FeCphasediagram}).
The Fe+C phase diagram is quite complicated with many phases, but is well known; while the proposed phase diagram of nuclear matter (Fig.~\ref{fig:phase_boundary}) seems much simpler, probably because it is largely unknown. 
\subsection{Latest results from the Magnet Division}
\begin{figure}[!h]
\begin{center}
\includegraphics[width=0.30\textwidth]{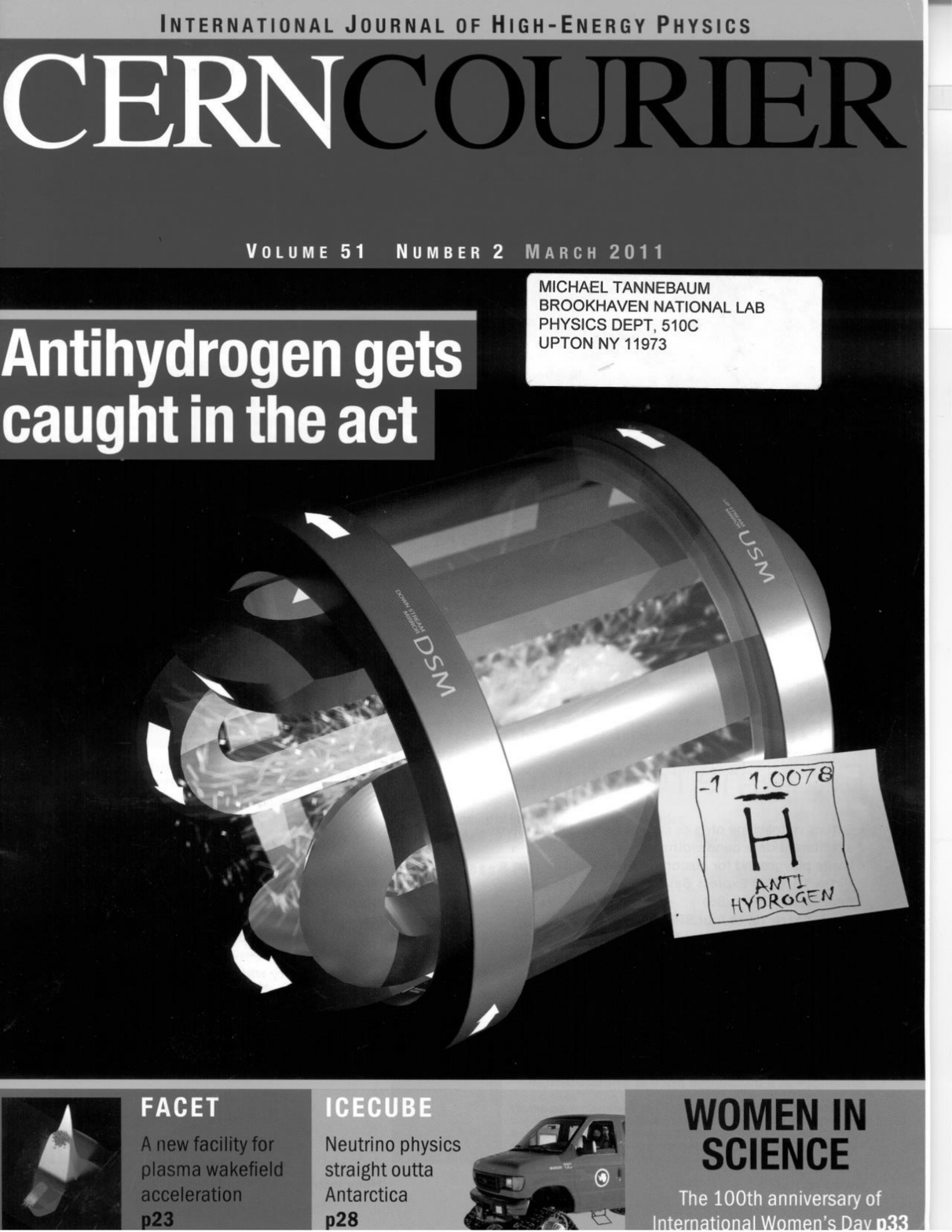}\hspace*{1pc}
\includegraphics[width=0.67\textwidth]{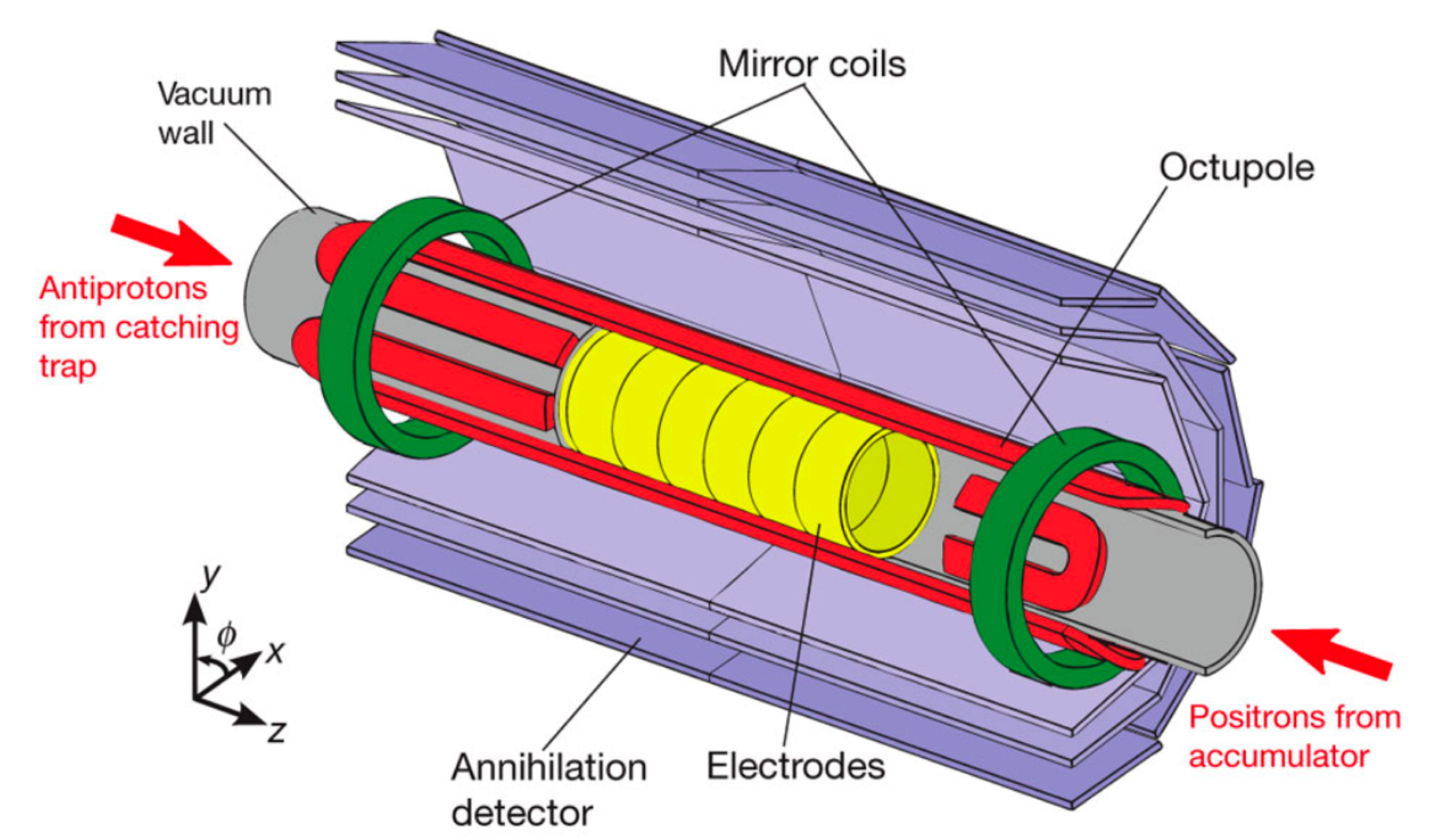}
\end{center}
\caption[]{a)(left) CERN Courier cover March 2011. b) (right) Antihydrogen synthesis and trapping region of the ALPHA apparatus~\cite{NaturePhysics7}. The atom-trap magnets, the modular annihilation detector and some of the Penning trap electrodes are shown (not to scale).  }
\label{fig:ALPHAexpt}
\end{figure}

BNL's superconducting magnet division is an international resource and actively participates in many projects, several in the news this year. The CERN Courier of March 2011 featured on its cover (Fig.~\ref{fig:ALPHAexpt}a) the beautiful magnet used to trap anti-hydrogen for 1000 seconds in the ALPHA experiment~\cite{NaturePhysics7} in the Antiproton Decelerator at CERN.
The octupole magnet (Fig.~\ref{fig:ALPHAexpt}b) was built by the BNL Magnet Division and provides a very pure octupole field which is zero on the axis and rises sharply near the vacuum wall to keep the anti-hydrogen confined radially. 

Another recent press release featuring BNL magnets was the ``Indication of Electron Neutrino Appearance'' at the T2K (Tokai to Kamioka) Experiment in Japan on June 15, 2011~\cite{T2KPR}. BNL provided 5 superconducting dipole corrector magnets in the proton beam at the Japan Proton Accelerator Research Complex (J-PARC) in Tokai which produced the $\mu$-neutrinos that were detected in the Super-Kamiokande detector after transforming to $e$-neutrinos over the 295 km flight path. 

Other work of the Magnet Division includes: NbTi magnets for the RHIC Electron Lens upgrade; NbTi final focus quadrupole for the ILC; Nb$_3$Sn 11.5 Tesla strand-test-barrel magnet and two coils for the LHC luminosity upgrade; High Temperature Superconductor (HTS) Quadrupole for the Facility for Rare Isotope Beams at Michigan State University; Spare NbTi dipole for LHC; HTS solenoid R\&D for muon collider and Energy Storage; Nb$_3$Sn Open Midplane Dipole for the Muon Collider.

\subsection{National Synchrotron Light Source-II}
In addition to accelerators for high energy particle and nuclear physics, Brookhaven has been innovative in synchrotron radiation light sources. The National Synchrotron Light Source (NSLS) which started operations in 1982 was the first to use the Chasman-Green double-bend achromat lattice~\cite{IEEENS22}, which is now the standard lattice at the major synchrotron light sources worldwide. A new third generation light source, NSLS-II, 4.66 times larger in circumference than NSLS, is now under construction at BNL, with unique design features of high brightness, small source size and long beam lines which will replace NSLS in 2014. 
NSLS-II is designed to deliver photons with high average spectral brightness in the 2 keV to 10 keV energy range exceeding $10^{21}$ ph/s/0.1\%BW/mm$^2$/mrad$^2$. The spectral flux density should exceed $10^{15}$ ph/s/0.1\%BW in all spectral ranges. This cutting-edge performance requires the storage ring to support a very high-current electron beam (I = 500 mA) with sub-nm-rad horizontal emittance (down to 0.5 nm-rad) and diffraction-limited vertical emittance at a wavelength of 1 \AA (vertical emittance of 8 pm-rad)~\cite{NSLSIIsource}. 

\section{\QGP\ Physics---Highlights from RHIC}

Given that I already said that the \QGP\ was discovered at RHIC, what further studies are important? 
   
\begin{figure}[h]
\begin{center}
\includegraphics[width=0.55\linewidth]{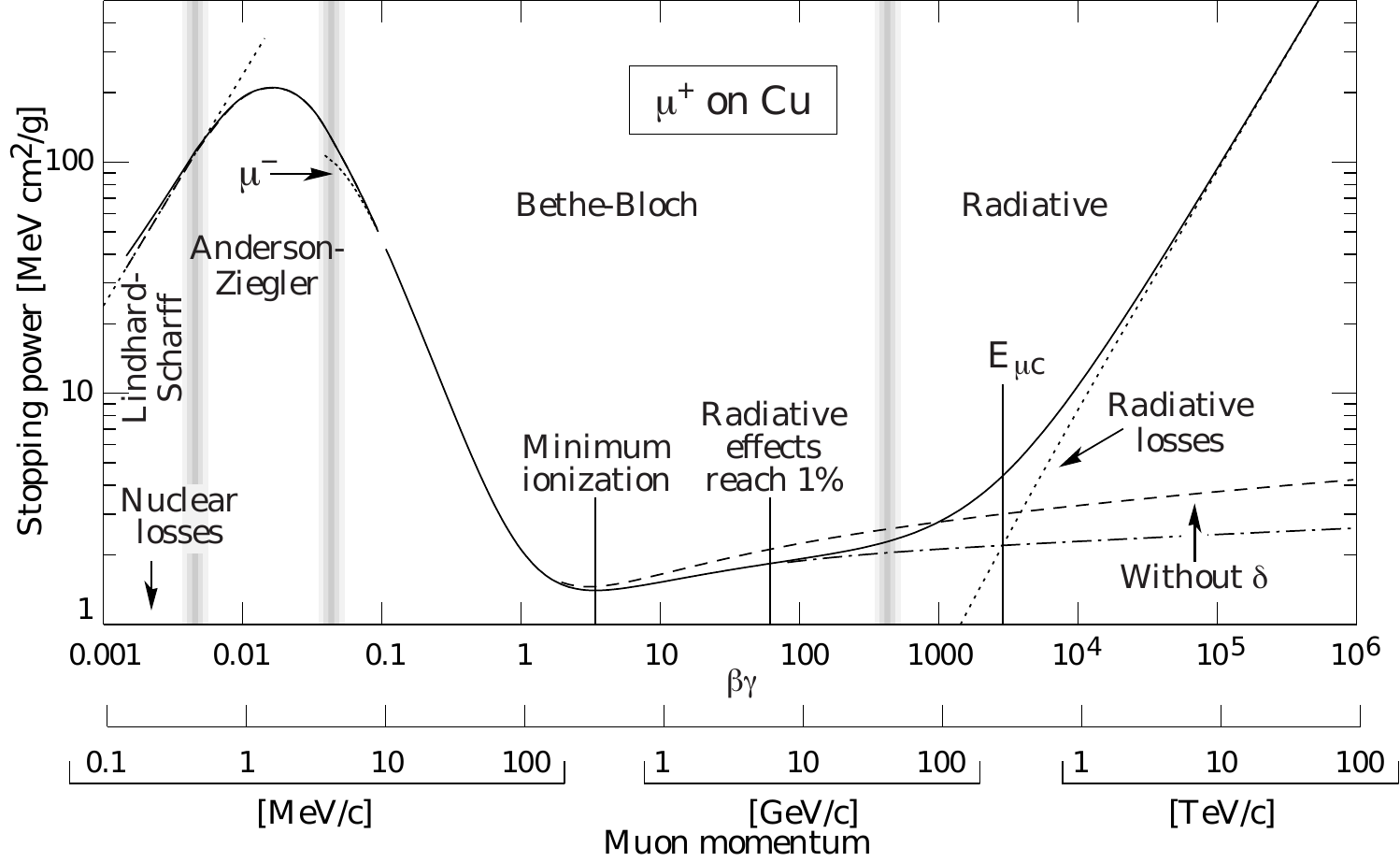}
\end{center}\vspace*{-0.25in}
\caption[]{$dE/dx$ of a $\mu^+$ in Copper as a function of muon momentum~\cite{PDG}.}
\label{fig:muCu}
\end{figure}

The \QGP\  is the only place in the universe where we can in principle and in practice study Quantum Chromo-Dynamics (\QCD) for color-charged quarks in a color-charged medium. For instance, how long will it take before we understand the passage of a quark through a \QGP\ medium as well as we understand the passage of a muon through Copper in QED (Fig.~\ref{fig:muCu})? 

Of course, in addition to understanding the behavior of \QCD\ in a medium, the central goal of our field is a quantitative study of the phases of nuclear matter. This requires a broad, quantitative study of the fundamental properties of the \QGP\  including the extraction of the transport coefficients of the medium such as critical temperature, $T_c$, speed of sound, $c_s$, the ratio of shear viscosity to entropy density, $\eta/s$, etc. To help understand how we shall proceed to address these issues, it is important to understand how we got to this point.  

This is the 11th year of RHIC operation, and the two major detectors PHENIX and STAR which study the \QGP\ at RHIC are basically first round detectors with a few incremental upgrades. Thus, the design of these detectors was heavily influenced by the c. 1990 view of the signatures of the \QGP, which as noted above were quite different from what was discovered.  

\subsection{$J/\Psi$ suppression---the original ``gold-plated" \QGP\ signature}

   Since 1986, the `gold-plated' signature of deconfinement was thought to be $J/\Psi$ suppression. Matsui and Satz~\cite{MatsuiSatz86} proposed that $J/\Psi$ production in A+A collisions would be suppressed by Debye screening of the quark
color charge in the \QGP. The $J/\Psi$ is produced when two gluons
interact to produce a $c, \bar c$ pair which then resonates to form the
$J/\Psi$. In the plasma the $c, \bar c$ interaction is screened so that the 
$c, \bar c$ go their separate ways and eventually pick up other quarks at
the periphery to become {\it open charm}. 

\begin{figure}[!tb]
\begin{center}
\includegraphics[width=0.44\linewidth]{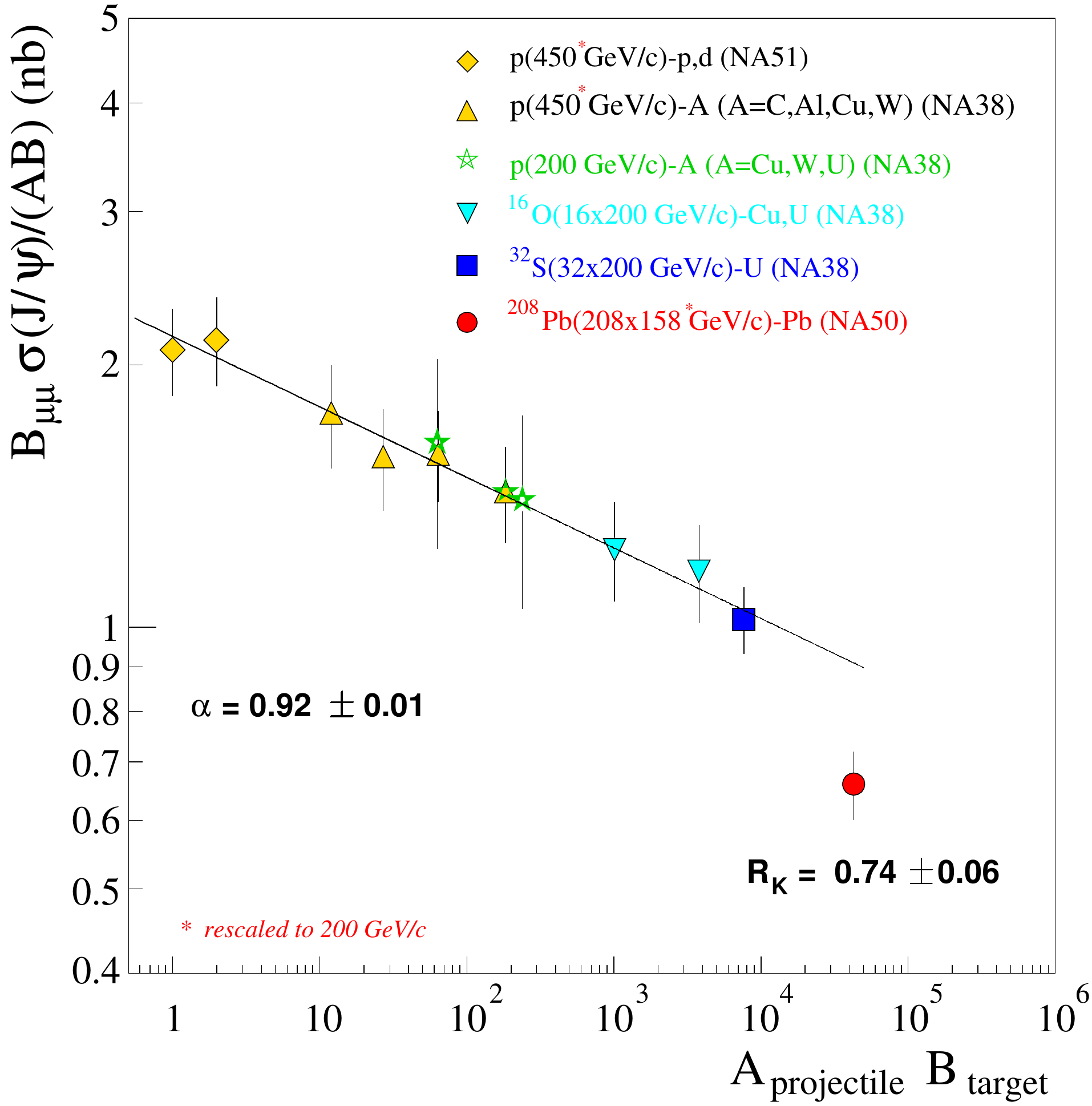}\hspace*{1pc}
\includegraphics[width=0.53\linewidth]{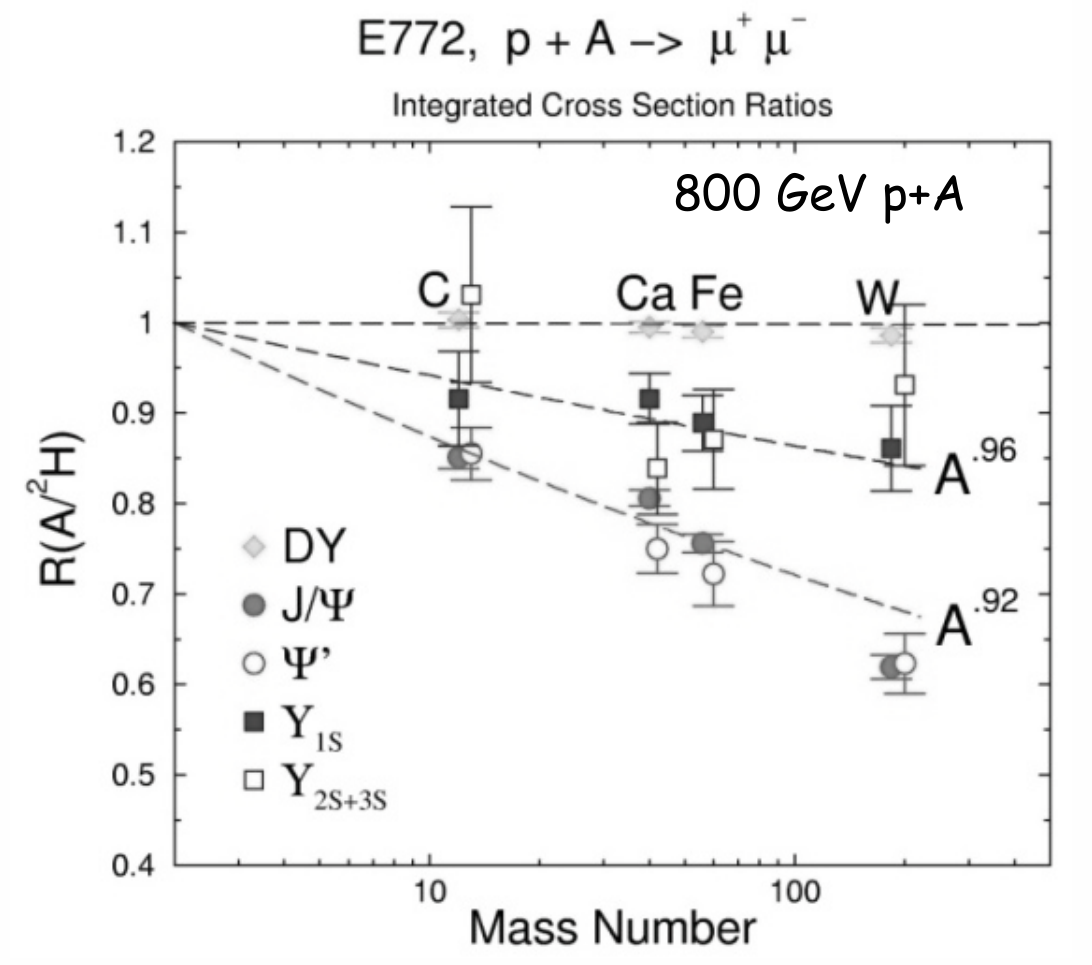}
\end{center}\vspace*{-0.25in}
\caption[]{\small a) (left) Total cross section for $J/\Psi$ production divided by $AB$ in A+B collisions at 158--200$A$ GeV~\cite{NA50EPJC39} (note the {\color{Red}*}). b) (right) $A$ dependence of charmonium and Drell-Yan pair production in 800 GeV p+A collisions~\cite{E772} expressed as the ratio of heavy nucleus to deuterium cross sections per nucleon. The dashed lines are fits to $A^\alpha$ for the CNM effect, with the values of $\alpha$ indicated. \label{fig:JPsiAB}}
\end{figure}

``Anomalous suppression'' of $J/\Psi$ was found in $\sqrt{s_{NN}}=17.2$ GeV Pb+Pb collisions at the CERN SpS~\cite{NA50EPJC39} (Fig.~\ref{fig:JPsiAB}a). This is the CERN fixed target heavy ion program's main claim to fame: but the situation has always been complicated because the $J/\Psi$ is suppressed in p+A collisions. For example, in $\sqrt{s_{NN}}=38.8$ GeV p+A collisions~\cite{E772} (Fig.~\ref{fig:JPsiAB}b) the Drell-Yan $\bar{q}q\rightarrow\mu^+ \mu^-$ cross-section per nucleon is constant as a function of mass number, A, which indicates the expected absence of shadowing in a nucleus for point-like production processes; while the $J/\Psi$ and $\Upsilon$ cross sections per nucleon are suppressed by an amount $A^\alpha$ with $\alpha=0.920\pm0.008$ for both $J/\Psi$ and $\Psi^{'}$ and $\alpha=0.96\pm0.01$ for both the $\Upsilon_{1s}$ and $\Upsilon_{2s+3s}$.  This is called a Cold Nuclear Matter or CNM effect and is shown as the line with $\alpha=0.92$ on Fig.~\ref{fig:JPsiAB}a. The ``Anomalous suppression'' is the difference between the data point at $AB=208^2$ and the line, provided that the CNM effect is the same at $\sqrt{s_{NN}}=17.2$ and 38.8 GeV.  

	The search for $J/\Psi$ suppression and thermal photon/dilepton radiation from the \QGP\ drove the design of the RHIC experiments. 
	\subsection{Detector issues in A+A compared to p-p collisions} 
\begin{figure}[!h]
\begin{center}
\begin{tabular}{cc}
\psfig{file=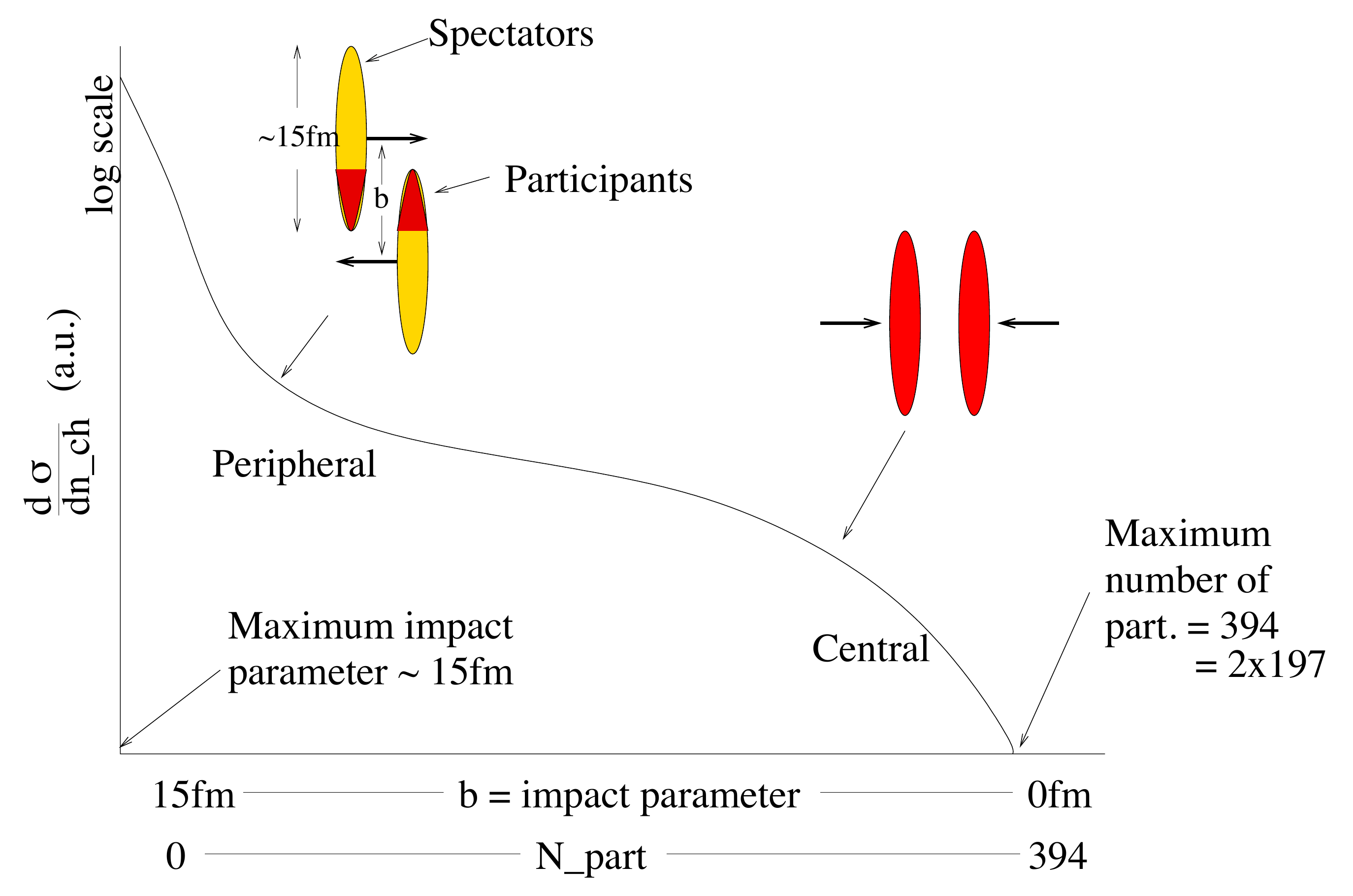,width=0.50\linewidth}
\psfig{file=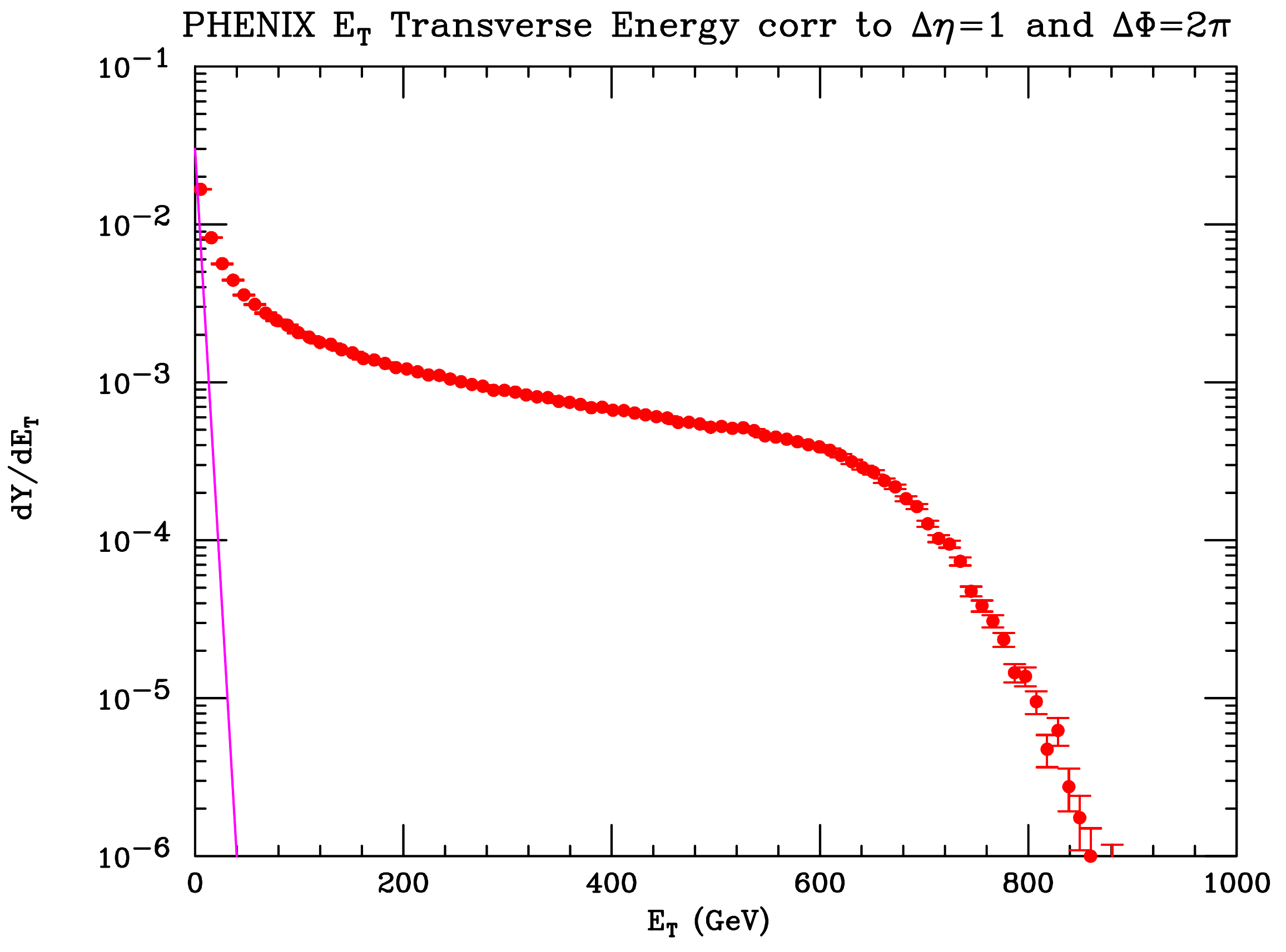,width=0.45\linewidth,angle=0}\end{tabular}
\end{center}\vspace*{-0.15in}
\caption[]{a) (left) Schematic of collision of two nuclei with radius $R$ and impact parameter $b$. The curve with the ordinate labeled $d\sigma/d n_{\rm ch}$ represents the relative probability of charged particle  multiplicity $n_{\rm ch}$ which is directly proportional to the number of participating nucleons, $N_{\rm part}$. b)(right) Transverse energy ($E_T$) distribution in Au+Au and p-p collisions at $\sqrt{s_{NN}}=200$ GeV from PHENIX~\cite{ppg019}.  
\label{fig:nuclcoll}}
\end{figure}

 	A main concern of experimental design in RHI collisions is the huge multiplicity in A+A central collisions compared to  p-p collisions. 
A schematic drawing of a collision of two relativistic Au nuclei is shown in Fig.~\ref{fig:nuclcoll}a. 
In the center of mass system of the nucleus-nucleus collision, the two Lorentz-contracted nuclei of radius $R$ approach each other with impact parameter $b$. In the region of overlap, the ``participating" nucleons interact with each other, while in the non-overlap region, the ``spectator" nucleons simply continue on their original trajectories and can be measured in Zero Degree Calorimeters (ZDC), so that the number of participants can be determined. The degree of overlap is called the centrality of the collision, with $b\sim 0$, being the most central and $b\sim 2R$, the most peripheral. The maximum time of overlap is $\tau_\circ=2R/\gamma\,c$ where $\gamma$ is the Lorentz factor and $c$ is the speed of light in vacuum.  

\begin{figure}[!h]
\begin{center}
\begin{tabular}{cc}
\psfig{file=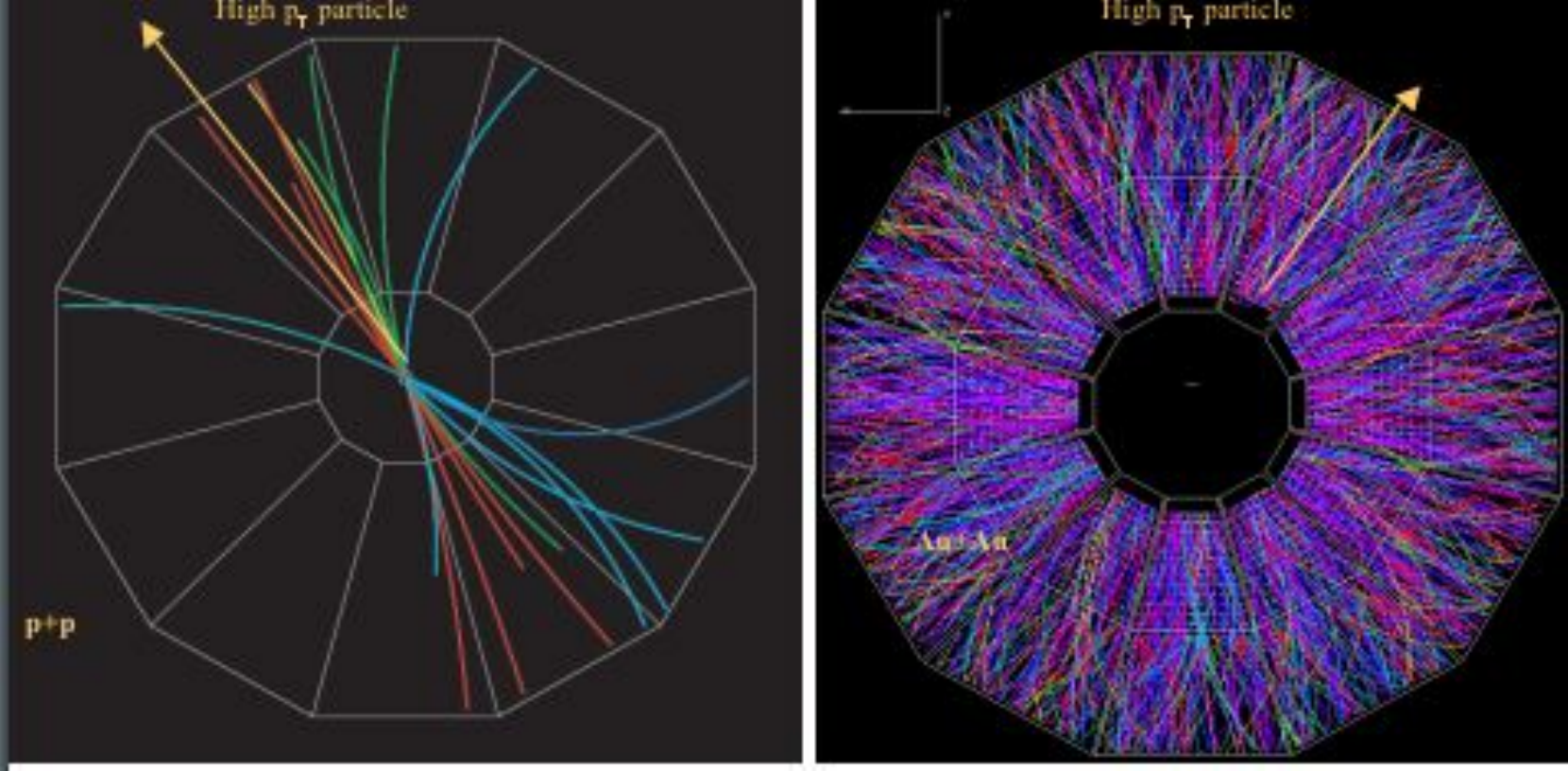,width=0.64\linewidth}&\hspace*{-0.025\linewidth}
\psfig{file=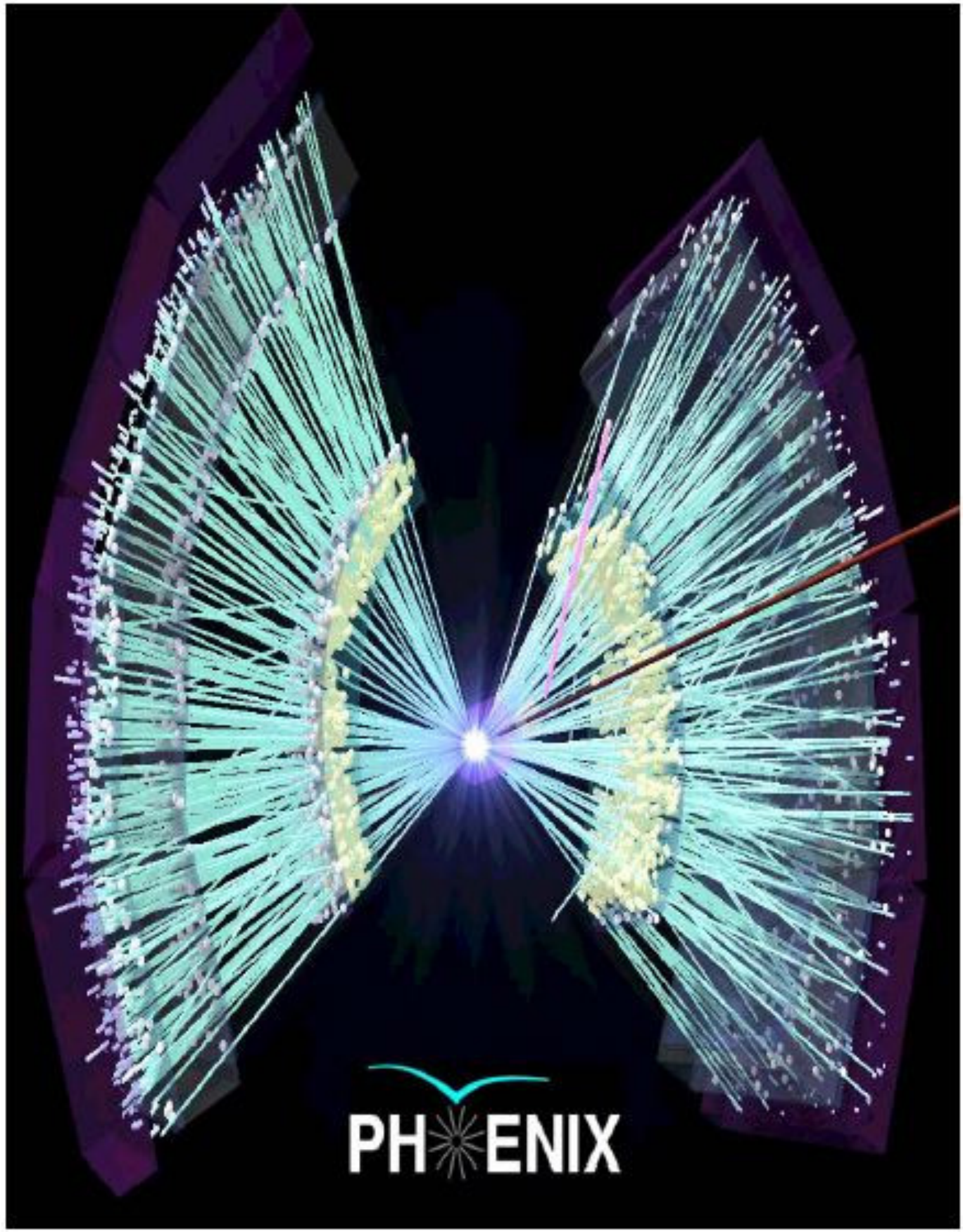,width=0.315\linewidth,height=0.315\linewidth}
\end{tabular}
\end{center}\vspace*{-0.35in}
\caption[]{ a) (left) A p-p collision in the STAR detector viewed along the collision axis; b) (center) Au+Au central collision at $\sqrt{s_{NN}}=200$ GeV in STAR;  c) (right) Au+Au central collision at $\sqrt{s_{NN}}=200$ GeV in PHENIX.  
\label{fig:collstar}}
\end{figure}

The energy of the inelastic collision is predominantly dissipated by multiple particle production, where $N_{\rm ch}$, the number of charged particles produced, is directly proportional~\cite{PXWP} to the number of participating nucleons ($N_{\rm part}$) as sketched on Fig.~\ref{fig:nuclcoll}a. Thus, $N_{\rm ch}$ or the total transverse energy $E_T$ in central Au+Au collisions is roughly $A$ times larger than in a p-p collision, as shown in the measured transverse energy spectrum in the PHENIX detector for Au+Au compared to p-p (Fig.~\ref{fig:nuclcoll}b) and in actual events from the STAR and PHENIX detectors at RHIC in Fig.~\ref{fig:collstar}. 
The impact parameter $b$ can not be measured directly, so the centrality of a collision is defined in terms of the upper percentile of $n_{\rm ch}$ or $E_T$ distributions, e.g. top 10\%-ile, upper $10-20$\%-ile. Unfortunately the ``upper'' and ``-ile'' are usually not mentioned which sometimes confuses the uninitiated.  

\begin{figure}[!h]
\begin{center}
\includegraphics[width=0.55\textwidth]{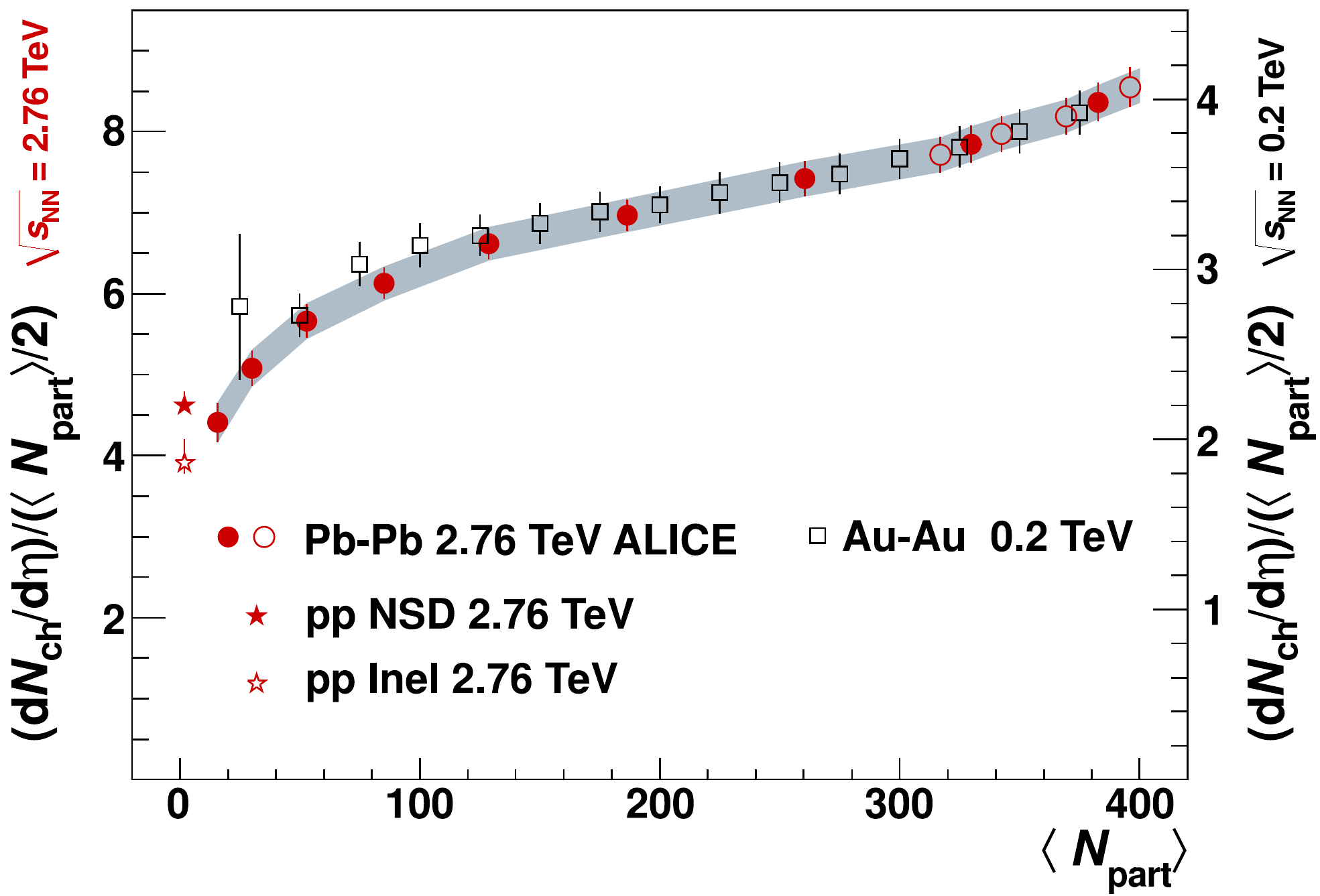}
\end{center}\vspace*{-2.1pc}
\caption[]{Dependence of $(dN_{\rm ch}/d\eta)/(N_{\rm part}/2)$ on the average number of participants $\mean{N_{\rm part}}$   in bins of centrality, for Pb+Pb collisions at $\sqrt{s_{NN}}=2.76$ TeV~\cite{ALICEmult} and Au+Au collisions at $\sqrt{s_{NN}}=0.200$ TeV~\cite{ppg019}. The scale for the lower-energy data (right side) differs by a factor of 2.1 from the scale for the higher-energy data (left side).          \label{fig:dNdeta}}
\end{figure}

In Fig.~\ref{fig:dNdeta}, measurements of the charged particle multiplicity density $dN_{\rm ch}/d\eta$ at mid-rapidity, $|\eta|<0.5$, relative to the number of participating nucleons, $N_{\rm part}$, are shown as a function of centrality for $\sqrt{s_{NN}}=200$ GeV Au+Au collisions at RHIC~\cite{ppg019} together with new results this year from \mbox{ALICE} in $\sqrt{s_{NN}}=2.76$ TeV Pb+Pb collisions at LHC~\cite{ALICEmult}. The results are expressed as $(dN_{\rm ch}/d\eta)/(N_{\rm part}/2)$ for easy comparison to p-p collisions. 

The LHC data show the effect well known from RHIC that $dN_{\rm ch}/d\eta$ does not depend linearly on $N_{\rm part}$, since $(dN_{\rm ch}/d\eta)/(N_{\rm part}/2)$ is not a constant for all $N_{\rm part}$. However the data also show the amazing effect that the ratio of $(dN_{\rm ch}/d\eta)/(N_{\rm part}/2)$ from LHC to RHIC is simply a factor of 2.1 in every  centrality bin.  Thus the LHC and RHIC data lie one on top of each other by simple scaling of the RHIC measurements by a factor of 2.1. This is an incredibly beautiful result which shows that in going from p-p to A+A collisions, the charged particle production is totally dominated by the nuclear geometry of the A+A collisions represented by the number of participating nucleons, $N_{\rm part}$, independently of the nucleon-nucleon c.m. energy, $\sqrt{s_{NN}}$.  

	Since it is a huge task to reconstruct the momenta and identity of all the particles produced in these events, the initial detectors at RHIC~\cite{RHICNIM} concentrated on the measurement of single-particle or multi-particle inclusive variables to analyze RHI collisions, with inspiration from the CERN ISR which emphasized those techniques before the era of jet reconstruction~\cite{egseeMJTISSP2009}. There are two major detectors in operation at RHIC, STAR and PHENIX, and there were also two smaller detectors, BRAHMS and PHOBOS, which have completed their program. As may be surmised from Fig.~\ref{fig:collstar}, STAR, which emphasizes hadron physics, is most like a conventional general purpose collider detector, a TPC to detect all charged particles over the full azimuth ($\Delta\phi=2\pi$) and  $\pm 1$ units of pseudo-rapidity ($\eta$); while PHENIX is a very high granularity high resolution special purpose detector covering a smaller solid angle at mid-rapidity, together with a muon-detector at forward rapidity~\cite{egseePT}. 
	
\begin{figure}[!h]\vspace*{-1.0pc}
\begin{center}
\includegraphics[width=0.40\textwidth]{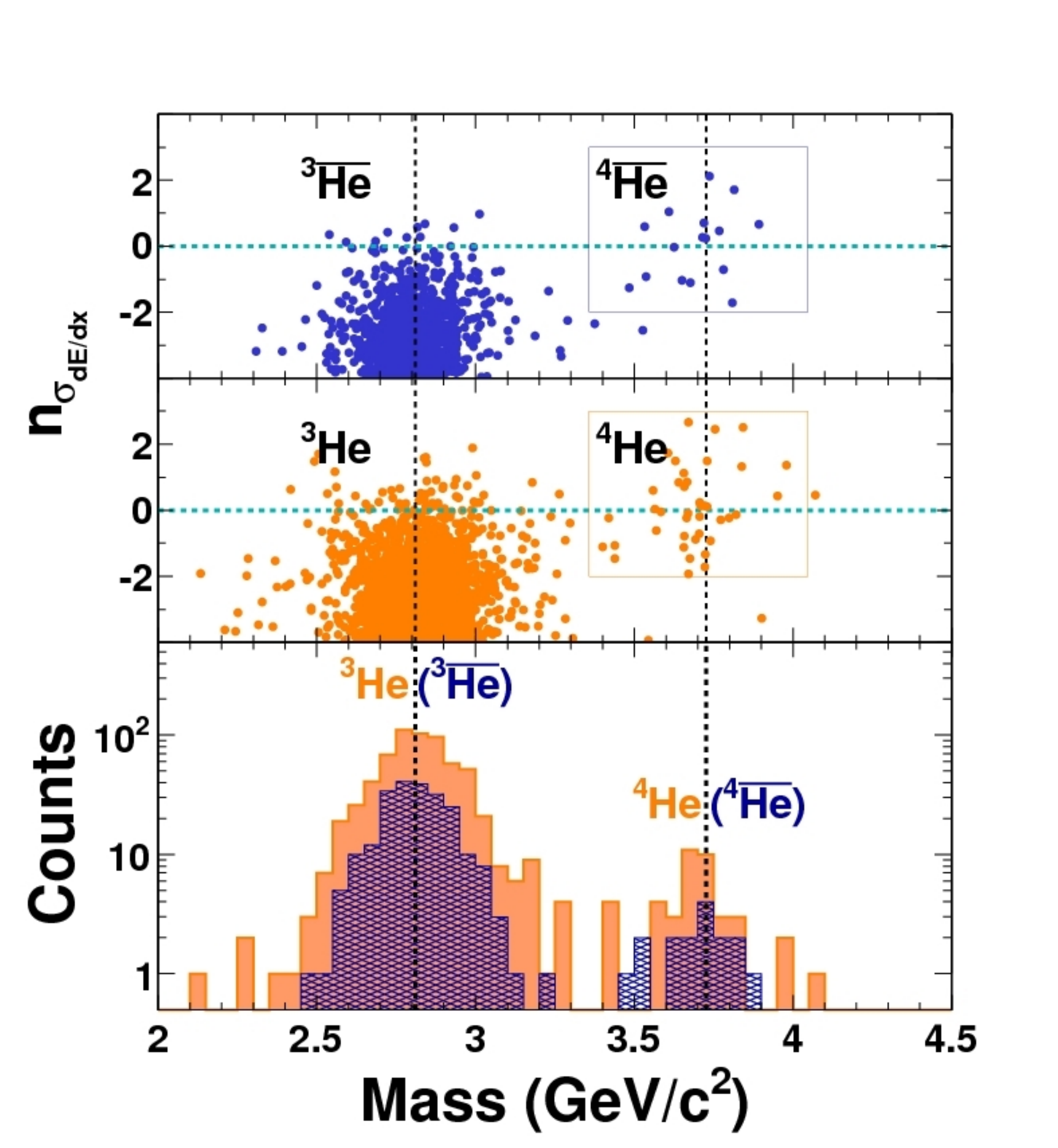}\hspace*{1pc}
\includegraphics[width=0.50\textwidth]{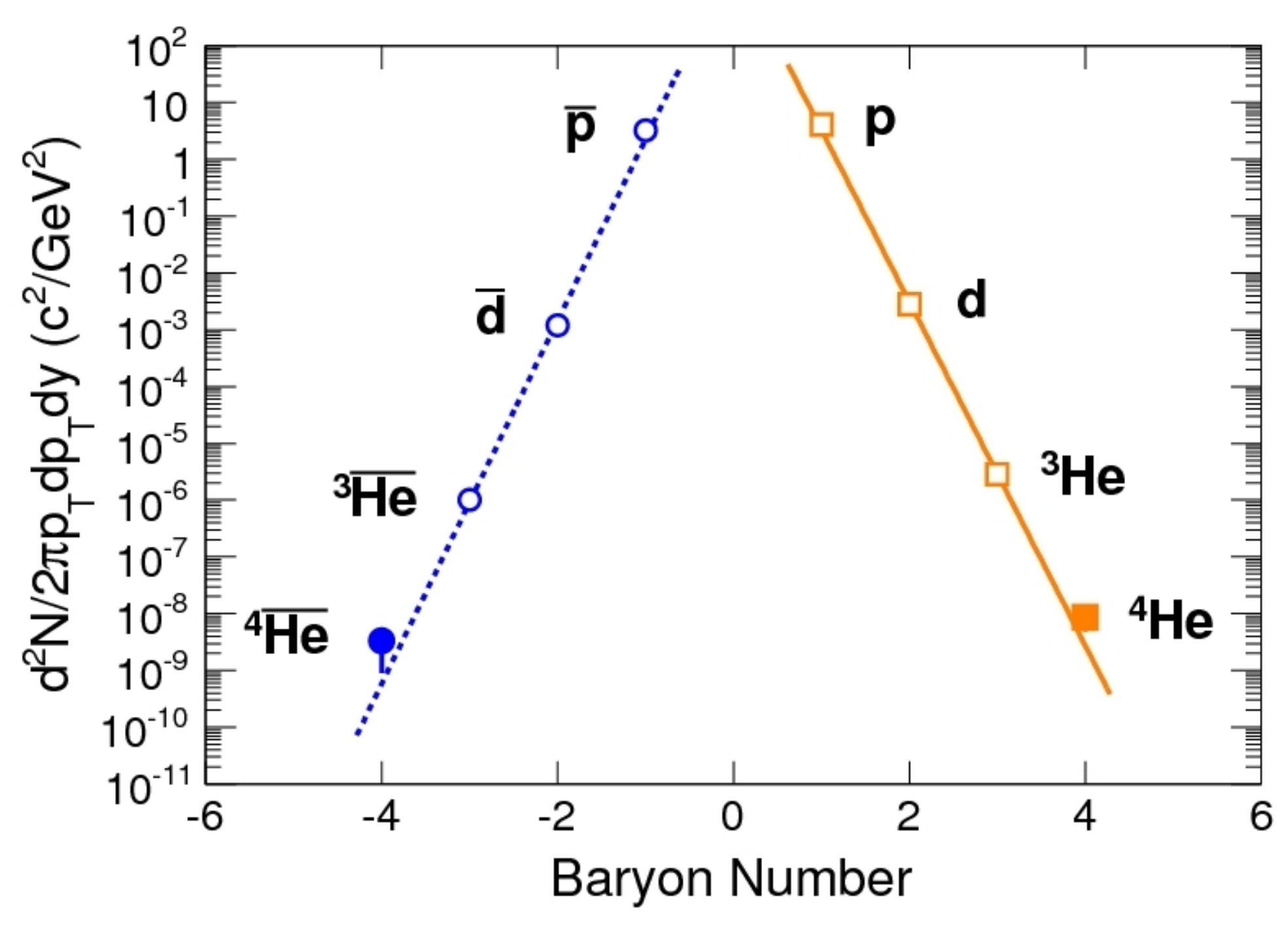}
\end{center}\vspace*{-2pc}
\caption[]{a)(left) Number of standard deviations, $n_{\sigma_{dE/dx}}$, of $dE/dx$ resolution from the expected value for $^4$He, for negative and positive particles as a function of calculated mass, and projected counts for $-2<n_{\sigma_{dE/dx}}<3$. b) Differential invariant yields as a function of baryon number, B.}
\label{fig:STARanti}\vspace*{-0.1pc}
\end{figure}

	One nice feature of the STAR detector is the ability to measure the mass/charge of a particle from its momentum/charge  and time of flight, and then use $dE/dx$ measured in the TPC to determine the charge. In this way STAR has observed many anti-nuclei, notably this year the ``Observation of the antimatter helium-4 nucleus'' in Au+Au collisions~\cite{STARNature473} (Fig.~\ref{fig:STARanti}a). 
The differential invariant yields $d^2N/(2\pi p_T dp_T dy)$ per central Au+Au collision at $\sqrt{s_{NN}}=200$ GeV as a function of baryon number B, evaluated at $p_T/|B|=0.875$ GeV/c, are shown in Fig.~\ref{fig:STARanti}b and show a steady exponential decrease with increasing $B$~\cite{0909.0566}. The anti-nuclei are made by coalescence of the large number of $\bar{n}$ and $\bar{p}$ produced, an advantage of the high multiplicity. 	
	
	PHENIX is designed to measure and trigger on rare processes involving leptons, photons and identified hadrons at the highest luminosities with the special features: i) a minimum of material (0.4\% $X_\circ$) in the aperture to avoid photon conversions; ii) possibility of zero magnetic field on axis to prevent de-correlation of $e^+ e^-$ pairs from photon conversions; iii) Electro-Magnetic Calorimeter (EMCal) and Ring Imaging Cherenkov Counter (RICH) for $e^{\pm}$ identification and level-1 $e^{\pm}$ trigger; iv) a finely segmented EMCal ($\delta\eta$, $\delta\phi=0.01 \times$ 0.01) to avoid overlapping showers due to the high multiplicity and for separation of single-$\gamma$ and $\pi^0$ up to $p_T\sim 25$ GeV/c; v) EMCal and precision Time of Flight measurement for particle identification. Some results uniquely possible with this detector such as measurements of direct photons via internal conversion to $e^+ e^-$ pairs will be discussed below.   	

	In addition to the large multiplicity, there are two other issues in RHI physics which are different from p-p physics: i) space-time issues, both in momentum space and coordinate space---for instance what is the spatial extent of fragmentation? is there a formation time/distance?; ii) huge azimuthal anisotropies of particle production in non-central collisions (colloquially collective flow) which are very interesting in their own right and provide much richer features than originally envisaged. 
	\section{Collective Flow} 
   A distinguishing feature of A+A collisions compared to either p-p or p+A collisions is the collective flow observed. This effect is seen over the full range of energies studied in heavy ion collisions, from incident kinetic energy of $100A$ MeV to c.m. energy of $\sqrt{s_{NN}}=200$ GeV~\cite{LaceyQM05}. Collective flow, or simply flow, is a collective effect which can not be obtained from a superposition of independent N-N collisions.  
      \begin{figure}[!thb]
   \begin{center}
\includegraphics[width=0.45\linewidth]{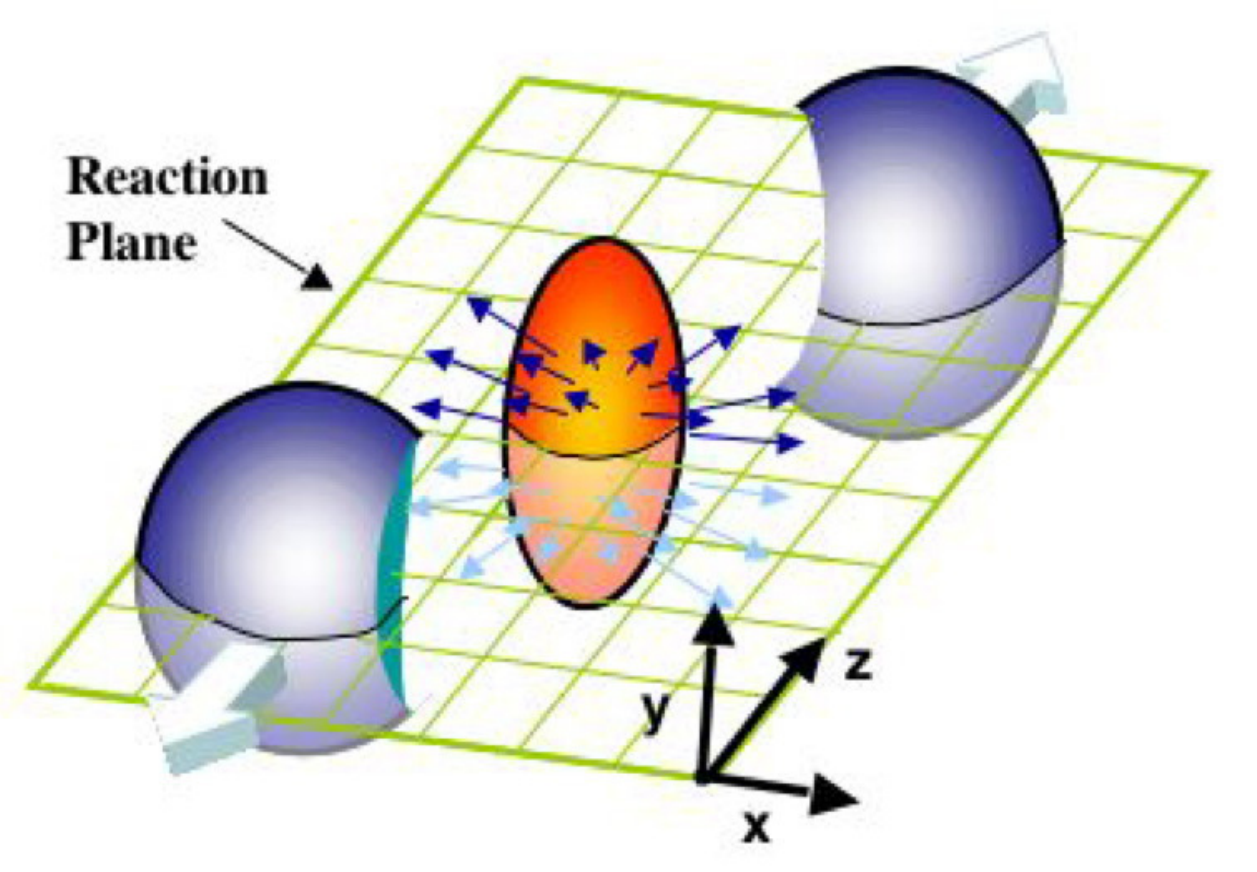}
\includegraphics[width=0.54\linewidth]{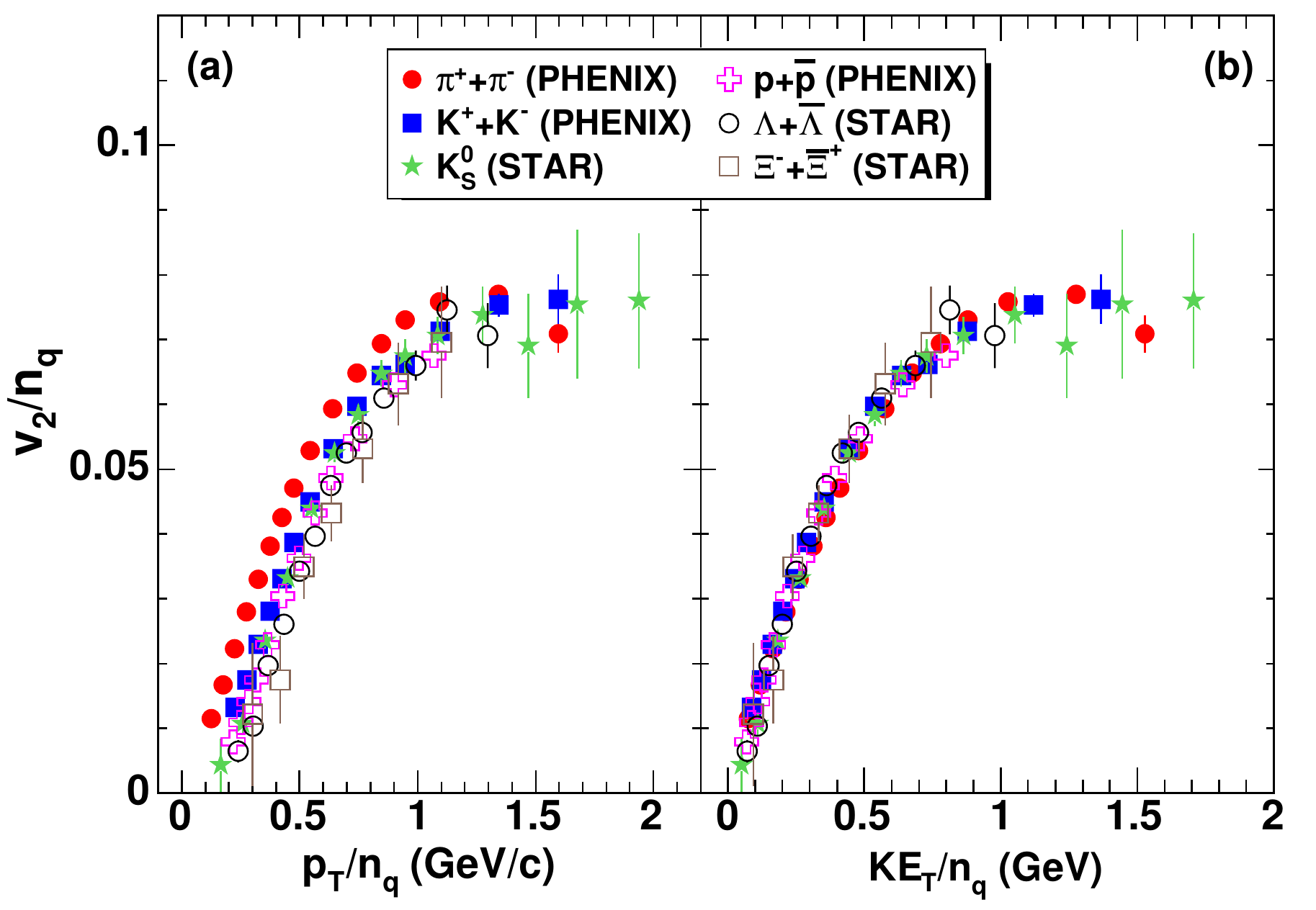}
\end{center}\vspace*{-0.25in}
\caption[]{(left) Almond shaped overlap zone generated just after an A+A collision where the incident nuclei are moving along the $\pm z$ axis. The reaction plane by definition contains the impact parameter vector (along the $x$ axis)~\cite{KanetaQM04}. (right) Measurements of elliptical-flow ($v_2$)  for identified hadrons plotted as $v_2$ divided by the number of constituent quarks $n_q$ in the hadron as a function of (a) $p_T/n_q$, (b) $KE_T/n_q$~\cite{PXArkadyQM06}.   
\label{fig:MasashiFlow}}
\end{figure}
Immediately after an A+A collision, the overlap region defined by the nuclear geometry is almond shaped (see Fig~\ref{fig:MasashiFlow}) with the shortest axis along the impact parameter vector. Due to the reaction plane breaking the $\phi$ symmetry of the problem, the semi-inclusive single particle spectrum is modified by an expansion in harmonics~\cite{Ollitrault} of the azimuthal angle of the particle with respect to the reaction plane, $\phi-\Phi_R$~\cite{HeiselbergLevy}, where the angle of the reaction plane $\Phi_R$ is defined to be along the impact parameter vector, the $x$ axis in Fig.~\ref{fig:MasashiFlow}: 
  \begin{equation}
\frac{Ed^3 N}{dp^3}=\frac{d^3 N}{p_T dp_T dy d\phi}
=\frac{d^3 N}{2\pi\, p_T dp_T dy} \left[ 1+\sum_n 2 v_n \cos n(\phi-\Phi_R)\right] .
\label{eq:siginv2} 
\end{equation} 
The expansion parameter $v_2$, called elliptical flow, is predominant at mid-rapidity. In general, the fact that flow is observed in final state hadrons  shows that thermalization is rapid, so that hydrodynamics comes into play at a time, $\tau_0$, which is before the spatial anisotropy of the overlap almond dissipates. At this early stage hadrons have not formed and it has been proposed that the constituent quarks flow~\cite{VoloshinQM02}, so that the flow should be proportional to the number of constituent quarks $n_q$, in which case $v_2/n_q$ as a function of $p_T/n_q$ would represent the constituent quark flow as a function of constituent quark transverse momentum and would be universal. However, in relativistic hydrodynamics, at mid-rapidity, the transverse kinetic energy, $m_T-m_0=(\gamma_T-1) m_0\equiv KE_T$, rather than $p_T$, is the relevant variable; and in fact $v_2/n_q$ as a function of $KE_T/n_q$ seems to exhibit nearly perfect scaling~\cite{PXArkadyQM06} (Fig.~\ref{fig:MasashiFlow}b). 

\begin{figure}[!thb]
   \begin{center}
\includegraphics[width=0.50\linewidth]{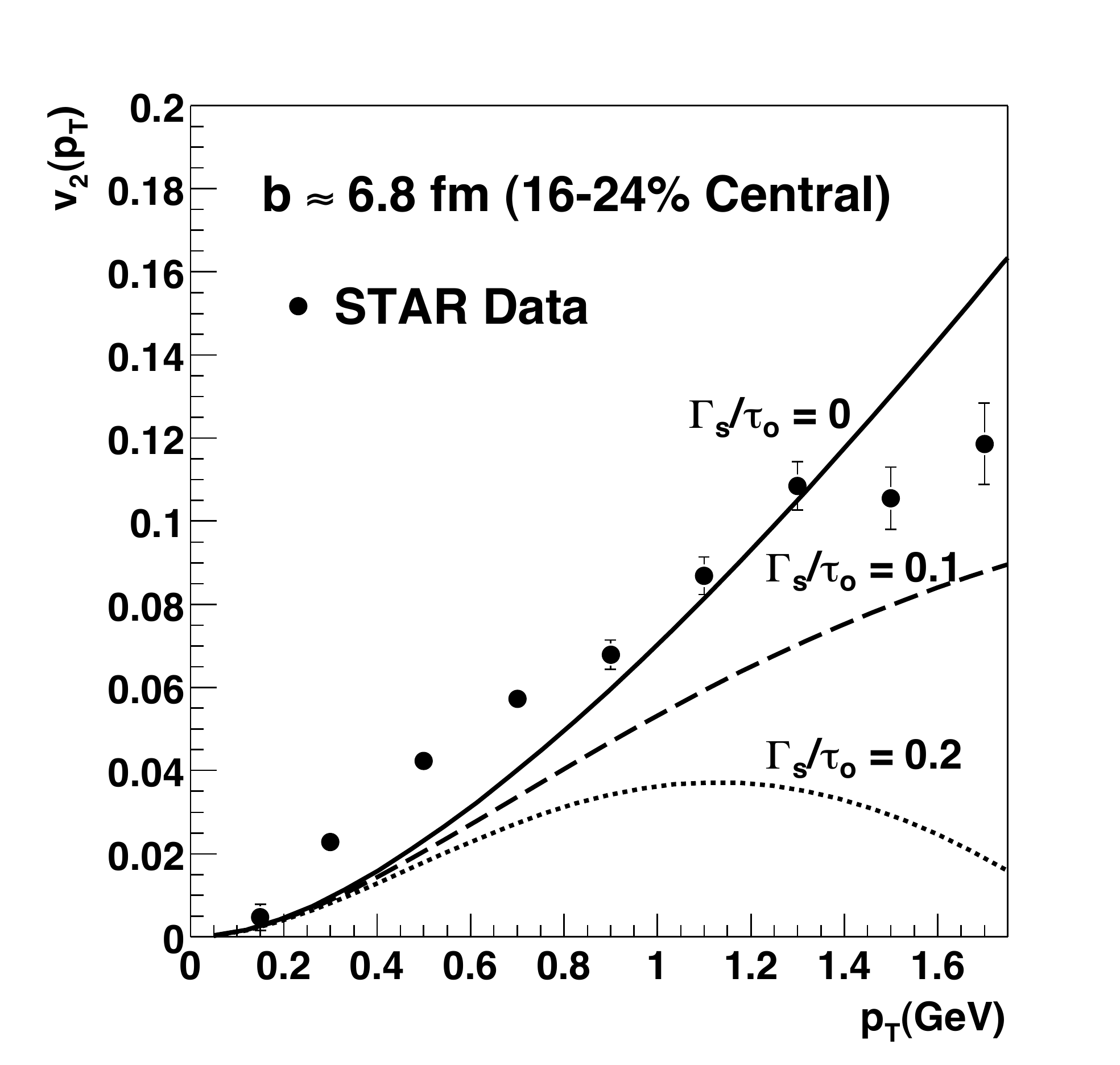}
\includegraphics[width=0.44\linewidth]{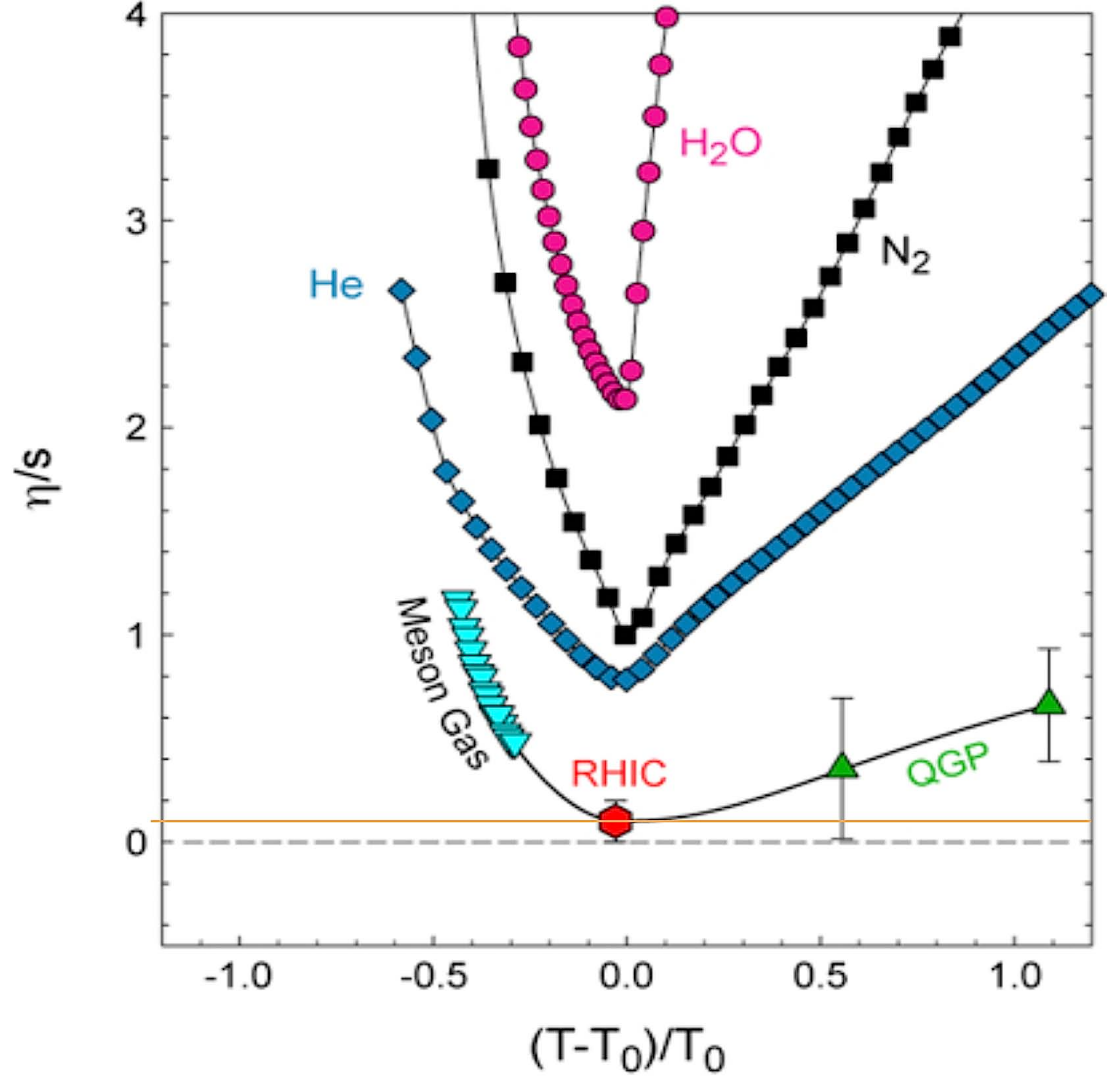}
\end{center}\vspace*{-0.25in}
\caption[]{a) (left) Teaney's~\cite{TeaneyPRC68} predictions for $v_2(p_T)$ for ideal ($\Gamma_s/\tau_0=0$) and viscous hydrodynamics, where $\Gamma_s=\frac{4}{3}\frac{\eta}{sT}$ is the sound attenuation length, and $\tau_0$ is the thermalization time. b) $\eta/s$ for various fluids at fixed pressure as a function of temperature $T$, where $T_0$ is the temperature of the critical point of the liquid-gas phase transition~\cite{LaceyPRL98,CKMPRL97}.  }
\label{fig:TeaneyFlow}
\end{figure}

    The fact that the flow persists for $p_T>1$ GeV/c (Fig.~\ref{fig:TeaneyFlow}a) implies that the viscosity is small~\cite{TeaneyPRC68}, perhaps as small as a quantum viscosity bound from string theory~\cite{Kovtun05}, $\eta/s=1/(4\pi)$ where $\eta$ is the shear viscosity and $s$ the entropy density per unit volume.  This has led to the description of the ``s\QGP'' produced at RHIC as ``the perfect fluid''~\cite{THWPS}. 
An estimate~\cite{LaceyPRL98} of $\eta/s$ for nuclear matter and for several common fluids, as a function of the fractional difference of the temperature from the critical temperature, at fixed pressure, is shown in Fig.~\ref{fig:TeaneyFlow}b. This particular estimate~\cite{LaceyPRL98} for the \QGP\ at RHIC is quite close to the quantum bound (solid line).  Also, empirically, for all common fluids $\eta/s$ is a minimum at or near the critical point~\cite{CKMPRL97} which might suggest that the conditions at RHIC energies are near the \QCD\ critical point.   
\subsection{Two-Particle Correlations and Flow}
In addition to measuring flow by the correlation of individual particles to the reaction plane, it is also possible to measure flow by the correlation of two-particles to each other. The advantage of this method is that one does not have to determine the reaction plane. Thus if two particles $A$ and $B$ are correlated to the reaction plane, but not otherwise correlated to each other, 
\[ \frac{dN^A}{d\phi^A}\propto 1+\sum_n 2 v_n^A \cos(n(\phi^A-\Psi_n)) ,\quad \frac{dN^B}{d\phi^B}\propto 1+\sum_n 2 v_n^B \cos(n(\phi^B-\Psi_n)) \]
then the correlation to the reaction plane induces a correlation of these two particles to each other which can be measured without knowledge of the reaction plane,
\begin{equation} \frac{dN^{AB}}{d\phi^A d\phi^B}\propto \left[1+2 v_2^A v_2^B \cos 2(\phi^A-\phi^B) + 2 v_3^A v_3^B \cos 3(\phi^A-\phi^B) +\ldots\right] \ .\label{eq:2partdphi}\end{equation}

In p-p collisions there is no collective flow but there are strong two-particle azimuthal correlations due to di-jet production in hard-scattering (see Fig.~\ref{fig:collstar}a), which also exist in A+A collisions but are obscured by the large multiplicity [e.g. can you find a jet in Fig.~\ref{fig:collstar}b]. Before the discovery of jets, two-particle correlations were used extensively at the CERN-ISR to study hard-scattering (the production of particles with large transverse momentum) in p-p collisions which was discovered there~\cite{egseeMJTISSP2009}. Due to the huge multiplicities in Au+Au collisions at RHIC, where for central Au+Au collisions at $\sqrt{s_{NN}}=200$ GeV, there is an estimated $\pi\Delta r^2\times{1\over {2\pi}} {dE_T\over{d\eta}}\sim 375$ GeV of energy 
in one unit of the nominal jet-finding cone,  $\Delta r=\sqrt{(\Delta\eta)^2 + (\Delta\phi)^2}$, two-particle correlations  were used exclusively for the first 10 years to study hard scattering at RHIC.

\begin{figure}[!thb]
\begin{center}
\begin{tabular}{cc}
\begin{tabular}[b]{c}
\hspace*{-0.10in}\includegraphics[width=0.52\linewidth]{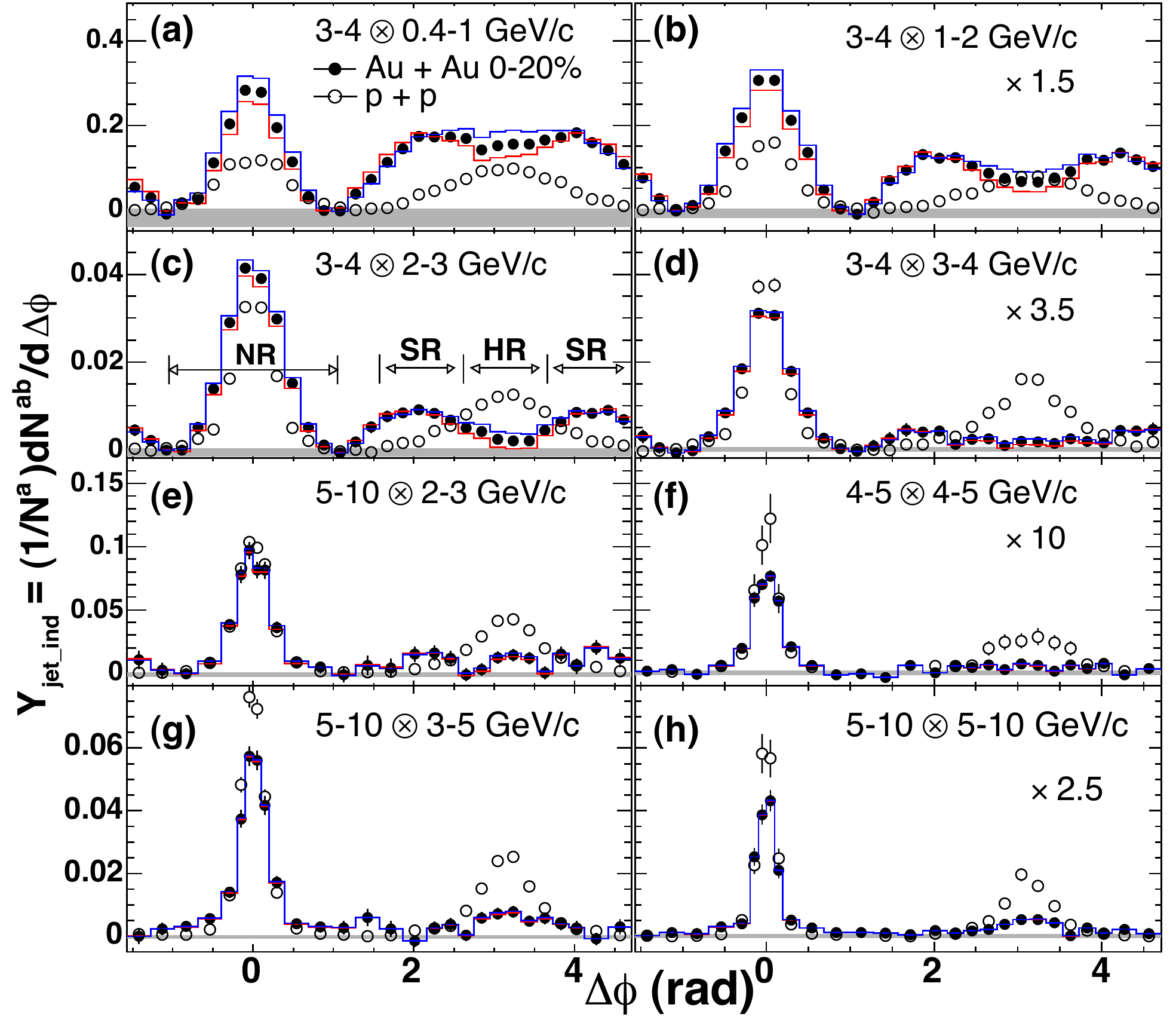}
\end{tabular}\hspace*{-0.2pc}
\begin{tabular}[b]{c}
\includegraphics[width=0.43\linewidth]{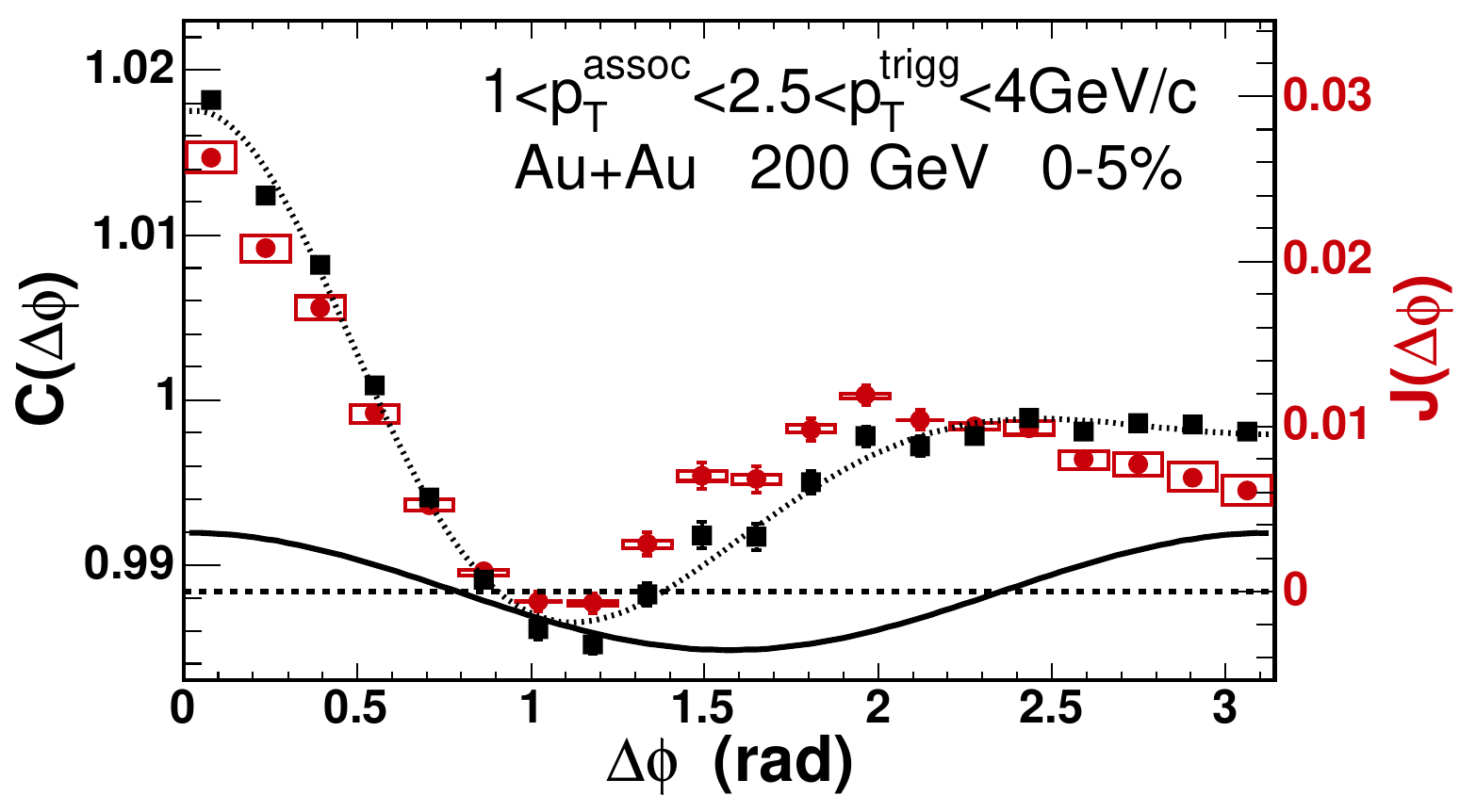}\\[0.3pc]
\includegraphics[width=0.42\linewidth]{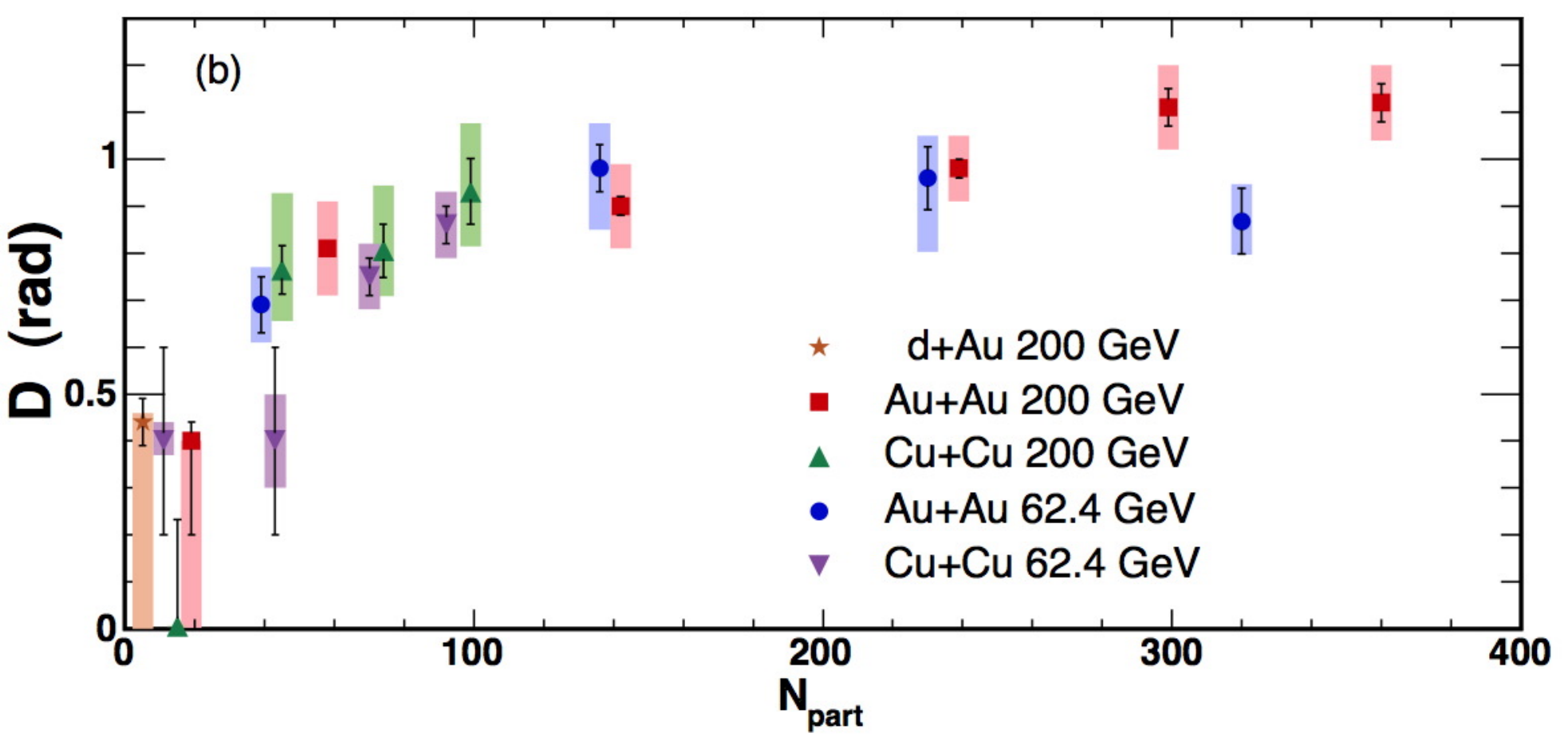}
\end{tabular}
\end{tabular}
\end{center}
\vspace*{-0.28in}
\caption[]
{a-h) (left) PHENIX~\cite{ppg083} azimuthal correlation conditional yield of associated $h^{\pm}$ particles with $p_{T_a}$ for trigger $h^{\pm}$ with $p_{T_t}$ for the various $p_{T_t} \otimes p_{T_a}$ combinations shown. i) (right)-(top)  PHENIX~\cite{ppg067} azimuthal correlation function $C(\Delta\phi)$ of $h^{\pm}$ with $1\leq p_{T_a}\leq 2.5$ GeV/c with respect to a trigger $h^{\pm}$ with $2.5\leq p_{T_t} \leq 4$ GeV/c in Au+Au central collisions, where the line with data points indicates $C(\Delta\phi)$ before correction for the azimuthally modulated ($v_2$) background, and the other line is the $v_2$ correction which is subtracted to give the jet correlation function $J(\Delta\phi)$ (data points). j) (right)-(bottom) PHENIX $D$ parameters~\cite{ppg067}, the angular distance of the apparently displaced peak of the $J(\Delta\phi)$ distribution from the angle $\Delta\phi=\pi$ as a function of centrality, represented as the number of participants $N_{\rm part}$, for the systems and c.m. energies indicated.
\label{fig:HSD} }
\end{figure}

Typical examples of the di-hadron measurements in p-p and Au+Au central (0--20\%) collisions at $\sqrt{s_{NN}}=200$ GeV are shown in Fig.~\ref{fig:HSD}~\cite{ppg083,ppg067} which are presented as azimuthal distributions of the conditional yields of associated particles, with $p_{T_a}$, with respect to trigger particles with $3\leq p_{T_t}\leq10$ GeV/c. The di-jet structure in p-p collisions 
is clearly indicated by the gaussian-like strong azimuthal correlation peaks on the same side ($\Delta\phi=\phi_a-\phi_t\sim0$) and away side ($\Delta\phi\sim\pi$ rad.) relative to the trigger particle for all ranges of $p_{T_t}$ and $p_{T_a}$ measured.      
However, one of the many interesting features in Au+Au collisions is that the away side azimuthal jet-like correlations (Fig.~\ref{fig:HSD}c) are much wider than in p-p collisions and show a two-lobed structure (``the shoulder'' (SR)) at lower $p_{T_t}$ with a dip at 180$^\circ$,  reverting to the more conventional structure of a peak at 180$^\circ$ (``the head'' (HR)) for larger $p_{T_t}$. 

\begin{figure}[!hbt]
\begin{center}
\includegraphics[width=0.44\linewidth]{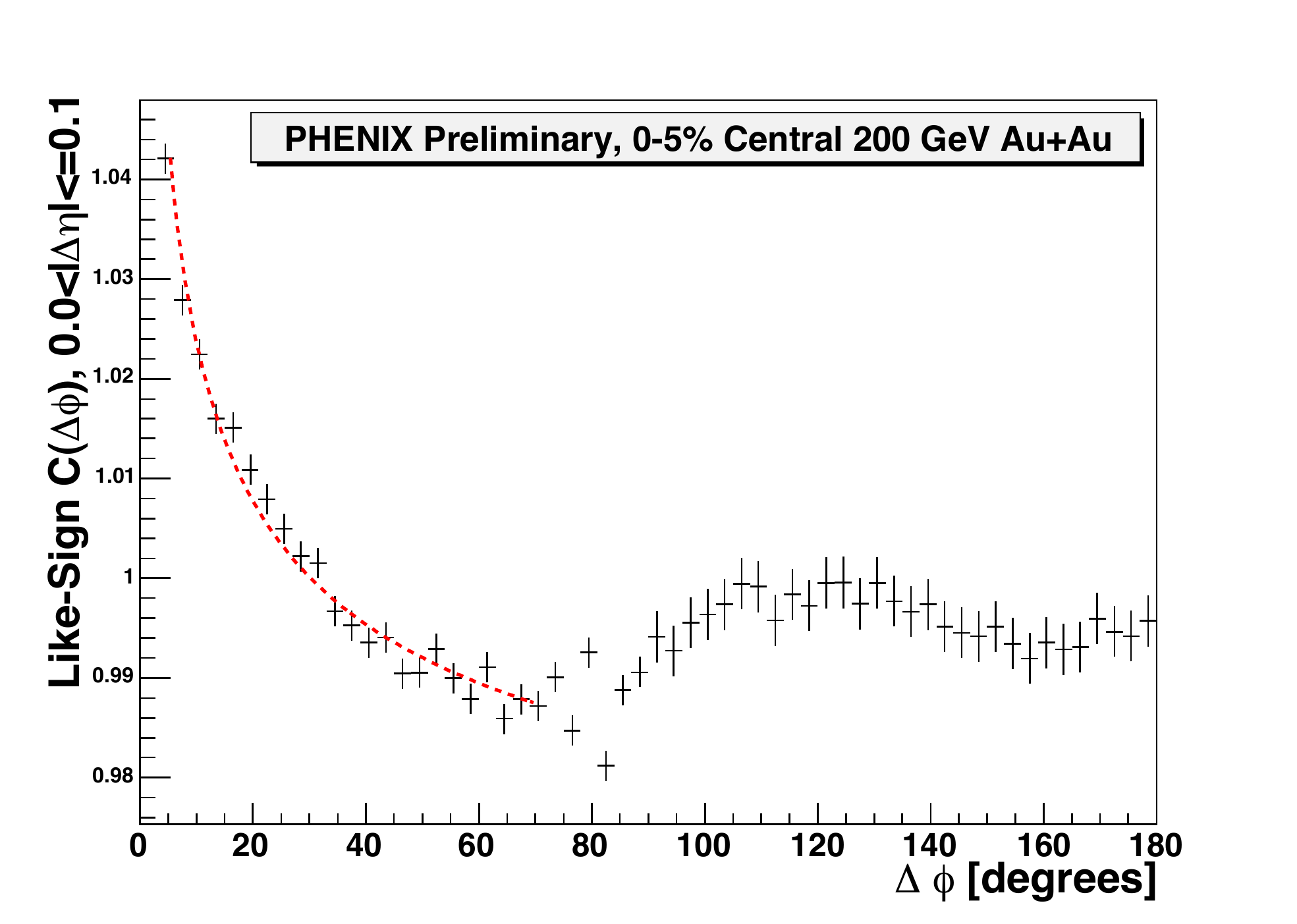}\hspace*{-0.5pc}
\raisebox{-0.06\linewidth}{\includegraphics[width=0.56\linewidth]{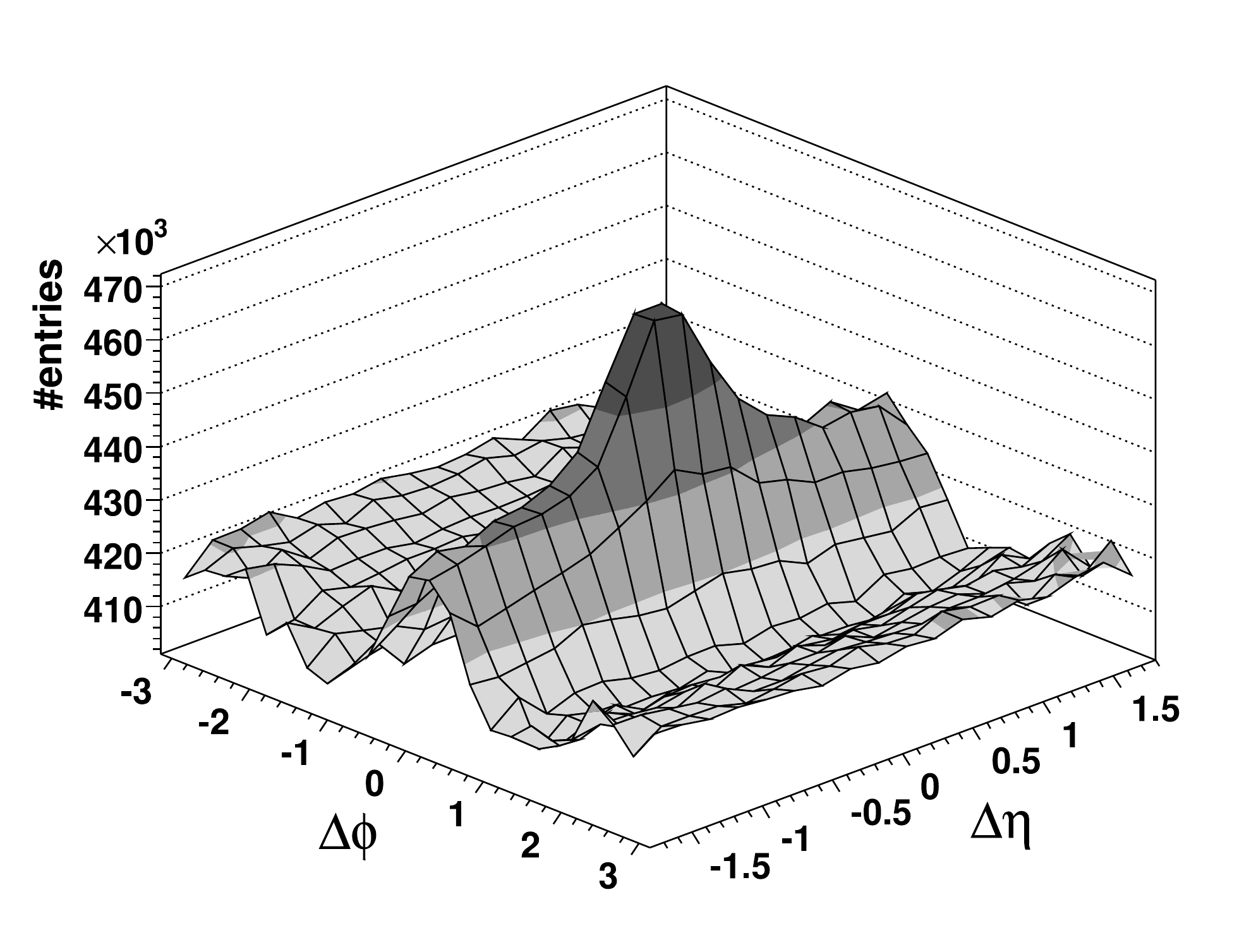}}
\end{center}
\vspace*{-0.28in}
\caption[]
{a) (left) Low $p_T$ like-sign pair azimuthal correlation function for 0-5\% central Au+Au collisions at $\sqrt{s_{NN}}=200$ GeV from charged hadrons with $0.2\leq p_{T_1}, p_{T_2}\leq 0.4$ GeV/c~\cite{JTMQM06}. b) (right) ``The Ridge''~\cite{PutschkeHP06,JacobsHP04}.\label{fig:moreartifacts} }
\end{figure}

The wide away-side correlation in central Au+Au collisions is significantly obscured by the large multiparticle background which is modulated in azimuth by the $v_2$ collective flow of a comparable width to the jet correlation (Fig.~\ref{fig:HSD}i). After the $v_2$ correction, the double peak structure $\sim \pm 1$ radian from $\pi$, with a dip at $\pi$ radians, becomes evident. The double-peak structure may indicate a reaction of the medium to a passing parton in analogy to a ``sonic boom'' or the wake of a boat, which was given the name ``Mach Cone''~\cite{CSST-Coney05}, and has been under active study both theoretically~\cite{egseeppg083} and experimentally. PHENIX characterizes this effect by the half-width $D$ ($\sim 1.1$ radian) of the Jet function, $J(\Delta\phi)$, the angular distance of the displaced peak of the distribution from the angle $\Delta\phi=\pi$. One of the striking features of the wide away side correlation is that the width $D$ (Fig.~\ref{fig:HSD}j)  does not depend on centrality, angle to the reaction plane,  $p_{T_a}$ and $\sqrt{s_{NN}}$, which always seemed problematic to me if the effect were due to a reaction to the medium. Another suspicious issue is that the same effect occurs even for auto-correlations of particles with very low $p_T$ between 0.2 and 0.4 GeV/c where any effect of hard-scattered partons should be submerged by the predominant soft physics (Fig.~\ref{fig:moreartifacts}a)~\cite{JTMQM06}.    

In addition to the Head/Shoulder or Mach Cone effect in two-particle correlations on the away-side, same-side correlations also show a new effect in A+A collisions called ``the Ridge''~\cite{STARridgePRC80}. This is seen in two-dimensional correlations in $\Delta\eta, \Delta\phi$ (Fig.~\ref{fig:moreartifacts}b)~\cite{PutschkeHP06,JacobsHP04} where the associated yield distribution can be decomposed into a narrow jet-like peak at small angular separation which has a similar shape to that found in p-p collisions, and a component that is narrow in $\Delta\phi$ but depends only weakly on $\Delta\eta$, the ``ridge.'' 
However, new results this past year have dramatically changed this picture.

\subsection{Triangular flow, odd harmonics}
For the first 10 years of RHIC running, and dating back to the Bevalac, all the experts thought that the odd harmonics in Eq.~\ref{eq:siginv2} would vanish by the symmetry $\phi\rightarrow \phi+\pi$ of the almond shaped overlap region~\cite{AlverOllitrault} (Fig.~\ref{fig:MasashiFlow}). However, last year, 2010, an MIT graduate student an his Professor in experimental physics, seeking (at least since 2006) how to measure the fluctuations of $v_2$ in the PHOBOS experiment at RHIC,  realized that fluctuations in the collision geometry on an event-by-event basis, i.e. the distribution of participants from event-to-event, did not respect the average symmetry. This resulted in what they called ``participant triangularity'' and ``triangular flow'', or $v_3$ in Eq.~\ref{eq:siginv2}, which they measured using both PHOBOS and STAR data~\cite{AlverRoland}.\footnote{It was pointed out by Leticia Palhares in the discussion that a Brazilian group showed in 2009 that  
that the ridge and the cone, i.e. $v_3$, does appear in an event-by-event hydrodynamics calculation without jets~\cite{BrazilNuXuv3}, but the MIT group~\cite{AlverRoland} was the first to show it with real data.}  

\begin{figure}[!h]
\begin{center}
\includegraphics[width=0.75\linewidth]{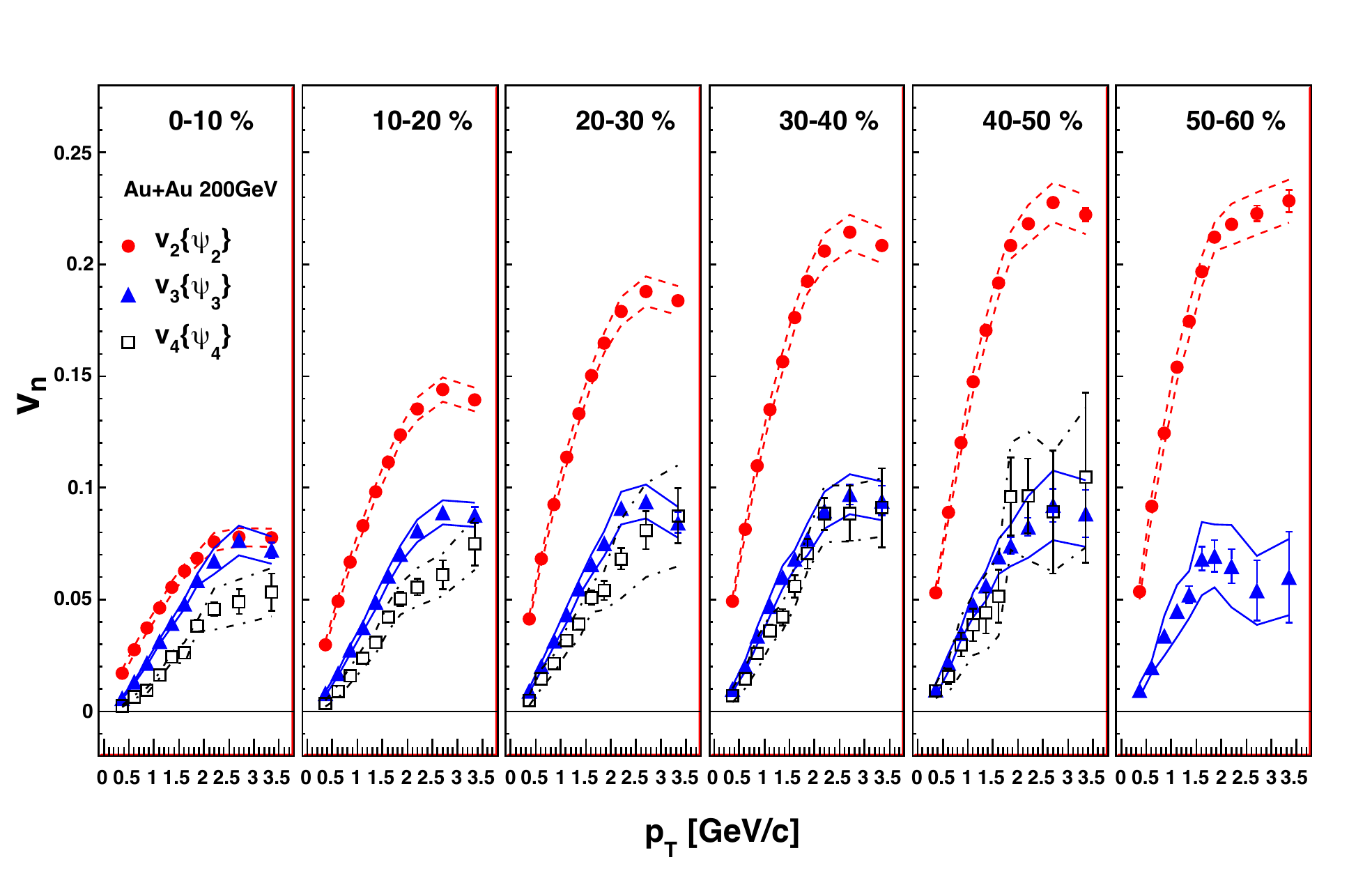}
\end{center}
\caption[]
{PHENIX~\cite{EsumiQM11} measurements of the $v_n$ parameters using Eq.~\ref{eq:siginv2} (with the appropriate reaction plane) as a function of $p_T$ for different centrality slices in $\sqrt{s_{NN}}=200$ GeV Au+Au collisions. \label{fig:PXv3} }
\end{figure}

Many experiments presented measurements of $v_3$ at Quark Matter 2011 this year, e.g. Fig.~\ref{fig:PXv3}~\cite{EsumiQM11},   and it was one of the most exciting results of the past year. 
There are two striking observations from Fig.~\ref{fig:PXv3} which indicate that fluctuations of the initial collision geometry are driving the observed $v_3$: i) the centrality dependence of $v_3(p_T)$ is weak as one would expect from fluctuations, but $v_2(p_T)$ which is most sensitive to the geometry of the ``almond''-shaped overlap region tracks the change in eccentricity with centrality; ii) for the most central collisions (0-10\%), where the overlap region is nearly circular so that all the $v_n$ are driven by fluctuations, $v_2(p_T)$, $v_3(p_T)$, $v_4(p_T)$ are comparable.  The fact that the observed collective flow of final state particles follows the fluctuations in the initial state geometry points to real hydrodynamic flow of a nearly perfect fluid (and convinces this author of the validity of hydrodynamics in RHI collisions, of which he was quite skeptical two years ago~\cite{egseeMJTISSP2009}). It is evident that $v_3$, a $\cos 3(\Delta\phi)$ term with lobes at $\Delta\phi=0, 2\pi/3$ and $4\pi/3 \approx 0, 2, 4$ radians, would explain the peaks at $\pi\pm D$ radian in the two-particle correlations (Fig.~\ref{fig:HSD}i,j) and also why $D\approx 1$ radian independent of centrality and kinematic variables; while the lobe at $\Delta\phi=0$ explains the ridge (Fig.~\ref{fig:moreartifacts}b). There is presently lots of activity to confirm in detail whether taking account of the odd harmonics in addition to $v_2$ and $v_4$ in the background of Fig.~\ref{fig:HSD}i will result in gaussian-like away-jet peaks in A+A collisions and the disappearance of the same-side ridge.

\section{RHIC beam energy scan---In search of the critical point}  
\begin{figure}[!thb]
   \begin{center}
\includegraphics[width=0.32\linewidth]{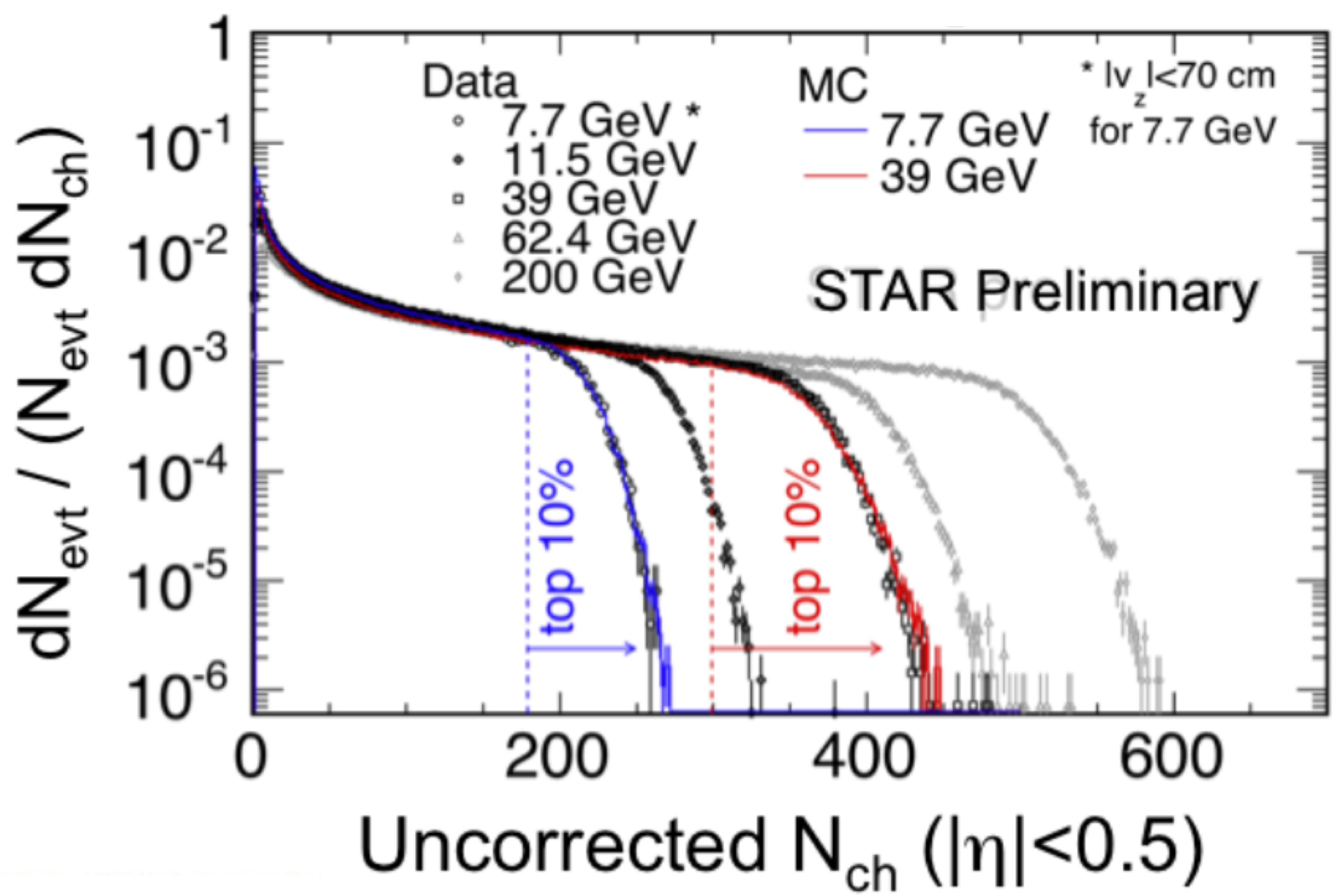}
\includegraphics[width=0.32\linewidth]{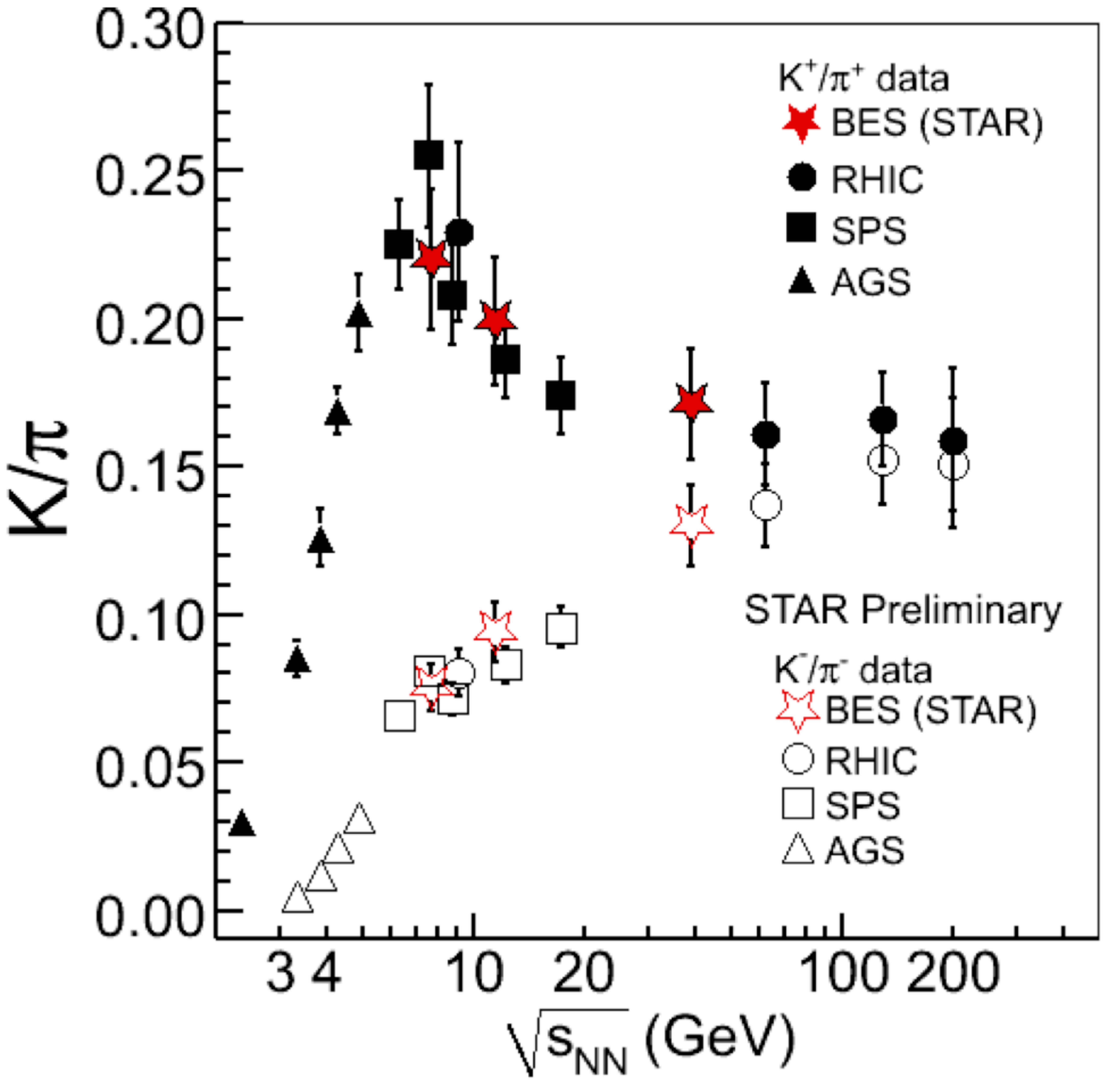}
\includegraphics[width=0.32\linewidth]{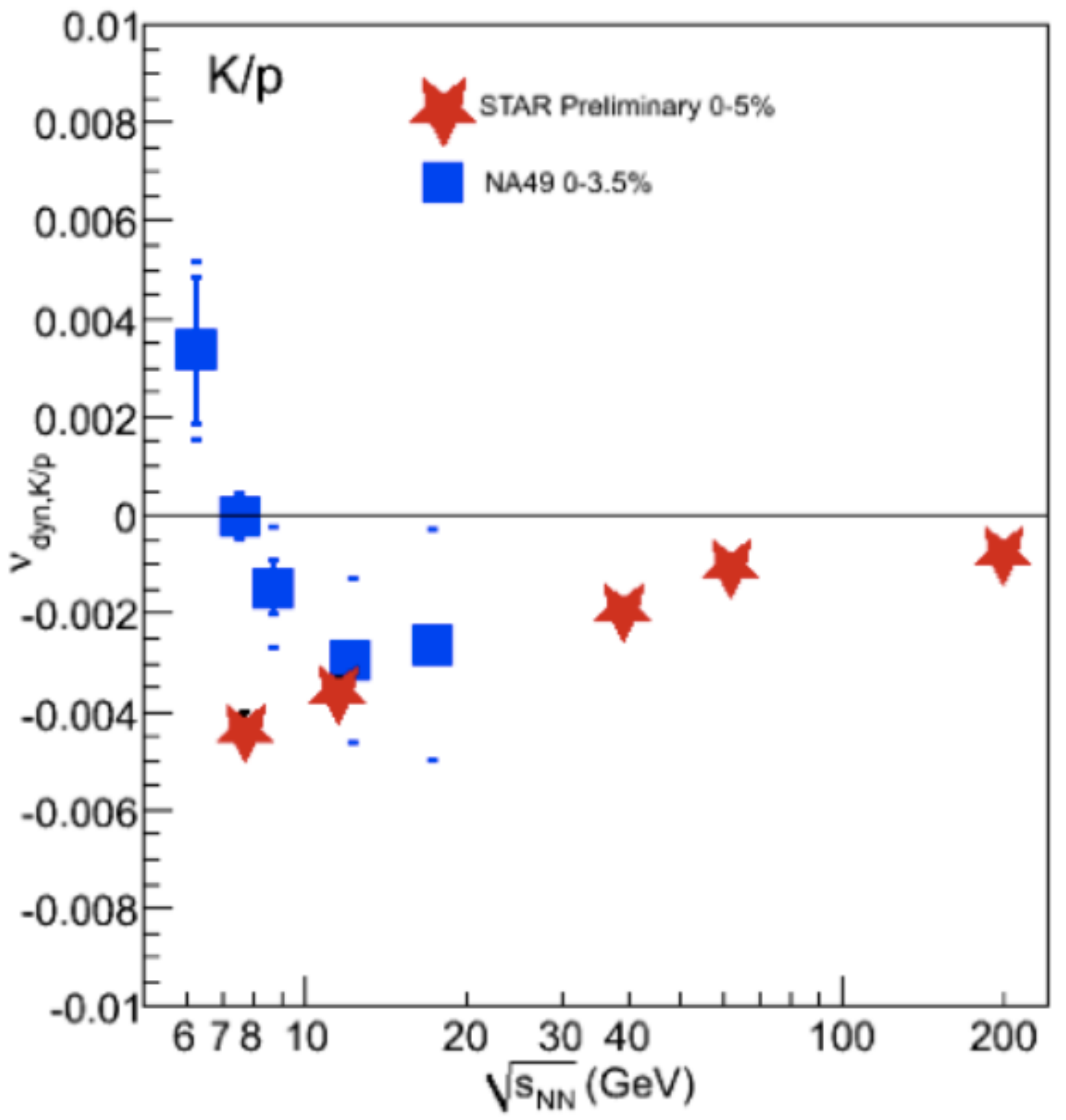}
\end{center}\vspace*{-0.25in}
\caption[]{a) (left) $N_{\rm ch}$ distribution in the STAR detector for 5 values of $\sqrt{s_{NN}}$~\cite{LKBMQM11}; b) (center) $K/\pi$ ratio vs $\sqrt{s_{NN}}$~\cite{LKBMQM11}; c) (right) Event-by-event fluctuation of $K/p$ ratio~~\cite{LKBMQM11}. }
\label{fig:byebyeMarek}
\end{figure}

The past year has seen runs at RHIC with Au+Au collisions at c.m. energies $\sqrt{s_{NN}}=7.7, 11.5$ and $39$ GeV, in addition to previous runs at 19.6, 62.4, 130 and 200 GeV, to search for the onset of large fluctuations which should occur near a critical point in the phase diagram (Fig.~\ref{fig:phase_boundary}). Such fluctuations in the $K/\pi$ and $K/p$ ratio had been claimed at the CERN SPS fixed target program near $\sqrt{s_{NN}}=8$ GeV and were presented~\cite{MarekQM11} as ``evidence of the onset of the deconfinement phase transition''. At QM2011, STAR~\cite{LKBMQM11} presented many excellent results on this subject of which I show a small selection in Fig.~\ref{fig:byebyeMarek}. 

Fig.~\ref{fig:byebyeMarek}a shows the multiplicity distribution $dN_{\rm ch}/d\eta$ which maintains the characteristic Nuclear Geometry dominated shape (as in Fig.~\ref{fig:nuclcoll}), stretching to higher multiplicity, $dN_{\rm ch}/d\eta$, with increasing $\sqrt{s_{NN}}$. Fig.~\ref{fig:byebyeMarek}b shows the $K^+/\pi^+$ and $K^-/\pi^-$ ratios over the entire range of $\sqrt{s_{NN}}$ measurements. The maximum of the $K^+/\pi^+$ ratio near $\sqrt{s_{NN}}=8$ GeV is naturally explained~\cite{CleymansOeschlerPLB615} by the change in dominant $K^+$ production from $K^+ \Lambda$ to $K^+ K^-$ whose smooth increase with $\sqrt{s_{NN}}$ can be seen from the $K^-/\pi^-$ ratio. The famous ``horn'', or apparent discontinuity, at $\sqrt{s_{NN}}=8$ GeV from the SPS data~\cite{MarekQM11} is greatly smoothed when the new STAR data are added. Fig.~\ref{fig:byebyeMarek}c shows a smooth variation of the the STAR measurements of the fluctuations of the event-by-event $K/p$ ratio as a function of $\sqrt{s_{NN}}$, which differs dramatically from the huge effect claimed by the SPS Fixed Target measurements below 12 GeV, notably the change from negative to positive~\cite{MarekQM11}. There is no doubt in this author's mind that one must prefer the collider measurements, where the detector position at mid-rapidity in the c.m. system is constant for all values of $\sqrt{s_{NN}}$, to the fixed target measurements, where the rapidity of the c.m. system moves dramatically with respect to the the detector as $\sqrt{s_{NN}}$ varies. This is a major strength of the RHIC Beam Energy Scan program.   
\subsection{A press release during the school}
On June 23, 2011, shortly before I was to give these lectures, a press release from LBL arrived claiming that ``By comparing theory with data from STAR, Berkeley Lab scientists and their colleagues map phase changes in the QGP''~\cite{LBLJune23}. Since I had criticized ``physics by press-release'' concerning the discovery of the \QGP\ (above), I felt that I was obliged to review the physics behind this latest example, presumably a ``Highlight from RHIC''.  

The subject is ``Fluctuations of conserved quantities'', in this case the net baryon distribution taken as $p-\bar{p}$. Since there can be no fluctuations of conserved quantities such as net charge or net baryon number in the full phase space, one has go to small intervals~\cite{AsakawaHMPRL85} to detect a small fraction of the protons and anti-protons which then fluctuates, i.e. varies from event to event. The argument is that, e.g. the fluctuation of one charged particle in or out of the considered interval produces a larger mean square fluctuation of the net electric charge if the system is in the hadron gas phase with integral charges than for the \QGP\ phase with fractional charges. 

However, while there are excellent statistical mechanical arguments about the utility of fluctuations of conserved quantities   such as net baryon number as a probe of a critical point~\cite{KochCFRNC06}, there are, so far, no adequate treatments of the mathematical statistics of the experimental measurements. Theoretical analyses tend to be made in terms of a Taylor expansion of the free energy $F=-T\ln Z$ around the critical temperature $T_c$ where $Z$ is the partition function, or sum over states, which is of the form $Z\propto e^{-(E-\sum_i \mu_i Q_i)/kT}$ and $\mu_i$ are chemical potentials associated with conserved charges $Q_i$~\cite{KochCFRNC06}.  The terms of the Taylor expansion, which are obtained by differentiation, are called susceptibilities, denoted $\chi$. The only connection of this method to mathematical statistics is that the cumulant generating function in mathematical statistics is also a Taylor expansion of the $\ln$ of an exponential:
\begin{equation}
g_x (t)=\ln\mean{e^{tx}}=\sum_{n=1}^\infty \kappa_n \frac{t^n}{n!} \qquad \kappa_m =\left.\frac{d^m g_x (t)}{dt^m}\right|_{t=0} \qquad .
\label{eq:cumgenfn}
\end{equation}
Thus, the susceptibilities are cumulants in mathematical statistics terms, where, in general, the cumulant $\kappa_m$ represents the $m^{\rm th}$ central moment with all $m$-fold combinations of the lower order moments subtracted.\footnote{Note that factorial cumulants, also known as ``Mueller moments''~\cite{AMuellerPRD4}, have been used previously for multiplicity distributions.} For instance, 
$\kappa_2=\mean{(x-\mu)^2}\equiv \sigma^2$, $\kappa_3=\mean{(x-\mu)^3}$, $\kappa_4=\mean{(x-\mu)^4}-3\kappa_2^2$, $\kappa_5=\mean{(x-\mu)^5}-10\kappa_3 \kappa_2$, where $\mu=\mean{x}$. Two so-called normalized or standardized cumulants are common in this field, the skewness, $S=\kappa_3/\sigma^3$ and the kurtosis, $\kappa=\kappa_4/\sigma^4=\mean{(x-\mu)^4}/\sigma^4-3$.  

\begin{figure}[!h]
   \begin{center}
\includegraphics[width=0.46\linewidth]{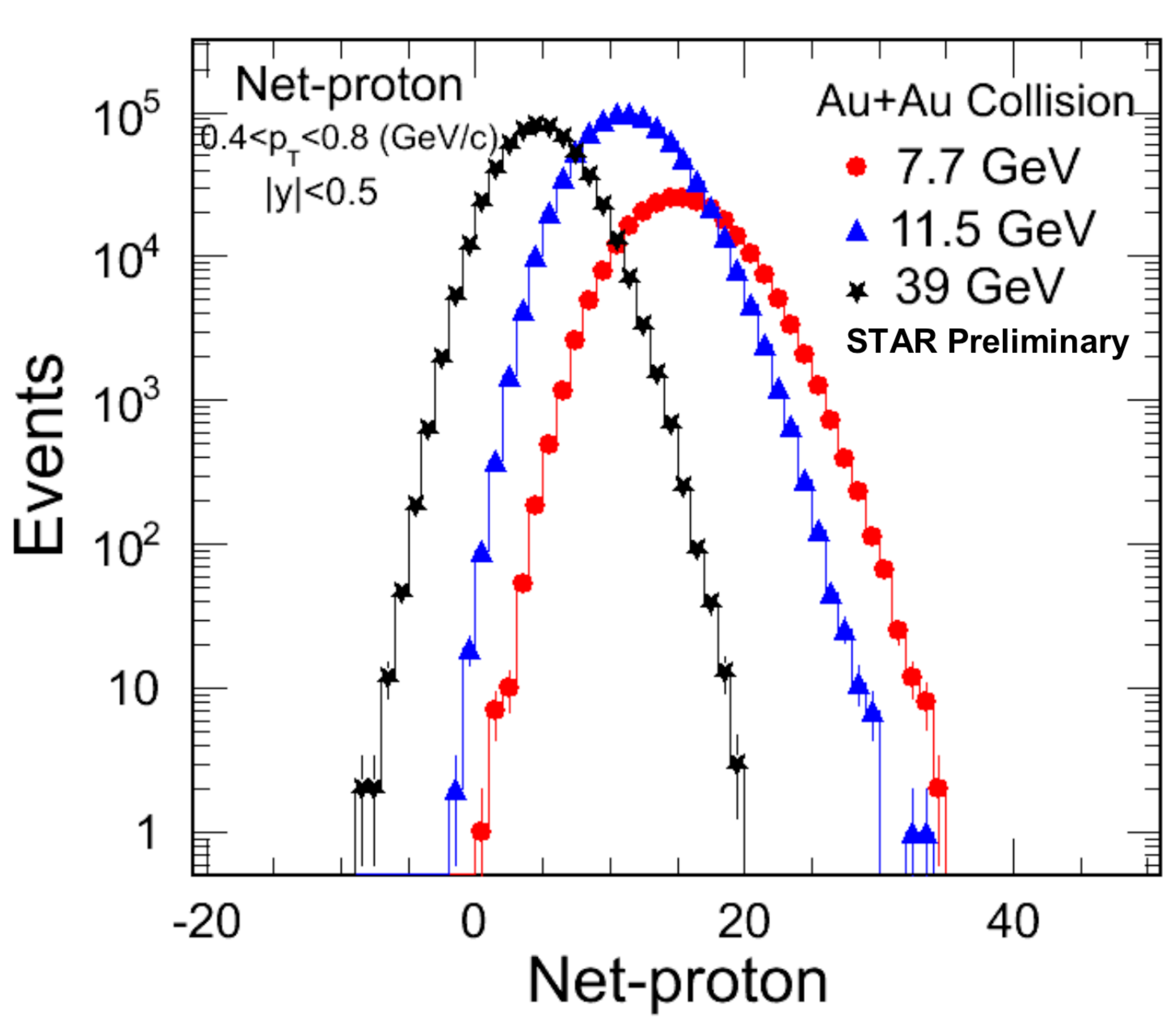}
\includegraphics[width=0.53\linewidth]{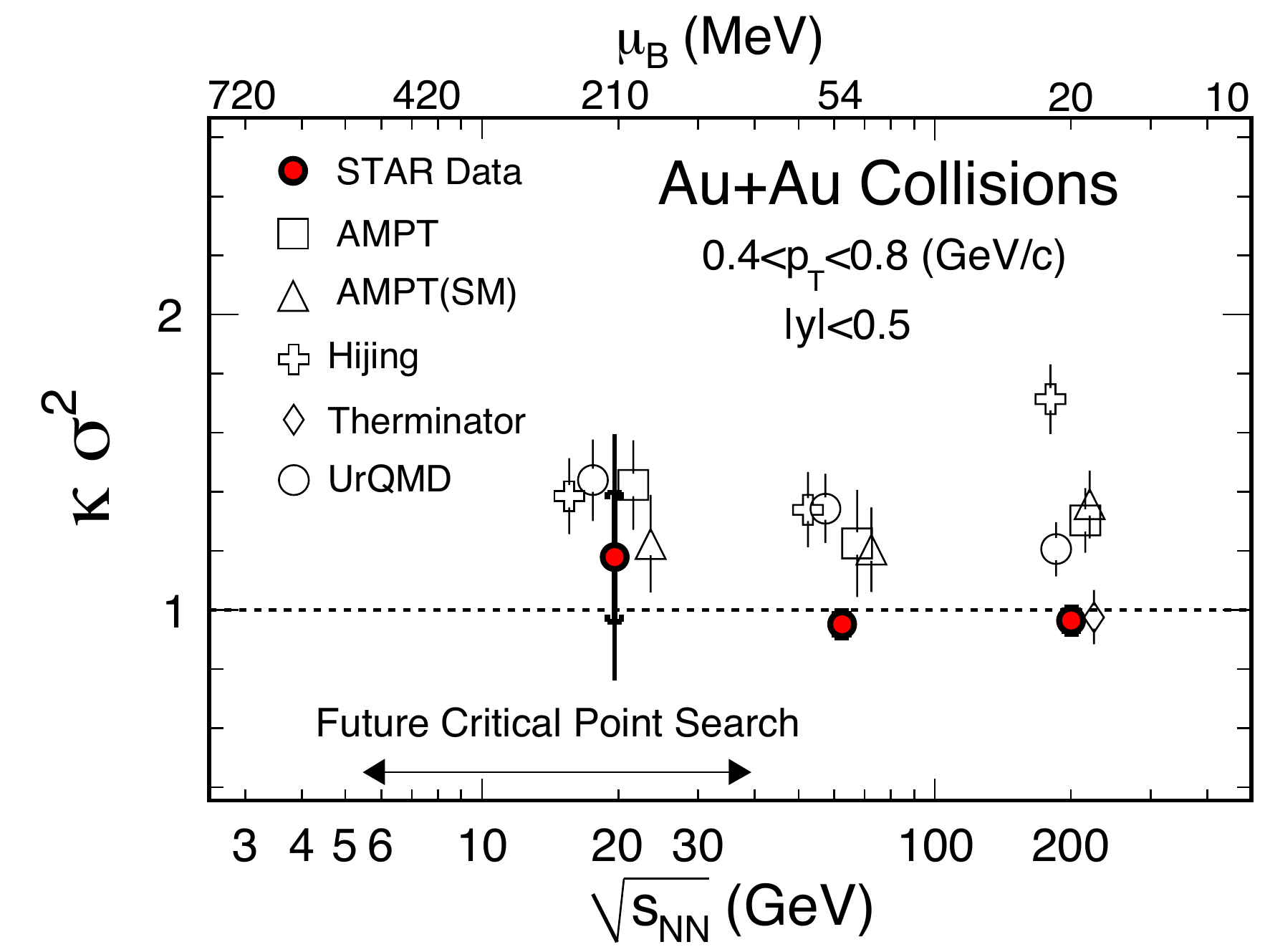}
\includegraphics[width=0.48\linewidth]{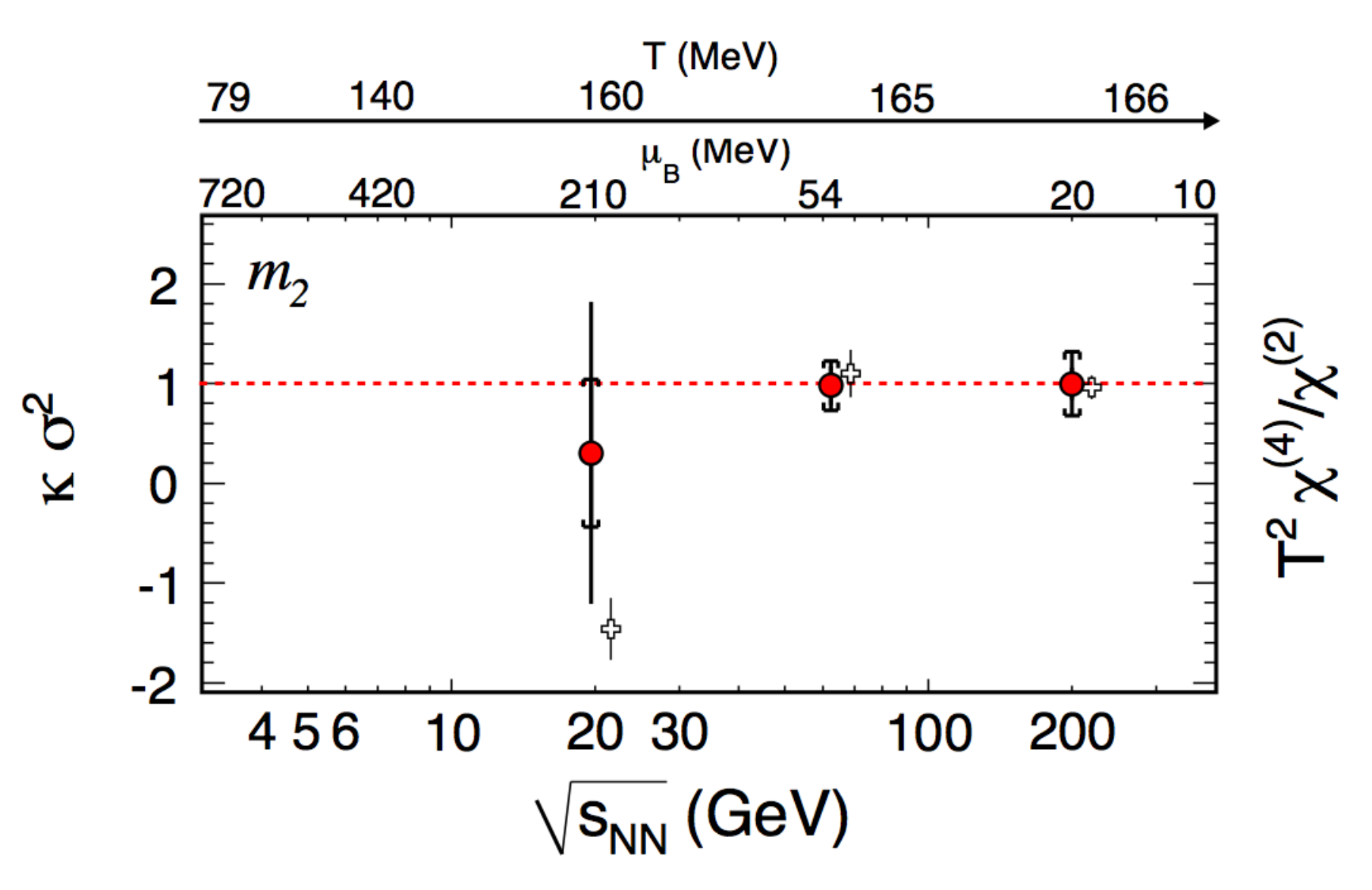}
\includegraphics[width=0.50\linewidth]{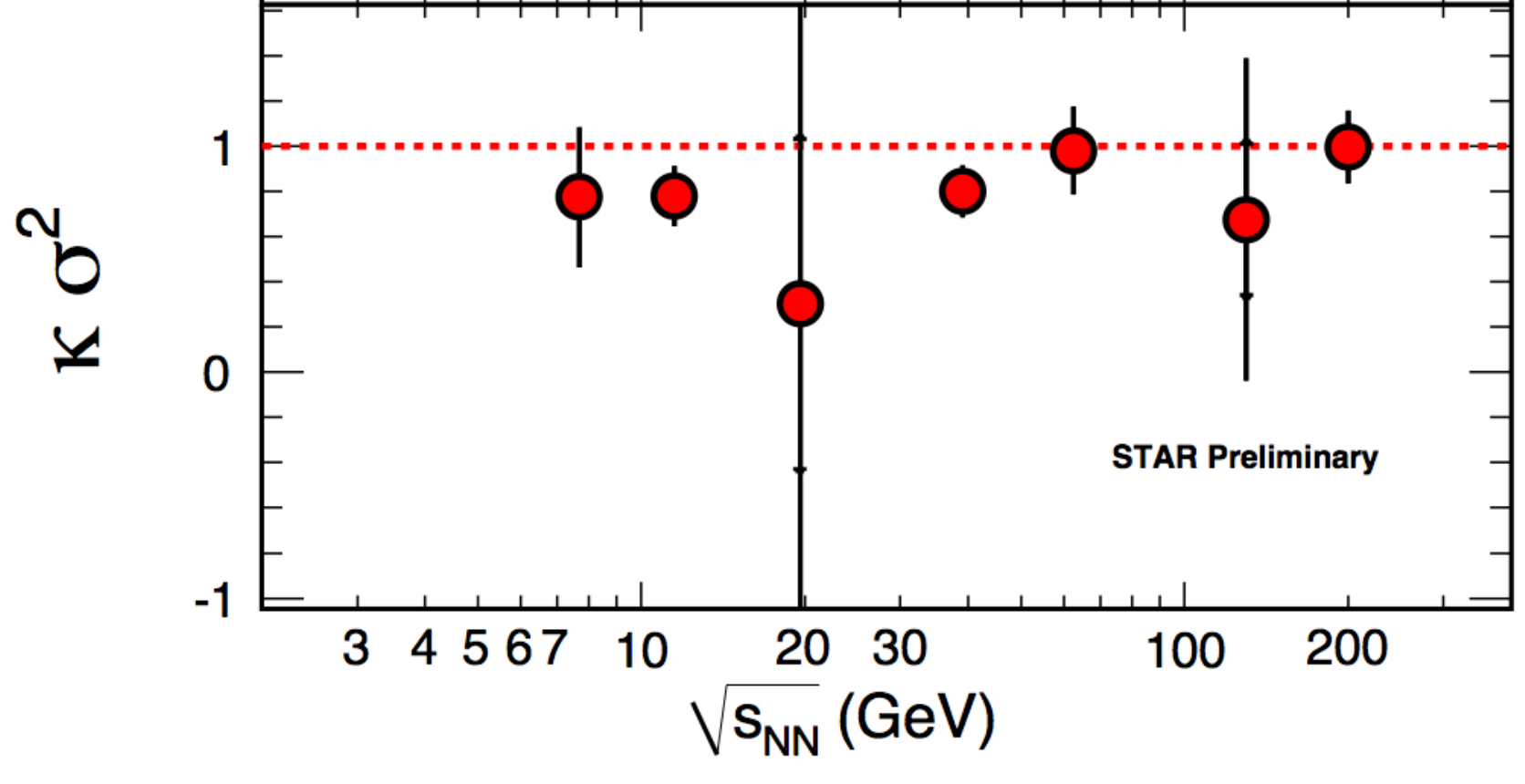}
\end{center}\vspace*{-0.25in}
\caption[]{a) (top-left) STAR~\cite{TarnowskyQM2011} distribution of event-by-event $p-\bar{p}$ at 3 values of $\sqrt{s_{NN}}$; b) (top-right) STAR published~\cite{STARnetPPRL105} measurements of $\kappa\sigma^2$; c) (bottom-left) Measurements from (b) as shown in Ref.~\cite{Science332} compared to the predicted ratio of susceptibilities (open crosses); d) (bottom-right) compilation~\cite{TarnowskyQM2011} of STAR measurements of $\kappa\sigma^2$.   }
\label{fig:STAR-NetP}
\end{figure}

A sample~\cite{TarnowskyQM2011} of STAR measurements of the distribution of net-protons in Au+Au collisions in the small interval $0.4\leq p_T\leq 0.8$ GeV/c, $|y|<0.5$ for different $\sqrt{s_{NN}}$ is shown in Fig.~\ref{fig:STAR-NetP}a. 

The moments in the form $\kappa\sigma^2=\kappa_4/\kappa_2$ are shown from a previous STAR publication~\cite{STARnetPPRL105} in  Fig.~\ref{fig:STAR-NetP}b while a plot, alleged to be of this same data, presented in the Lattice \QCD\ theory publication that generated the press-release, is shown in Fig.~\ref{fig:STAR-NetP}c~\cite{Science332}; and a plot of the $\kappa\sigma^2$ from the data of Fig.~\ref{fig:STAR-NetP}c, combined with the results from Fig.~\ref{fig:STAR-NetP}b, is shown in Fig.~\ref{fig:STAR-NetP}d~\cite{TarnowskyQM2011}. 
There are many interesting issues to be gleaned from Fig.~\ref{fig:STAR-NetP}. 

The data point at 20 GeV in Fig.~\ref{fig:STAR-NetP}c is not the published one from (b), as stated in the caption~\cite{Science332}, but the one from (d), which is different and with a much larger error. This, in my opinion, makes the data point look better compared to the predicted discontinuous value of $\kappa\sigma^2=-1.5$ for the critical point at 20 GeV (open crosses) in contrast to the predictions of 1.0 for both 62.4 and 200 GeV. The published measurements in (b) together with the new measurements in (d) are all consistent with $\kappa\sigma^2=1$; but clearly indicate the need for a better measurement at $\sqrt{s_{NN}}=20$ GeV. Apart from these issues, the main problem of comparing Lattice \QCD\ ``data'' to experimental measurements is that it is like comparing peaches to a fish, since the prediction is the result of derivatives of the log of the calculated partition function of an idealized system, which may have little bearing on what is measured using finite sized nuclei in an experiment with severe kinematic cuts. Maybe this is too harsh a judgement; but since this is the first such comparison (hence the press release), perhaps the situation will improve in the future. 

When I first saw the measured distributions in Fig.~\ref{fig:STAR-NetP}a, my immediate reaction was that STAR should fit them to Negative Binomial distributions so that they would know all the cumulants. However, I later realized that my favorite 3 distributions for integer random variables, namely, Poisson, Binomial, and Negative Binomial, are all defined only for positive integers, while the number of net-protons on an event can be negative as well as positive, especially at higher c.m. energies. This is why somebody should work out the mathematical statistics of the net proton distribution as we did in PHENIX for the distribution of the difference in foreground (opposite charge) and background (like charge) di-lepton events when both are Poisson distributed~\cite{ppg104}. Until then, it is instructive to compare the values of $\kappa\sigma^2$ in Fig.~\ref{fig:STAR-NetP} to those from the well-known distributions: Poisson, $\kappa\sigma^2=1$; Binomial, $\kappa\sigma^2=1-6p+6p^2<1$; Negative Binomial, $\kappa\sigma^2=1+6\mu/k+6\mu^2/k^2>1$; Gaussian, all cumulants=0 for $k>2$, so $\kappa\sigma^2=0$. The data favor Poisson (i.e. no correlation) everywhere, with some hint of Binomial. Nevertheless, if a future measurement would show a significant huge discontinuity of $\kappa\sigma^2$ similar to the theoretical prediction at $\sqrt{s_{NN}}=20$ GeV, then even I would admit that such a discovery would deserve a press release, maybe more!

\section{Hard Scattering as a probe of the \QGP}

\begin{wrapfigure}{l}{0.28\textwidth}
\begin{center}\vspace*{-2pc}
\includegraphics[width=0.30\textwidth]{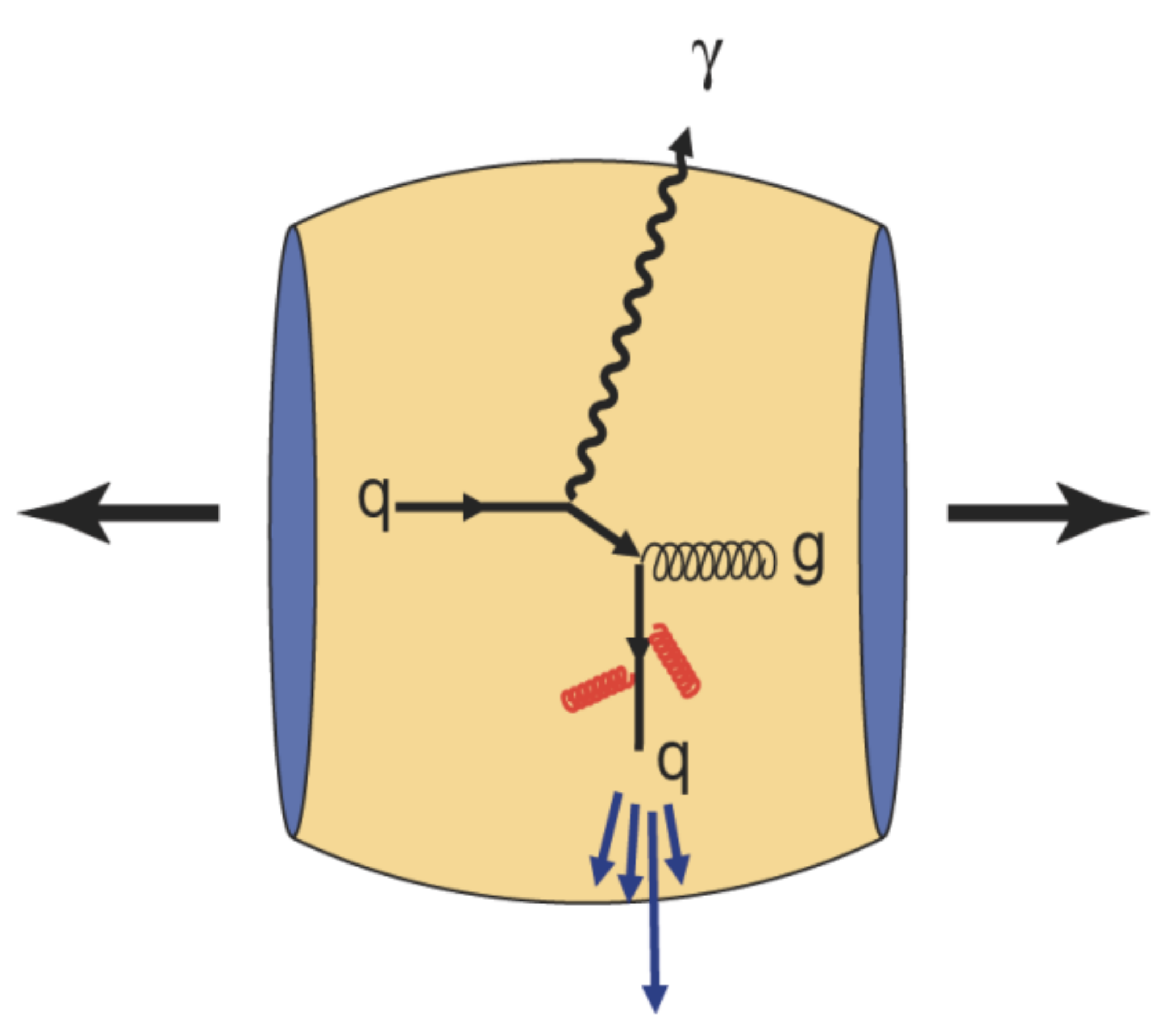}
\end{center}\vspace*{-2pc}
\caption[]{$g+q\rightarrow \gamma+q$ reaction with outgoing $\gamma$ and $q$ passing through the medium}
\label{fig:colldwg}
\end{wrapfigure}

One of the best probes found at RHIC to study the \QGP\ is the hard-scattering of quarks and gluons (the constituents of the nucleon) which can be observed via inclusive single particle production at large transverse momentum ($p_T$) or by two-particle correlations with a high $p_T$ trigger. The hard-scattering takes place in the initial collision of the highly Lorentz contracted nuclei.  The scattered partons which emerge near 90$^\circ$ to the collision axis (the sweet-spot for such observations) pass through the medium formed and then fragment to jets of particles which are detected (Fig.~\ref{fig:colldwg}). The most likely  
constituent reactions are: $g+g\rightarrow g+g$, $g+q\rightarrow g+q$, $q+q\rightarrow q+q$, and $g+q\rightarrow \gamma+q$ (shown). This last reaction~\cite{QCDCompton}, called direct-$\gamma$ production or the inverse QCD Compton effect, is the most beautiful because the $\gamma$-ray participates directly in the hard scattering, then emerges from the medium without interacting and can be measured with high precision. 
No fragmentation is involved and, in fact, these direct-$\gamma$ are distinguished from e.g. $\gamma$ rays from fragmenting partons because they are isolated, with no accompanying fragments. Triggering on a direct-$\gamma$ of a given $\hat{p}_{T_t}$ provides a `beam' of partons (8/1 u-quark) with exactly opposite (thus well-known) initial $p_T=-\hat{p}_{T_t}$, so that the effect of the medium can be determined by measuring, for instance, the ratio of the transverse momentum $\hat{p}_{T_a}$ of the jet from the away-parton to that of the direct-$\gamma$ trigger, denoted $\hat{x}_h=\hat{p}_{T_a}/\hat{p}_{T_t}$, or equivalently, the fractional jet imbalance, $1-\hat{x}_h$, as used by CMS at LHC~\cite{CMSdijet}.

Since hard-scattering at high $p_T >2$ GeV/c is point-like, with distance scale $1/p_T < 0.1$ fm, the cross section in p+A (A+A) collisions, compared to p-p, should be larger by the relative number of possible point-like encounters, a factor of $A$ ($A^2$) for p+A (A+A) minimum bias collisions. When the impact parameter or centrality of the collision is defined, the proportionality factor becomes $\mean{T_{AA}}$, the average overlap integral of the nuclear thickness functions.

\subsection{Jet quenching---suppression of high $p_T$ particles}
\begin{figure}[!t]
\includegraphics[height=0.25\textheight]{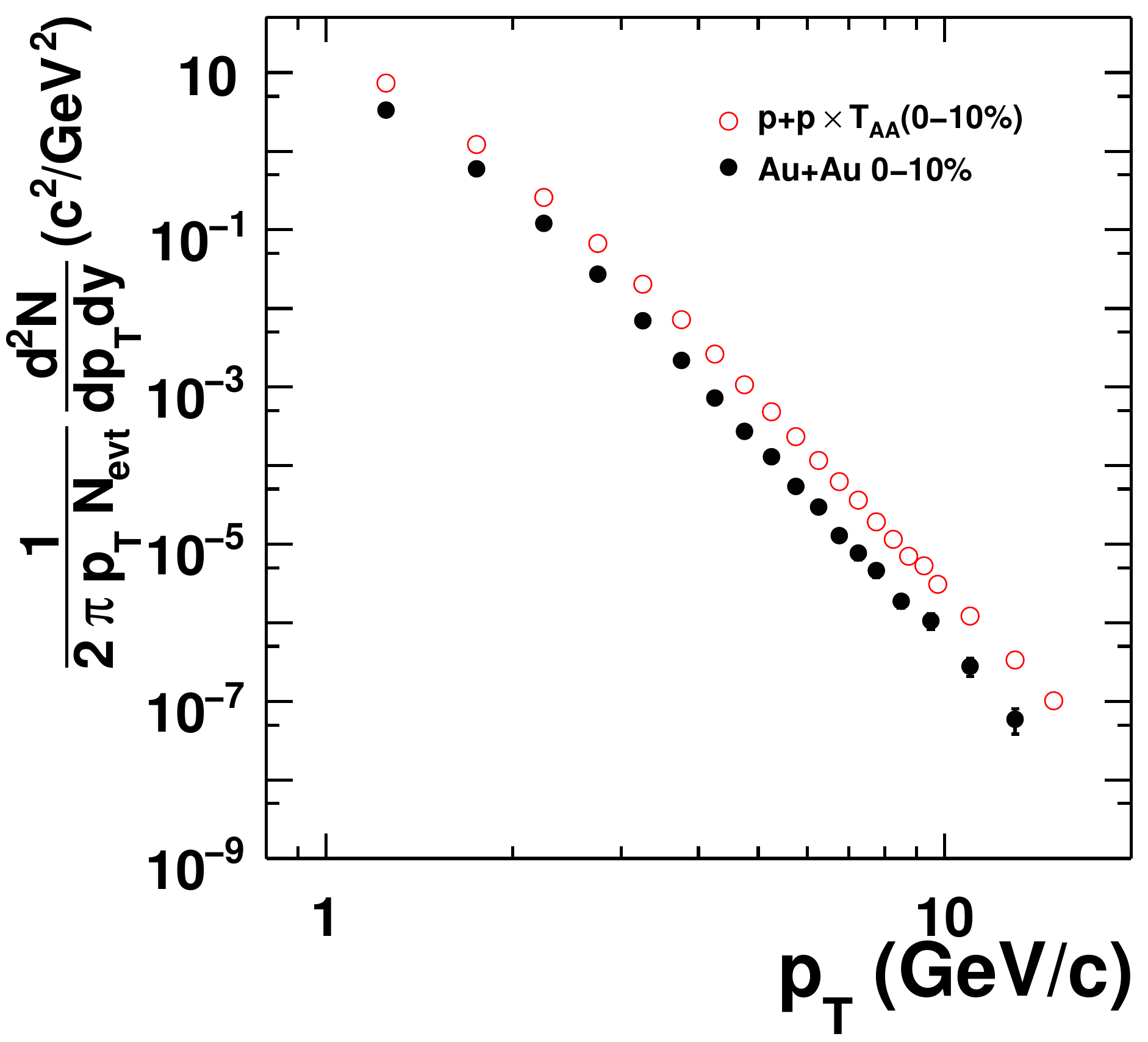}
\hspace*{0.04\textwidth} \includegraphics[height=0.25\textheight]{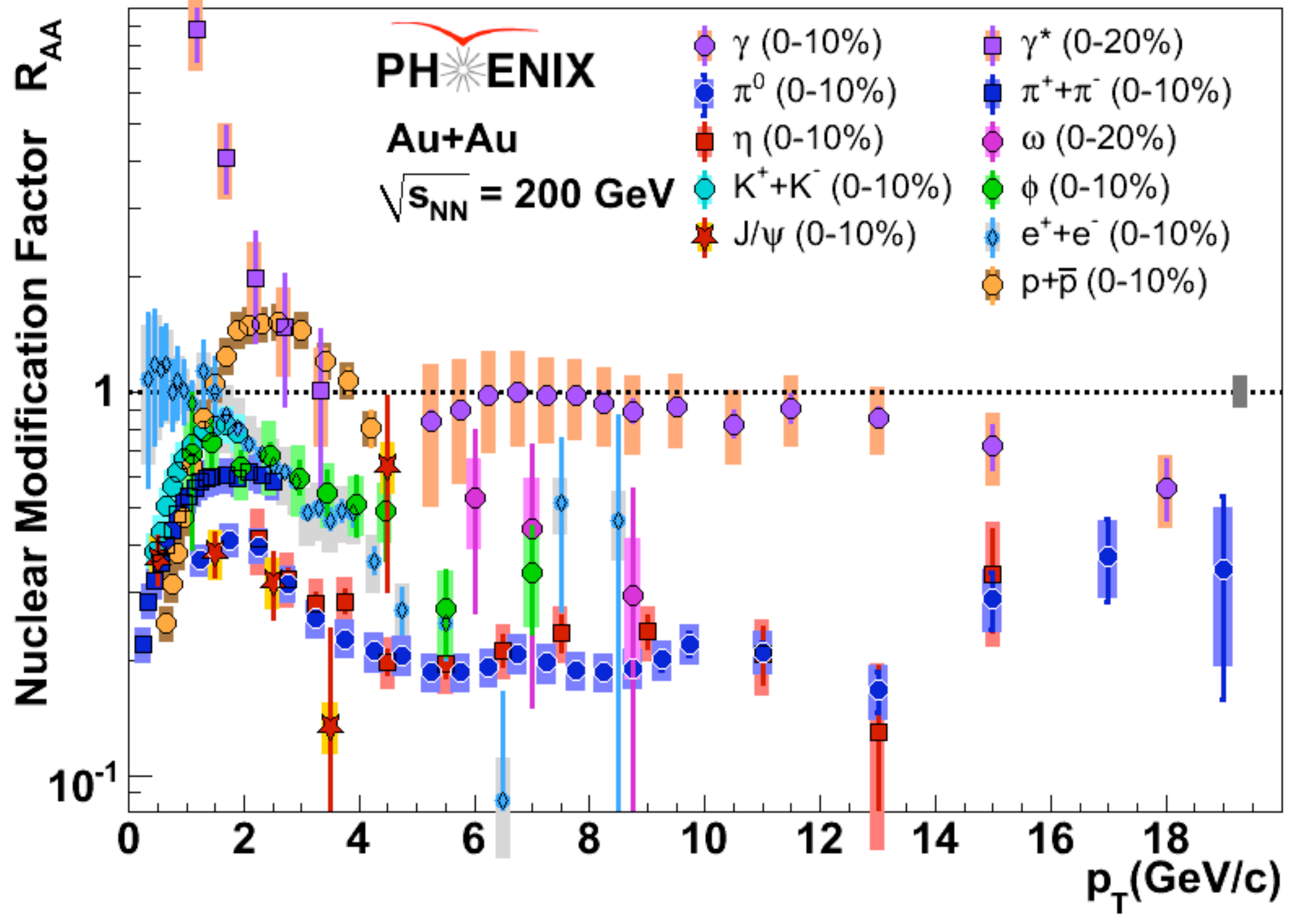}
\caption{a) (left) Invariant yield of $\pi^0$ at $\sqrt{s_{NN}}=200$ GeV as a function of transverse momentum $p_T$ in p-p collisions multiplied by $\mean{T_{AA}}$ for Au+Au central (0--10\%) collisions compared to the Au+Au measurement~\cite{ppg054}. b) (right) $R_{AA}(p_T)$ for all identified particles so far measured by PHENIX in Au+Au central collisions at $\sqrt{s_{NN}}=200$ GeV.}
\label{fig:Tshirt}
\end{figure}

   The discovery, at RHIC~\cite{ppg003}, that $\pi^0$'s produced at large transverse momenta are suppressed in central Au+Au collisions by roughly a factor of 5 compared to point-like scaling from p-p collisions is arguably {\em the}  major discovery in Relativistic Heavy Ion Physics. For $\pi^0$ (Fig.~\ref{fig:Tshirt}a)~\cite{ppg054} the hard-scattering in p-p collisions is indicated by the power law behavior $p_T^{-n}$ for the invariant cross section, $E d^3\sigma/dp^3$, with $n=8.1\pm 0.05$ for $p_T\geq 3$ GeV/c.  The Au+Au data can be characterized either as shifted lower in energy relative to the point-like scaled p-p data, or down in magnitude, i.e. suppressed. In Fig.~\ref{fig:Tshirt}b, the suppression of the many identified particles measured by PHENIX at RHIC is presented as the Nuclear Modification Factor, 
$R_{AA}(p_T)$, the ratio of the yield of  per central Au+Au collision (upper 10\%-ile of observed multiplicity)  to the point-like-scaled p-p cross section:
   \begin{equation}
  R_{AA}(p_T)=\frac{{d^2N^{\pi}_{AA}/dp_T dy N_{AA}}} { \mean{T_{AA}} d^2\sigma^{\pi}_{pp}/dp_T dy} \quad . 
  \label{eq:RAA}
  \end{equation}
 
The striking differences of $R_{AA}(p_T)$ in central Au+Au collisions for the many particles measured by PHENIX  (Fig.~\ref{fig:Tshirt}b) illustrates the importance of particle identification for understanding the physics of the medium produced at RHIC. Most notable are the equal suppression by a constant factor of 5 of $\pi^0$ and $\eta$ for $4\leq p_T \leq 15$ GeV/c, the equality of suppression of direct-single $e^{\pm}$ (from heavy quark ($c$, $b$) decay) and $\pi^0$ at $p_T\gsim 5$ GeV/c, the non-suppression of direct-$\gamma$ for $p_T\geq 4$ GeV/c and the exponential rise of $R_{AA}$ of direct-$\gamma$ for $p_T<2$ GeV/c~\cite{ppg086}, which is totally and dramatically different from all other particles and attributed to thermal photon production by many authors (e.g. see citations in reference~\cite{ppg086}). For $p_T\gsim 4$ GeV/c, the hard-scattering region,  the fact that all hadrons are suppressed but direct-$\gamma$ are not suppressed, indicates that suppression is a medium effect on outgoing color-charged partons.  The suppression can be explained by the energy loss of the outgoing partons in the dense color-charged medium due to coherent Landau-Pomeranchuk-Migdal radiation of gluons, predicted in QCD~\cite{BDMPS}, which is sensitive to properties of the medium. Measurements of two-particle correlations (discussed below) confirm the loss of energy of the away-jet relative to the trigger jet in Au+Au central collisions compared to p-p collisions. However, as we shall see, lots of details remain to be understood. 
  
The most interesting new results this year concern: i) direct-$\gamma$-hadron correlations at large $p_T$ to measure the fragmentation function in p-p collisions and to search for a possible modification of the fragmentation function in Au+Au; ii) $\pi^0$-hadron correlations to compare the fractional jet imbalance at RHIC to the new LHC measurement~\cite{CMSdijet}; iii) measurement of the flow ($v_2$) of direct-$\gamma$ at low and high $p_T$; and iv) confirmation at the LHC of the suppression of heavy quarks comparable to that of light quarks. 
\subsection{Fragmentation Function and Jet Imbalance} 
The key to measuring the fragmentation function of the jet of particles from a hard-scattered parton  is to know the energy of the original parton which fragments, as pioneered at LEP~\cite{Tingpizff}. Thus, in p-p collisions, a measurement of the direct-$\gamma-h$ correlation from $g+q\rightarrow \gamma+q$, where the $h$ represents charged hadrons opposite in azimuth to the direct-$\gamma$,  is (apart from the low rate) excellent for this purpose since both the transverse momentum and identity of the jet (8/1 $u$-quark, maybe 8/2 if the $\bar{q}+q\rightarrow \gamma+g$ channel is included) are known to high precision. 
		Two particle correlations are analyzed in terms of the two variables~\cite{ppg029}: $p_{\rm out}=p_T \sin(\Delta\phi)$, the out-of-plane transverse momentum of a track;  
 and $x_E$, where:\\ 
 \begin{minipage}[c]{0.5\textwidth}
\vspace*{-0.30in}
\[
x_E=\frac{-\vec{p}_T\cdot \vec{p}_{Tt}}{|p_{Tt}|^2}=\frac{-p_T \cos(\Delta\phi)}{p_{Tt}}\simeq \frac {z}{z_{\rm trig}}  
\]
\vspace*{0.06in}
\end{minipage}
\hspace*{0.01\textwidth}
\begin{minipage}[b]{0.50\textwidth}
\vspace*{0.06in}
\includegraphics[scale=0.5]{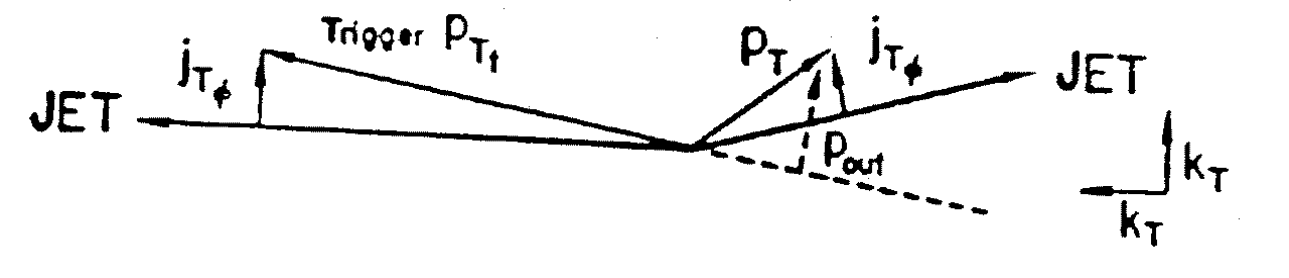}
\vspace*{-0.12in}
\label{fig:mjt-poutxe} \nonumber
\end{minipage}

\noindent $z_{\rm trig}\simeq p_{Tt}/p_{T{\rm jet}}$ is the fragmentation variable of the trigger jet, and $z$ is the fragmentation variable of the away jet. Note that $x_E$ would equal the fragmenation fraction $z$ of the away jet, for $z_{\rm trig}\rightarrow 1$, if the trigger and away jets balanced transverse momentum. The beauty of direct-$\gamma$ for this purpose is that $z_{\rm trig}\equiv 1$. 

\begin{figure}[t]
\includegraphics[height=0.225\textheight]{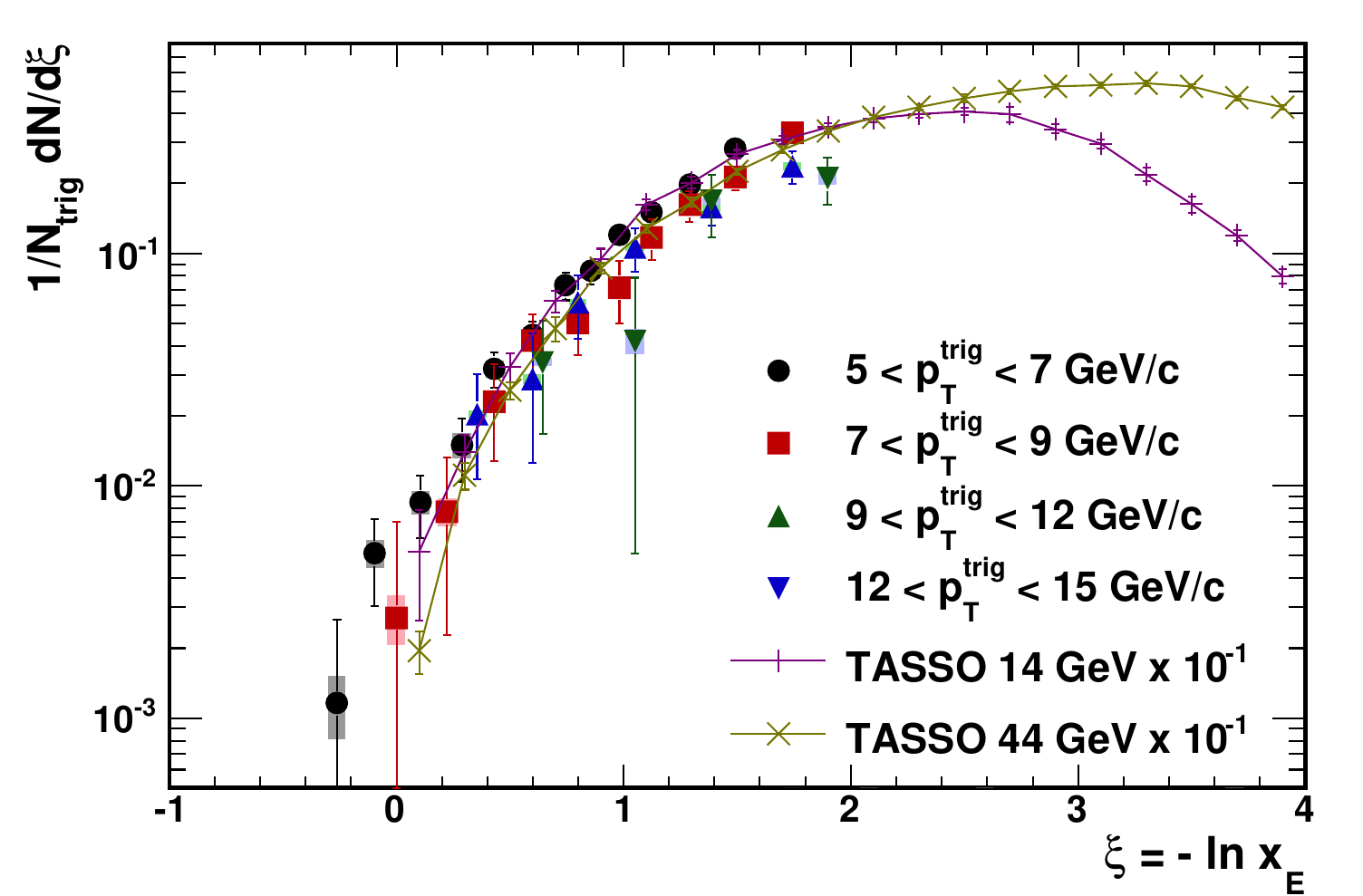}
\includegraphics[height=0.225\textheight]{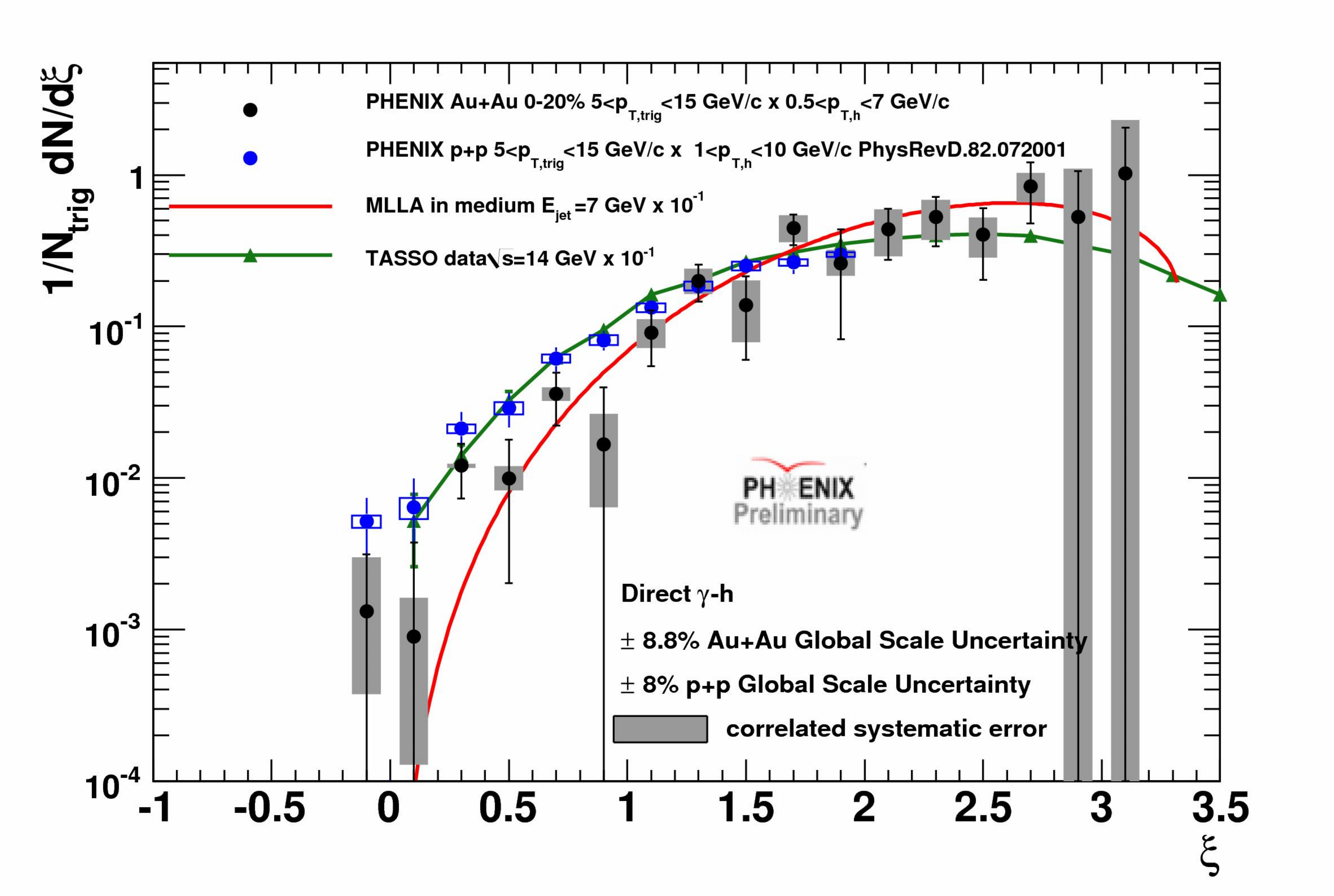}\vspace*{-0.5pc}
\caption{a) (right) $\xi=-\ln\, x_E$ distributions for PHENIX isolated direct-$\gamma$ p-p data~\cite{ppg095} for all $p_{T_t}$ ranges combined, compared to $e^+ e^-$ collisions at $\sqrt{s}=14$ and 44 GeV. b) (left) $\xi=-\ln\, x_E$ distribution in Au+Au (0--20\%)~\cite{ConnorsHP2010} }
\label{fig:BorgWied}\vspace*{-0.5pc}
\end{figure}

Following the approach of
Borghini and Wiedemann~\cite{BW06} who predicted the medium modification of fragmentation functions in the hump-backed or $\xi=\ln(1/z)$ representation, PHENIX measured $x_E$ distributions in p-p collisions~\cite{ppg095} and converted them to the $\xi=-\ln\, x_E$ representation (Fig.~\ref{fig:BorgWied}a) which are in quite excellent agreement with the dominant $u$-quark fragmentation functions measured in $e^+ e^-$ collisions at $\sqrt{s}/2=7$ and 22 GeV, which cover a comparable range in jet energy. 
PHENIX preliminary measurements~\cite{ConnorsHP2010} of the $\xi=-\ln\, x_E$ distribution in central (0--20\%) Au+Au collisions, which suggest a modification consistent with Ref.~\cite{BW06} are shown in Fig.~\ref{fig:BorgWied}b. A firm conclusion awaits final results with improved statistics. Fragmentation functions from full jet reconstruction in A+A collisions are not yet available at RHIC. 

\begin{figure}[!h]
\includegraphics[height=0.45\textwidth]{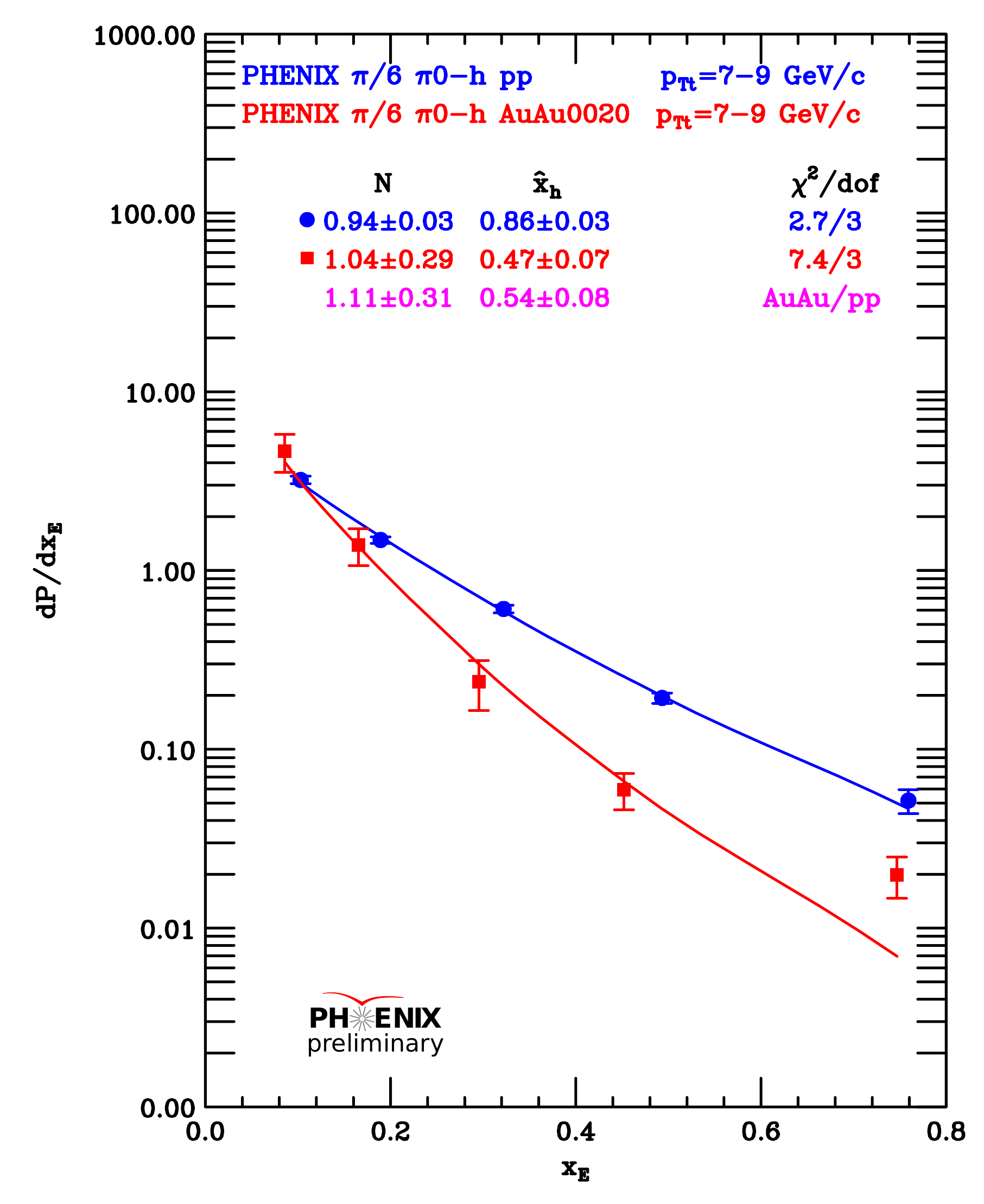}
\includegraphics[height=0.45\textwidth]{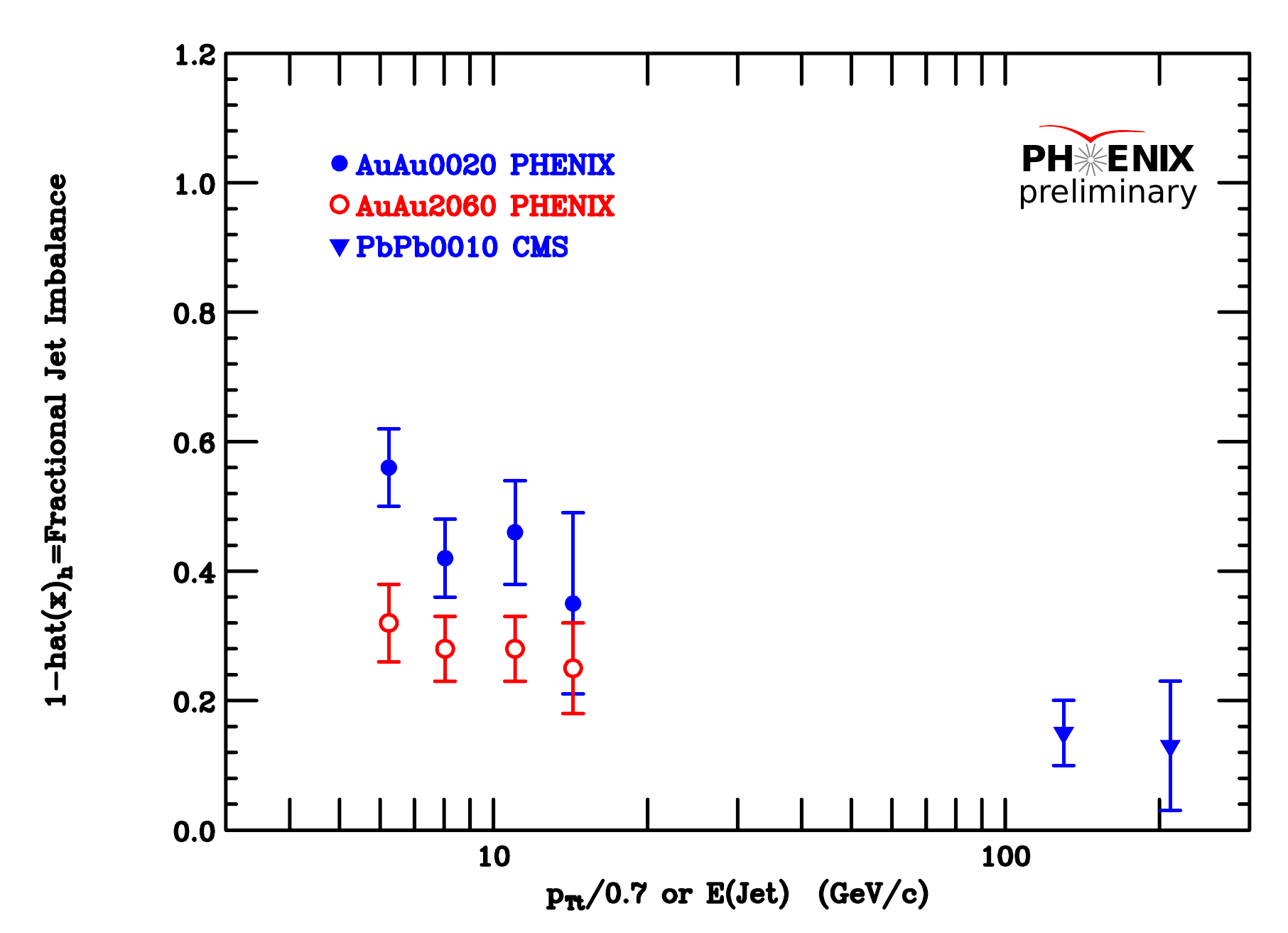}
\caption{(left) a) $x_E$ distributions~\cite{MJT-Utrecht} from p-p (circles) and AuAu 0-20\% centrality (squares) for $p_{T_t}=7-9$ GeV/c, together with fits to Eq.~\ref{eq:condxeN2} (solid lines) with parameters indicated.  The ratios of the fitted parameters for AuAu/pp are also given. b) (right) Fractional jet imbalance~\cite{MJT-Utrecht}, \hbox{$1-\hat{x}_h^{AA}/\hat{x}_h^{pp}$}, for RHIC and CMS data. }
\label{fig:AuAupp79}
\end{figure}

One of the important lessons learned at RHIC~\cite{ppg029} about fragmentation functions is that the away-side $x_E$ distribution of particles opposite to a trigger particle (e.g. a $\pi^0$), which is itself the fragment of a jet, does not measure the fragmentation function, but, instead, measures the ratio of $\hat{p}_{T_a}$ of the away-parton to $\hat{p}_{T_t}$ of the trigger-parton and depends only on the same power $n$ as the invariant single particle spectrum:  
		     \begin{equation}
\left.{dP \over dx_E}\right|_{p_{T_t}}\approx {N\,(n-1)}{1\over\hat{x}_h} {1\over
{(1+ {x_E \over{\hat{x}_h}})^{n}}} \, \qquad . 
\label{eq:condxeN2}
\end{equation}
This equation gives a simple relationship between the ratio, $x_E\approx p_{T_a}/p_{T_t}\equiv z_T$, of the transverse momenta of the away-side particle to the trigger particle, and the ratio of the transverse momenta of the away-jet to the trigger-jet, $\hat{x}_{h}=\hat{p}_{T_a}/\hat{p}_{T_t}$. PHENIX measurements~\cite{ppg106} of the $x_E$ distributions of $\pi^0$-h correlations in p-p and Au+Au collisions at $\sqrt{s_{NN}}=200$ GeV were fit to Eq.~\ref{eq:condxeN2} (Fig.~\ref{fig:AuAupp79}a)~\cite{MJT-Utrecht}. The steeper distribution in Au+Au shows that the away parton has lost energy. 
The results for the fitted parameters are shown on the figure. In general the values of $\hat{x}^{pp}_h$ do not equal 1 but range between $0.8<\hat{x}^{pp}_h<1.0$ due to $k_T$ smearing and the range of $x_E$ covered. In order to take account of the imbalance ($\hat{x}^{pp}_h <1$) observed in the p-p data, the ratio $\hat{x}_h^{AA}/\hat{x}_h^{pp}$ is taken as the measure of the energy of the away jet relative to the trigger jet in A+A compared to p-p collisions. The fractional jet imbalance was also measured directly with reconstructed di-jets by the CMS collaboration at the LHC in Pb+Pb central collisions at $\sqrt{s_{\rm NN}}=2.76$ TeV~\cite{CMSdijet}. They also showed a large effect in p-p collisions. I calculated $\hat{x}_h$ from their p-p and Pb+Pb results~\cite{MJT-Utrecht} and compared these LHC values of $1-\hat{x}_h^{AA}/\hat{x}_h^{pp}$ to those from PHENIX (Fig.~\ref{fig:AuAupp79}b). The large difference in fractional jet imbalance between RHIC and LHC c.m. energies could be due to the difference in jet $\hat{p}_{T_t}$ between RHIC ($\sim 20$ GeV/c) and LHC ($\sim 200$ GeV/c), or different sensitivity of the direct and indirect methods, or the difference in $n$ for the different $\sqrt{s}$, or to a difference in the properties of the medium. Future measurements will need to sort out these issues by extending both the RHIC and LHC measurements to overlapping regions of $p_T$.  

\section{Anisotropic flow ({$v_2$}) of direct-{$\gamma$} }
Although direct-$\gamma$ production~\cite{QCDCompton} is the most beautiful \QCD\  subprocess, it has a very serious problem: an overwhelming background of photons from high $p_T$ $\pi^0\rightarrow \gamma+\gamma$ and $\eta\rightarrow \gamma+\gamma$  decays  makes it a very difficult experiment.  One must measure all the background sources: $\pi^0$, $\eta$, \ldots, and calculate their contributions to the inclusive $\gamma$-ray spectrum. In principle, the background can be calculated whatever the $p_T$ distribution of the $\pi^0$ and $\eta$. However nature has been kind in that the invariant cross section for hard-scattering is a power law, $d\sigma/p_T dp_T\propto 1/p_T^n$, with $n=8.1\pm0.05$ for $\pi^0$ with $p_T\geq 3$ GeV/c (Fig.~\ref{fig:Tshirt}a)~\cite{ppg054}; also for $p_T\geq 3$ GeV/c, $\eta/\pi^0=0.48\pm 0.03$ is a constant. This implies that for $\pi^0\rightarrow \gamma +\gamma$, the spectrum of decay photons has the same power as the parent $\pi^0$ so that the ratio at any $p_T$ is a constant: 
\begin{equation}
\frac{\gamma}{\pi^0} \biggm |_{\pi^0}=\frac{2}{n-1} \qquad.
\label{eq:2nm1}
\end{equation}
The resulting background inclusive $\gamma$ spectrum from $\pi^0$ and $\eta$ decays at $\sqrt{s_{NN}}=200$ GeV is: 
\begin{equation} \gamma_{\rm background}/\pi^0 \approx (1+0.48\times0.39)\times 2/7.1 = 1.19\times2/7.1 = 0.334 \label{eq:gbkg} 
\end{equation}
where 0.39 is the branching ratio for $\eta\rightarrow \gamma +\gamma$. 
In PHENIX we plot what we call the double ratio: 
 \[R_{\gamma}=(\gamma_{\rm inclusive}/\pi^0)/(\gamma_{\rm background}/\pi^0)=\gamma_{\rm inclusive}/\gamma_{\rm background} \]
where it is important to see the calculated $\gamma_{\rm background}/\pi^0$ ratio, which usually comes from some opaque Monte Carlo program, to understand whether it makes sense according to Eqs.~\ref{eq:2nm1} and \ref{eq:gbkg}. 

\begin{figure}[!h]
\begin{center}
\includegraphics[width=0.95\textwidth]{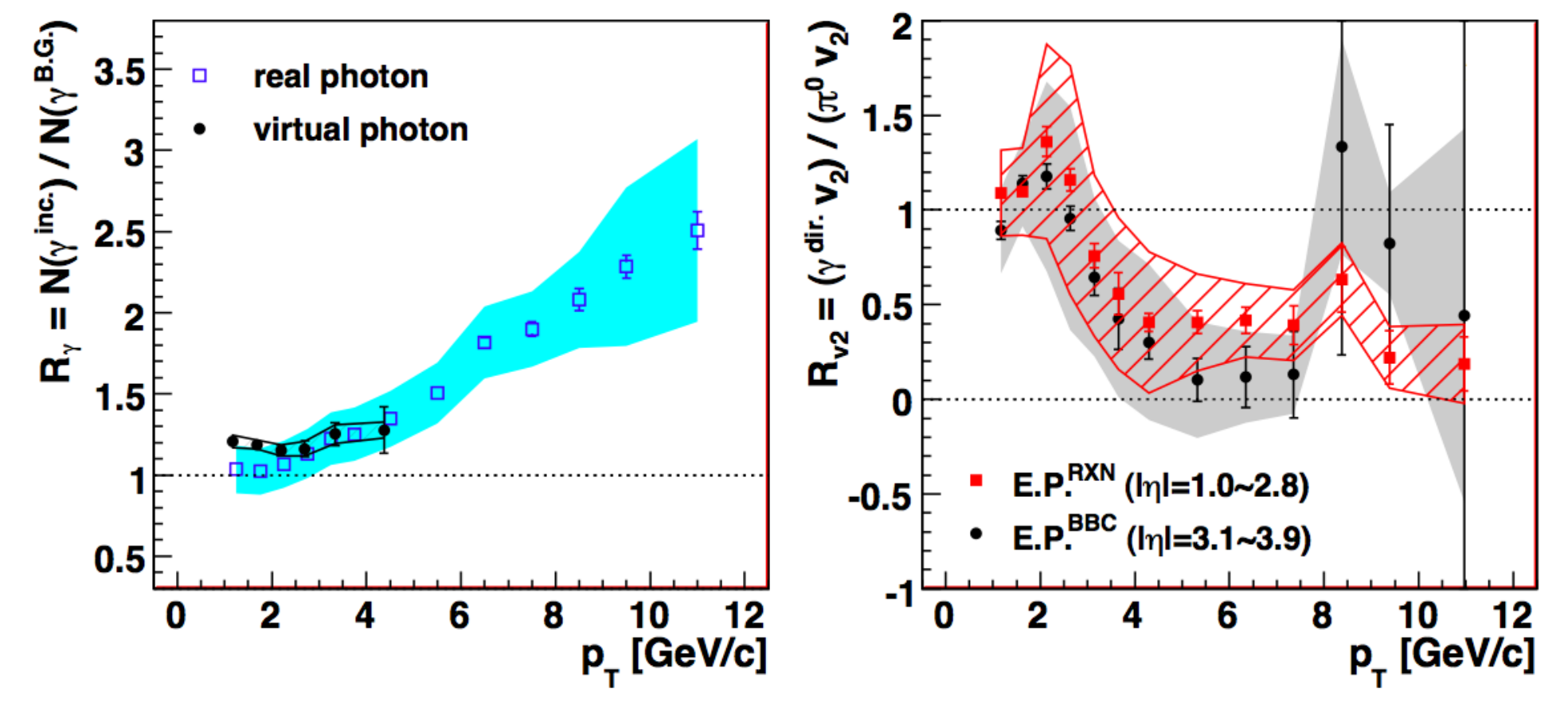}
\end{center}\vspace*{-2.0pc}
\caption[]{a) (left) $R_{\gamma}$ vs $p_T$~\cite{ppg126} for 
virtual photons (solid circles)~\cite{ppg086} and  
real photons (open squares)~\cite{ppg042} for minimum bias Au+Au collisions at $\sqrt{s_{NN}}=200$ GeV. b) (right) Ratio of direct-$\gamma$ $v_2$ to $\pi^0$ $v_2$ for two reaction plane detectors~\cite{ppg126}. }
\label{fig:dirgv2}
\end{figure}

Fig.~\ref{fig:dirgv2}a~\cite{ppg126}  
shows $R_{\gamma}$ for real photons, measured in an EM calorimeter, and virtual photons, which are $e^+ e^-$ pairs from internal conversion of the direct-$\gamma$, with $0.12< m_{ee}<0.30$ GeV/c$^2$ where there is no background from $\pi^0$ Dalitz decay. This reduces the background by a factor of $\geq1.19/0.19\approx 6$~\cite{egseeMJTISSP2009}, and allows the precision of $R_{\gamma}$ to be greatly improved as shown. 
Then, using the precise virtual photon $R_{\gamma}$     
with the much higher statistics inclusive real-$\gamma$ data, one can derive $v_2$ for direct-$\gamma$ from the measured $v_2$ of inclusive real-$\gamma$ compared to the measured $v_2$ of $\gamma$'s from $\pi^0$ and $\eta$ decay. 

The result (Fig.~\ref{fig:dirgv2}b)~\cite{ppg126} is that the $v_2$ of direct-$\gamma$ is large in the range $1\leq p_T\leq 3$ GeV/c but drops to zero for $p_T \geq 5$ GeV/c where the photons are produced from initial hard-scattering and do not interact with the medium so that they do not flow. Since thermal radiation is produced in the flowing medium, the observed large $v_2$ in the range $1\leq p_T\leq 3$ GeV/c, where the direct-$\gamma$ $p_T$ spectrum is exponential (Fig.~\ref{fig:Tshirt}b), confirms that these $\gamma$ are thermal radiation from the medium. What is very surprising is that the $v_2$ of the thermal photons is so large, the same or slightly greater than that of $\pi^0$'s. 

\section{A charming surprise}
\begin{figure}[!ht]
\begin{center} 
\begin{tabular}{cc}
\hspace*{-0.02\linewidth}\includegraphics*[width=0.51\linewidth]{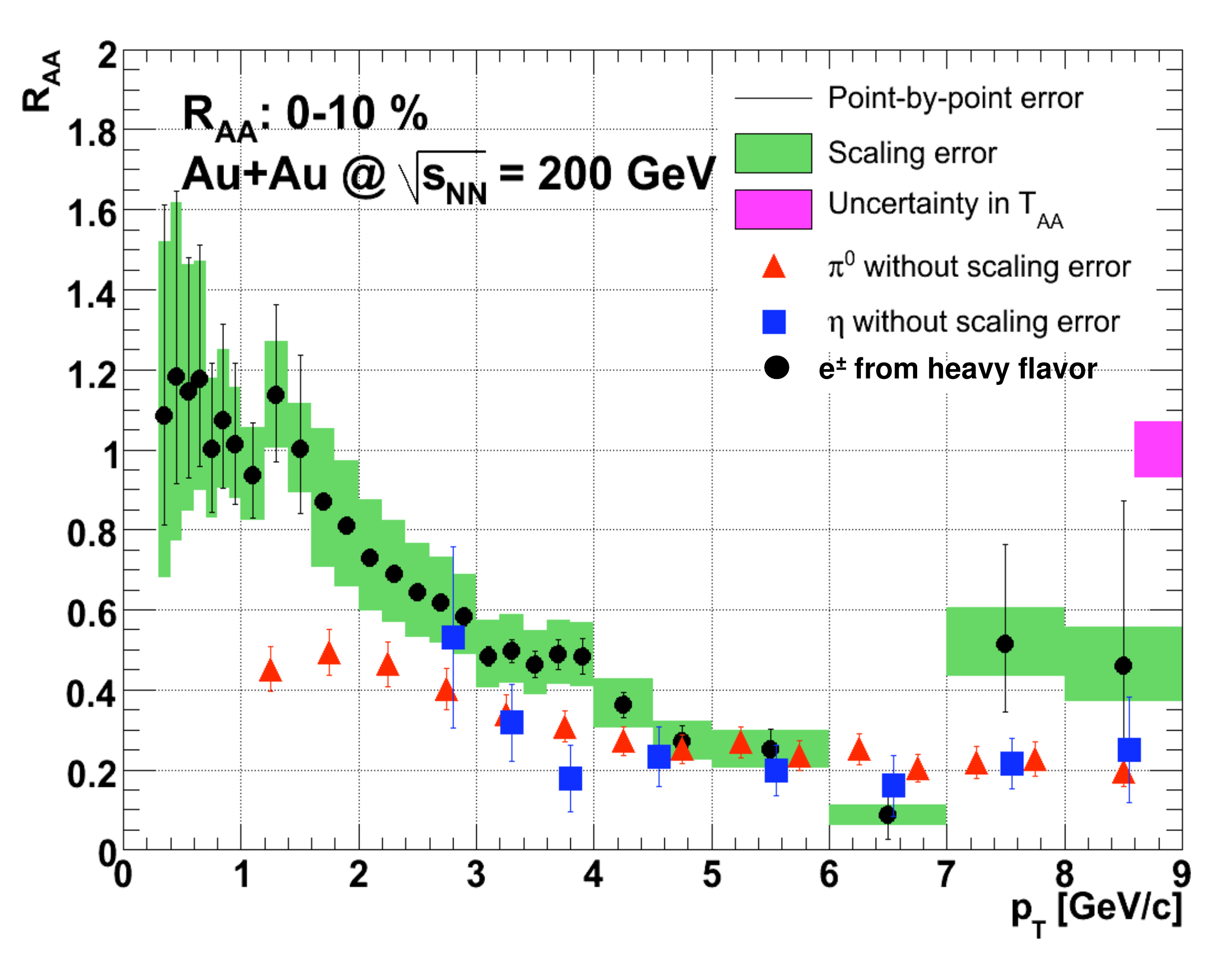} & 
\hspace*{-0.02\linewidth}\includegraphics[width=0.50\linewidth]{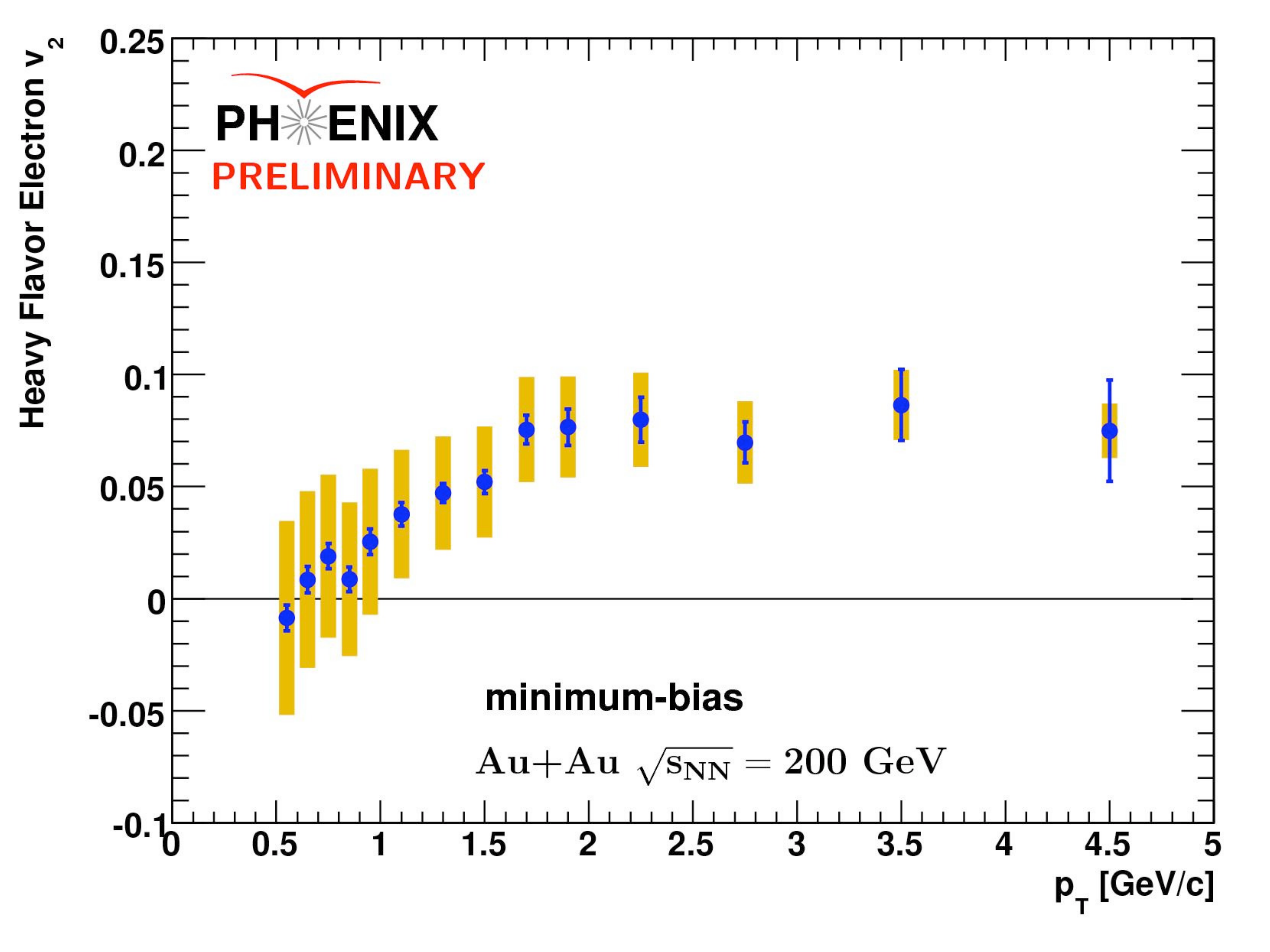} 
\end{tabular}
\end{center}\vspace*{-1.5pc}
\caption[]{PHENIX~\cite{PXcharmAA06}: a) (left) $R_{AA}$ (central Au+Au) b) (right) $v_2$ (minimum bias Au+Au) as a function of $p_T$ for direct-$e^{\pm}$ at $\sqrt{s_{NN}}=200$ GeV. }
\label{fig:fcrisis}
\end{figure}

  This past year, the ALICE experiment at LHC~\cite{ALICE-charm11} confirmed, with reconstructed charm mesons, the suppression of heavy quarks comparable to that of $\pi$ (from light quarks) for $p_T\gsim 4$ GeV/c as previously observed at RHIC using direct-single-$e^{\pm}$ from heavy quark ($c$, $b$) decay (Fig.~\ref{fig:fcrisis}a)~\cite{PXcharmAA06}. Also seen at RHIC is that heavy quarks exhibit collective flow ($v_2$) (Fig.~\ref{fig:fcrisis}b), another indication of a very strong interaction with the medium. 
The fact that heavy quarks are suppressed the same as light quarks strongly disfavors the \QCD\  energy-loss explanation of jet-quenching because, naively, heavy quarks should radiate much less than light quarks and gluons in the medium; but opens up a whole range of new possibilities including string theory~\cite{egsee066}. 

\section{Zichichi to the rescue?}
  In September 2007, I read an article by Nino, ``Yukawa's gold mine'' in the CERN Courier taken from his talk at the 2007 International Nuclear Physics meeting in Tokyo, Japan, in which he proposed:``We know that confinement produces masses of the order of a giga-electron-volt. Therefore, according to our present understanding, the QCD colourless condition cannot explain the heavy quark mass. However, since the origin of the quark masses is still not known, it cannot be excluded that in a QCD coloured world, the six quarks are all nearly massless and that the colourless condition is `flavour' dependent.'' 
  
  Nino's idea really excited me even though, or perhaps because, it appeared to overturn two of the major tenets of the Standard Model since it seemed to imply that: QCD isn't flavor blind;  the masses of quarks aren't given by the Higgs mechanism.  Massless $b$ and $c$ quarks in a color-charged medium would be the simplest way to explain the apparent equality of gluon, light quark and heavy quark suppression indicated by the equality of $R_{AA}$ for $\pi^0$ and direct single-$e^{\pm}$ in regions where both $c$ and $b$ quarks dominate. Furthermore RHIC and LHC-Ions are the only place in the Universe to test this idea. 
  
  Nino's idea seems much more reasonable to me than the string theory explanations of heavy-quark suppression (especially since they can't explain light-quark suppression). Nevertheless, just to be safe, I asked some distinguished theorists what they thought, among others~\cite{egseeMJTISSP2009}, Steve Weinberg. He said that he and Lenny Susskind had a model Technicolor (or Hypercolor) that worked well in the vector boson sector but didn't give mass to the fermions.

	 Nino proposed to test his idea by shooting a proton beam through a \QGP\ formed in a Pb+Pb collision at the LHC and seeing the proton `dissolved' by the \QGP. My idea is to use the new PHENIX VTX detector, installed in 2011, to  map out, on an event-by-event basis, the di-hadron correlations from identified $b-\overline{b}$ di-jets and identified $c-\overline{c}$ di-jets, which do not originate from the vertex, and light quark and gluon di-jets, which originate from the vertex and can be measured with $\pi^0$-hadron correlations. A steepening of the slope of the $x_E$ distribution of heavy-quark correlations as in Fig.~\ref{fig:AuAupp79}a will confirm in detail (or falsify) whether the different flavors of quarks behave as if they have the same energy loss (hence mass) in a color-charged medium. If Nino's proposed effect is true, that the masses of fermions are not given by the Higgs particle and all quarks are nearly massless in the \QGP, and we can confirm the effect at RHIC or LHC-Ions, this would be a case where we Relativistic Heavy Ion Physicists may have something unique to contribute at the most fundamental level to the Standard Model, which would constitute a ``transformational discovery.'' Of course the LHC could falsify this idea by finding the Higgs decay to $b-\bar{b}$ at the expected rate in p-p collisions. Clearly, there are exciting years ahead of us!


\begin{thebibliography}{99}
\bibitem{BearMountain} \href{http://www.osti.gov/energycitations/product.biblio.jsp?query_id=0&page=0&osti_id=4061527}{\it Report of the Workshop on BeV/Nucleon Coliisions of Heavy Ions---How and Why}, Bear Mountain, NY, 29 November--1 December 1974. (BNL-50445, Upton NY, 1975). 
\bibitem{seeMJTROP} See Ref.~\cite{MJTROP} for a more extensive list of references. 
\bibitem{MJTROP} M.~J.~Tannenbaum, \Journal{\RPP}{69}{2005--2059}{2006}.
\bibitem{Shuryak80} E.~V.~Shuryak, \Journal{\PLC} {61} {71--158}{1980}. 
\bibitem{LRP07} {\em The Frontiers of Nuclear Science}, \href{http://science.energy.gov/np/nsac/}{NSAC Long Range Plan 2007.}
\bibitem{BRWP} I.~Arsene {\it et al.}, BRAHMS Collab.  \Journal{\NPA}{757}{1--27}{2005}.
\bibitem{PHWP} B.~B.~Back {\it et al.}, PHOBOS Collab. \Journal{\NPA}{757}{28--101}{2005}.
\bibitem{STWP} J.~Adams {\it et al.}, STAR Collab. \Journal{\NPA}{757}{102--183}{2005}.
\bibitem{PXWP} K.~Adcox {\it et al.}, PHENIX Collab. \Journal{\NPA}{757}{184--283} {2005}.
\bibitem{THWPS} D.~Rischke and G.~Levin, eds., \Journal{\NPA}{750}{1--171}{2005}. 
\bibitem{CERNBaloney} Press Conference, {\it A New State of Matter Created at CERN} \href{http://newstate-matter.web.cern.ch/newstate-matter/}{http://newstate-matter.web.cern.ch/newstate-matter/Story.html}
\bibitem{NYT02102000} New York Times, February 10, 2000, Page 1, \href{http://www.nytimes.com/2000/02/10/world/particle-physicists-getting-closer-to-the-bang-that-started-it-all.html?scp=4&sq=A%20new%20state%20of%20matter&st=cse}  {\em Particle Physicists Getting Closer To the Bang That Started It All}. 
\bibitem{NSRL} NASA/BNL Space Radiation Program, \url{http://www.bnl.gov/medical/NASA/LTSF.asp}
\bibitem{CBAmagnetsNIMA235} E.~J.~Bleser, {\it et al.}, \Journal{\NIMA}{235}{435}{1985}.
\bibitem{Bozorth} R.~M.~Bozorth, {\it Ferromagnetism} (VanNostrand, New York, 1951)
\bibitem{NaturePhysics7} The ALPHA Collaboration, \href{http://dx.doi.org/10.1038/nphys2025}{\Journal{Nature Physics\ }{7}{558--564}{2011}}; G.~B.~Andresen, {\it et al.}, \Journal{Nature\ }{468}{673--676}{2010}.
\bibitem{IEEENS22} R.~Chasman, G.~K.~Green and E.~M.~Rowe, \Journal{IEEE Trans Nucl. Sci.\ }{NS-22}{1765--1767}{1975}.
\bibitem{T2KPR} Press Release, ``Indication of Electron Neutrino Appearance at the T2K Experiment'' June 15, 2011, \url{http://www.kek.jp/intra-e/press/2011/J-PARC_T2Kneutrino.html}
\bibitem{NSLSIIsource} NSLS-II Source Properties and Floor Layout \url{http://www.bnl.gov/ps/docs/pdf/SourceProperties.pdf}
\bibitem{PDG} C.~Amsler {\it et al.} (Particle Data Group), \Journal{\PLB}{667}{1}{2008}.
\bibitem{MatsuiSatz86} 
T.~Matsui and H.~Satz, \Journal{\PLB}{178}{416}{1986}. 
\bibitem{NA50EPJC39} B.~Alessandro, {\it et al.}, NA50 Collab., \Journal{\EPJC}{39}{335--345}{2005}. See also, F.~Prino, Proc. XXX International Symposium on  Multiparticle Dynamics, \href{http://arxiv.org/abs/hep-ex/0101052}{arXiv:hep-ex/0101052v1}.
\bibitem{E772} D.~M.~Alde, {\it et al.} E772 Collab., \Journal{\PRL}{66}{2285--2288}{1991}. See, also, M.~J.~Leitch, \Journal{\EPJC}{43}{157--160}{2005} and references therein.  
\bibitem{UA1} C.~Albajar, {\it et al.} UA1 Collab., \Journal{\PLB}{186}{237}{1987}. 
\bibitem{ppg019} S.~S.~Adler, {\it et al.} PHENIX Collab., \Journal{\PRC}{71}{034908}{2005}.
\bibitem{RHICNIM} {\it The Relativistic Heavy Ion Collider Project: RHIC and its Detectors}, \Journal{\NIMA}{499}{235--880}{2003}.
\bibitem{egseeMJTISSP2009} See, for example, M.~J.~Tannenbaum, Proc. Int. School Subnuclear Physics, ``The most  unexpected at LHC and the status of high energy frontier'', 47th Course, Erice-Sicily: 29 August - 7 September 2009, \href{http://arxiv.org/abs/1006.5701}{arXiv:1006.5701v1 [nucl-ex]}.
\bibitem{ALICEmult} K.~Aamodt, {\it et al.} ALICE Collab., \Journal{\PRL}{106}{032301}{2011}.
\bibitem{egseePT} The detector is so non-conventional that it made the cover of \href{http://www.phenix.bnl.gov/phenix/WWW/docs/covers/phystoday/2003oct/phystoday03oct.jpg}{Physics Today, October 2003}.
\bibitem{STARNature473} H.~Agakishiev, {\it et al.} STAR Collab., \Journal{Nature\ }{473}{353--356}{2011}. 
\bibitem{0909.0566} B.~I.~Abelev, {\it et al.} STAR Collab., \href{http://arxiv.org/abs/0909.0566v1}{arXiv:0909.0566v1 [nucl-ex]}.
 
\bibitem{LaceyQM05} R.~A.~Lacey, \Journal{\NPA}{774}{199--214}{2006}.
\bibitem{KanetaQM04} M.~Kaneta {\it et al.}, PHENIX Collab., \Journal{\JPG}{30}{S1217--S1220}{2004}.
\bibitem{PXArkadyQM06} A.~Adare, {\it et al.}, PHENIX Collab., \Journal{\PRL}{98}{162301}{2007}.

\bibitem{Ollitrault} J.-Y.~Ollitrault, \Journal{\PRD}{46}{229--245}{1992}; \Journal{\NPA}{638}{195c--206c}{1998}.
\bibitem{HeiselbergLevy} H.~Heiselberg and A.-M.~Levy, \Journal{\PRC}{59}{2716--2727}{1999}.
\bibitem{VoloshinQM02} S.~A.~Voloshin, \Journal{\NPA}{715}{379c--388c}{2003}.
\bibitem{TeaneyPRC68} D.~Teaney, \Journal{\PRC}{68}{034913}{2003}.
\bibitem{Kovtun05} P.~K.~Kovtun, D.~T.~Son, and A.~O.~Starinets, \Journal{\PRL}{94}{111601}{2005}. 
\bibitem{LaceyPRL98} R.~A.~Lacey, {\it et al.}, \Journal{\PRL}{98}{092301}{2007}. 
\bibitem{CKMPRL97} L.~P.~Csernai, J.~I.~Kapusta, and L.~D.~McLerran, \Journal{\PRL}{97}{152303}{2006}.
\bibitem{ppg083} A.~Adare, {\it et al.} (PHENIX Collab.), \Journal{\PRC}{78}{014901}{2008}; \Journal{\PRC}{77}{011901(R)}{2008}.
\bibitem{ppg067} A.~Adare, {\it et al.} (PHENIX Collab.), \Journal{\PRL}{98}{232302}{2007}.
\bibitem{CSST-Coney05} J.~Casalderrey-Solana, E.~V.~Shuryak and D. Teaney, \Journal{\JPCS}{27}{22-31}{2005}.
\bibitem{egseeppg083} For example, see Ref.~\cite{ppg083} for a discussion and list of references. 
\bibitem{JTMQM06} J.~T.~Mitchell, {\it et al.} (PHENIX Collab.), \Journal{\JPG}{34}{S911--S914}{2007}. 
\bibitem{STARridgePRC80} B.~I.~Abelev, {\it et al.} (STAR Collab.), \Journal{\PRC}{80}{064912}{2009}.
\bibitem{PutschkeHP06} J.~Putschke, {\it et al.} (STAR Collab.), \Journal{\NPA}{783}{507c--510c}{2007}.
\bibitem{JacobsHP04} P.~Jacobs, \Journal{\EPJC}{43}{467--473}{2005}.
\bibitem{AlverOllitrault} B.~H.~Alver, C.~Gombeaud, M.~Luzum, and J.~Y.~Ollitrault, \Journal{\PRC}{82}{034913}{2010}.
\bibitem{AlverRoland} B.~Alver and G.~Roland, \Journal{\PRC}{81}{054905}{2010}.
\bibitem{BrazilNuXuv3} J.~Takahashi, {\it et al.}, \Journal{\PRL}{103}{242301}{2009}. Also, see A.~P.~Mishra, {\it et al.}, \Journal{\PRC}{77}{064902}{2008}.  
\bibitem{EsumiQM11} S.~Esumi, {\it et al.} (PHENIX Collab.), \href{http://arxiv.org/abs/1110.3223v1}{arXiv:1110.3223v1 [nucl-ex]}.
\bibitem{MarekQM11} Presentation by M.~Ga\'zdzicki at \href{http://indico.cern.ch/confSpeakerIndex.py?confId=30248}{Quark Matter 2011}. Also see M.~Ga\'zdzicki, {\it et al.} (NA49 Collab.), \Journal{\JPG}{30}{S701--S708}{2004}.
\bibitem{LKBMQM11} Presentations by L.~Kumar and B.~Mohanty at \href{http://indico.cern.ch/confSpeakerIndex.py?confId=30248}{Quark Matter 2011}.
\bibitem{CleymansOeschlerPLB615} J.~Cleymans, H.~Oeschler, K.~Redlich and S.~Wheaton, \Journal{\PLB}{615}{50--54}{2005}.
\bibitem{LBLJune23} \url{http://newscenter.lbl.gov/news-releases/2011/06/23/when-matter-melts/}.
\bibitem{KochCFRNC06} e.g. see V.~Koch, \href{http://pos.sissa.it/archive/conferences/030/008/CFRNC2006_008.pdf}{PoS(CFRNC2006)008}
\bibitem{AsakawaHMPRL85} M.~Asakawa, U.~Heinz, and B.~M\"uller, \Journal{\PRL}{85}{2072}{2000}.
\bibitem{AMuellerPRD4} A.~H.~Mueller, \Journal{\PRD}{4}{150}{1971}. 
\bibitem{TarnowskyQM2011} Presentation by T.~J.~Tarnowsky at \href{http://indico.cern.ch/confSpeakerIndex.py?confId=30248}{Quark Matter 2011}. Also see T.~J.~Tarnowsky, {\it et al.} (STAR Collab.), \href{http://arxiv.org/abs/1106.6110v1}{arXiv:1106.6110v1 [nucl-ex]}.
\bibitem{STARnetPPRL105} M.~M.~Aggarwal, {\it et al.} (STAR Collab.), \Journal{\PRL}{105}{022302}{2010}.
\bibitem{Science332} S.~Gupta, X.~Luo, B.~Mohanty, H.~G.~Ritter and N.~Xu, \Journal{Science\ }{332}{1525}{2011}. 
\bibitem{ppg104} A.~Adare, {\it et al.} (PHENIX Collab.), \href{http://arxiv.org/abs/1105.1966v1}{arXiv:1105.1966v1 [nucl-ex]}.
\bibitem{QCDCompton} H.~Fritzsch and P.~Minkowski, \Journal{\PLB}{69}{316}{1977}.
\bibitem{CMSdijet} S.~ Chatrchyan, {\it et al.} (CMS Collab.), \Journal{\PRC}{84}{024906}{2011}.
\bibitem{ppg003} K.~Adcox, {\it et al.} (PHENIX Collab.), \Journal{\PRL}{88}{022301}{2002}.
\bibitem{ppg054} S.~S.~Adler, {\it et al.} (PHENIX Collab.), \Journal{\PRC}{76}{034904}{2007}.
\bibitem{ppg086} A.~Adare, {\it et al.} (PHENIX Collab.), \Journal{\PRL}{104}{132301}{2010}.
\bibitem{BDMPS} See R.~Baier, D.~Schiff and B.~G.~Zakharov, \Journal{\ARNPS}{50}{37--69}{2000}, and references therein. 
\bibitem{Tingpizff} B.~Adeva, {\it et al.} (L3 Collab.), \Journal{\PLB}{259}{199--208}{1991}.
\bibitem{BW06} N.~Borghini and U.~A.~Wiedemann, \Journal{\NPA}{774}{549--522}{2006}; see also {arXiv:hep-ph/0506218v1}.
\bibitem{ppg095} A.~Adare, {\it et al.} PHENIX Collab., \Journal{\PRD}{82}{072001}{2010}.
\bibitem{ConnorsHP2010} M.~Connors, {\it et al.} (PHENIX Collab.), \Journal{\NPA}{855}{335--338}{2011}. 
\bibitem{ppg029} S.~S.~Adler, {\it et al.} (PHENIX Collab.), \Journal{\PRD}{74}{072002}{2006}.
\bibitem{ppg106} A.~Adare, {\it et al.} (PHENIX Collab.), \Journal{\PRL}{104}{252301}{2010}.
\bibitem{MJT-Utrecht} M.~J.~Tannenbaum, {\it et al.} (PHENIX Collab.), \href{http://arxiv.org/abs/1109.0760v1}{arXiv:1109.1760v1 [nucl-ex]}. 
\bibitem{ppg042} S.~S.~Adler, {\it et al.} (PHENIX Collab.), \Journal{\PRL}{94}{232301}{2005}.
\bibitem{ppg126} A.~Adare, {\it et al.} (PHENIX Collab.), \href{http://arxiv.org/abs/1105.4126v2}{arXiv:1109.1105.4126v2  [nucl-ex]}.

\bibitem{ALICE-charm11} A.~Dainese, {\it et al.} (ALICE Collab.), \href{http://arxiv.org/abs/1106.4042v2}{arXiv:1109.6.4042v2  [nucl-ex]}.
\bibitem{PXcharmAA06} A.~Adare, {\it et al.}, PHENIX Collab., \Journal{\PRL}{98}{172301}{2007}.
\bibitem{egsee066} e.g. see Ref.~\cite{PXcharmAA06} for a list of references. 


\end{thebibliography}
\end{document}